\documentclass[paper]{JHEP3}

\usepackage{epsfig}
\usepackage{graphicx}
\usepackage{feynmp}

\usepackage{amsmath}
\usepackage{amsfonts}
\usepackage{graphicx}

\def\beq{\begin{equation}}
\def\eeq{\end{equation}}
\newcommand{\bea}{\begin{eqnarray}}
\newcommand{\eea}{\end{eqnarray}}

\def\Eqn#1{Eq.~(\ref{#1})}
\def\Ref#1{Ref.~\cite{#1}}

\def\sec#1{Section~{\ref{#1}}}

\def\bsp#1\esp{\begin{split}#1\end{split}}

\newcommand{\eps}{\epsilon}
\newcommand{\lam}{\lambda}

\newcommand{\ord}{\begin{cal}O\end{cal}}
\newcommand{\cI}{\begin{cal}I\end{cal}}
\newcommand{\cJ}{\begin{cal}J\end{cal}}

\newcommand{\cD}{{\cal D}}
\newcommand{\cP}{{\cal P}}
\newcommand{\cQ}{{\cal Q}}
\newcommand{\cM}{{\cal M}}

\newcommand{\mfunc}{\cM}

\newcommand{\halfD}{\frac{D}{2}}
\def\Dx{\int {\cal D}\alpha}
\def\Q{{\cal Q} }
\def\P{{\cal P} }
\def\M{{\cal M} }
\def\Id{I_n^D\Big(\{\nu_i\}; \{Q^2_i\}; \{M_i\}\Big)}

\newcommand{\eug}{\gamma_E}

\newcommand{\li}{\mathrm{Li}_2}

\newcommand{\ie}{\emph{i.e.}~}
\newcommand{\eg}{\emph{e.g.}~}
\newcommand{\etal}{\emph{et al.}~}
\newcommand{\etc}{\emph{etc.}~}
\newcommand{\rd}{\mathrm{d}}
\newcommand{\infsum}[1]{\sum_{#1 = 0}^{\infty}}
\newcommand{\ph}[2]{(#1)_{#2}}

\newcommand{\mbint}{\int_{-i\infty}^{+i\infty}}

\def\bit#1\eit{\begin{itemize}#1\end{itemize}}
\def\ben#1\een{\begin{enumerate}#1\end{enumerate}}

\newenvironment{sloppyequation}[0]{\sloppy\begin{flushleft}\hspace*{0.75cm}\(}{\)\end{flushleft}\fussy}
\newenvironment{sloppytext}[0]{\sloppy\begin{flushleft}}{\end{flushleft}\fussy}

\newcommand{\beqsloppy}{\begin{sloppyequation}}
\newcommand{\eeqsloppy}{\end{sloppyequation}}
\newcommand{\btxtsloppy}{\begin{sloppytext}}
\newcommand{\etxtsloppy}{\end{sloppytext}}

\title{The one-loop pentagon to higher orders in $\eps$}

\author{Vittorio Del Duca\\
Istituto Nazionale di Fisica Nucleare\\
Laboratori Nazionali di Frascati\\
00044 Frascati (Roma), Italy\\
       E-mail: \email{delduca@lnf.infn.it}}

\author{Claude Duhr\\
Institut de Physique Th\'eorique \&
Centre for Particle Physics and Phenomenology (CP3)\\
Universit\'e Catholique de Louvain\\
Chemin du Cyclotron 2,
B-1348 Louvain-la-Neuve, Belgium\\
E-mail: \email{claude.duhr@uclouvain.be}}

\author{E.~W.~Nigel~Glover\\
Institute for Particle Physics Phenomenology, 
University of Durham\\ Durham, DH1 3LE, U.K.\\
E-mail: \email{E.W.N.Glover@durham.ac.uk}}

\author{Vladimir A. Smirnov\\
Nuclear Physics Institute of Moscow State University\\
Moscow 119992, Russia\\
E-mail: \email{smirnov@theory.sinp.msu.ru}}

\abstract{
We compute the one-loop scalar massless pentagon integral $I_5^{6-2\eps}$ 
in $D=6-2\eps$ dimensions  in the limit of multi-Regge kinematics.  This integral first contributes to the parity-odd part of the one-loop $\begin{cal}N\end{cal}=4$ five-point MHV amplitude $m_5^{(1)}$ at ${\cal O}(\eps)$. In the high energy limit defined by
$s \gg s_1, ~s_2 \gg -t_1, -t_2$, the pentagon integral reduces to double sums or equivalently two-fold 
Mellin-Barnes integrals. 
By determining the ${\cal O}(\eps)$ contribution to  $I_5^{6-2\eps}$, one therefore gains knowledge of $m_5^{(1)}$ through to ${\cal O}(\eps^2)$ which is necessary for studies of the iterative structure of $\begin{cal}N\end{cal}=4$ SYM amplitudes beyond one-loop. One immediate application is the extraction of the one-loop gluon-production vertex through to ${\cal O}(\eps^2)$ and the iterative construction of the two-loop gluon-production vertex through to finite terms which is described in a companion 
paper~\cite{noi}.
The analytic methods we have used for evaluating the pentagon integral in the high energy limit may also be applied to the hexagon integral and may ultimately give information on the form of the $R_6^{(2)}$ remainder function.
}

\keywords{QCD, MSYM, small $x$}
\preprint{IPPP/09/25\\ CP3-09-15}

\begin{document}

\section{Introduction}
\label{sec:intro}

In the planar $\begin{cal}N\end{cal}=4$ supersymmetric Yang-Mills theory,
Anastasiou, Bern, Dixon and Kosower (ABDK)~\cite{Anastasiou:2003kj} proposed an iterative 
structure for the colour-stripped two-loop scattering amplitude with an arbitrary number $n$ of external 
legs in a maximally-helicity violating (MHV) configuration. 
The proposed iteration formula for the two-loop MHV amplitude $m_n^{(2)}(\eps)$ is
\beq
m_n^{(2)}(\eps) = \frac{1}{2} \left[m_n^{(1)}(\eps)\right]^2
+ f^{(2)}(\eps)\, m_n^{(1)}(2\eps) + C^{(2)} + \ord(\eps)\, ,\label{eq:ite2bds}
\eeq
thus the two-loop amplitude is determined in terms of 
the one-loop MHV amplitude $m_n^{(1)}(\eps)$ evaluated through to $\ord(\eps^2)$
in the dimensional-regularisation parameter $\eps$,
a constant, $C^{(2)}$, and a function, $f^{(2)}(\eps)$, which is related to the 
cusp~\cite{Korchemsky:1987wg,Beisert:2006ez} and 
collinear~\cite{Magnea:1990zb,Sterman:2002qn} anomalous dimensions.

Subsequently, Bern, Dixon and Smirnov (BDS) proposed an all-loop resummation 
formula~\cite{Bern:2005iz} for the colour-stripped $n$-point MHV amplitude, which implies a tower
of  iteration formul\ae, allowing one to determine the $n$-point amplitude at a given number of
loops in  terms of amplitudes with fewer loops, evaluated to higher orders of $\eps$. BDS checked
that the ansatz was correct for the three-loop four-point amplitude, by evaluating analytically 
$m_4^{(3)}(\eps)$ through to finite terms, as well as $m_4^{(2)}(\eps)$ through to $\ord(\eps^2)$ and
$m_4^{(1)}(\eps)$ through to $\ord(\eps^4)$. 
The BDS ansatz has been proven to be correct also for the
two-loop five-point  amplitude~\cite{Bern:2006vw,Cachazo:2008vp}, for which $m_5^{(2)}(\eps)$ has
been computed numerically through to finite terms, as well as $m_5^{(1)}(\eps)$ through to $\ord(\eps^2)$.

In the strong-coupling limit, Alday and Maldacena  showed that the ansatz is violated for amplitudes with six or more legs~\cite{Alday:2007he}. This provoked the numerical calculation of $m_6^{(2)}(\eps)$ through to finite terms and
of $m_6^{(1)}(\eps)$ through to $\ord(\eps^2)$, where
the BDS ansatz was demonstrated to fail in Ref.~\cite{Bern:2008ap}, and where
it was shown that the finite pieces of the parity-even part of $m_6^{(2)}(\eps)$
are incorrectly determined by the ansatz\footnote{It was subsequently shown that
the parity-odd part of $m_6^{(2)}(\eps)$ does satisfy the
ansatz~\cite{Cachazo:2008hp}.}. In particular, it was shown that the remainder function,
\beq
R_n^{(2)} = m_n^{(2)}(\eps) - \frac{1}{2} \left[m_n^{(1)}(\eps)\right]^2
- f^{(2)}(\eps)\, m_n^{(1)}(2\eps) - C^{(2)}\, ,\label{eq:discr}
\eeq
is different from zero for $n = 6$,
where $R_n^{(2)}$ may be a function of the kinematical parameters of the $n$-point amplitude,
but a constant with respect to $\eps$. Because the calculation of Ref.~\cite{Bern:2008ap}
is numerical, the analytic form of $R_6^{(2)}$ is unknown.

Using the AdS/CFT correspondence, Alday and Maldacena~\cite{Alday:2007hr} showed that in the
strong-coupling limit planar scattering amplitudes exponentiate like in the BDS ansatz, and 
suggested that in the weak-coupling regime the vacuum expectation value of the $n$-edged Wilson
loop could be related to the $n$-point MHV amplitude in $\begin{cal}N\end{cal}=4$ super  Yang-Mills
theory. The agreement (up to a constant) between the light-like Wilson loop and the  (parity-even
part of the) MHV amplitude has been verified for the one-loop four-edged~\cite{Drummond:2007aua}
and $n$-edged~\cite{Brandhuber:2007yx} Wilson loops, and for the two-loop 
four-edged~\cite{Drummond:2007cf}, five-edged~\cite{Drummond:2007au} and 
six-edged~\cite{Drummond:2007bm,Drummond:2008aq} Wilson loops. Furthermore, it was shown that the
light-like Wilson loop exhibits a conformal symmetry, and that the BDS ansatz is a solution of
the ensuing Ward identities, up to functions of conformally invariant cross-ratios, which are
present for $n\ge 6$~\cite{Drummond:2007au}. Recently, also the two-loop seven-edged and
eight-edged Wilson loops have been computed  numerically~\cite{Anastasiou:2009kn}, although no
corresponding $\begin{cal}N\end{cal}=4$  MHV amplitudes are known. Thus, also the remainder
functions $R_7^{(2)}$ and $R_8^{(2)}$  (for the Wilson loops, as well as for the MHV amplitudes, if
we suppose that the agreement between them and the Wilson loops holds for seven or more points) are
known numerically, and the numerical evidence~\cite{Anastasiou:2009kn} seems to confirm that they
are functions of conformally invariant cross-ratios only. However, their analytic form is unknown.

Through to finite terms, one-loop $\begin{cal}N\end{cal}=4$ MHV amplitudes, $m_n^{(1)}(\eps)$, are
parity-even and can be expressed in terms of box functions, which are massless for $n=4$,
one-mass for $n=5$ and two-mass easy box functions for $n\ge 6$~\cite{Bern:1994zx}.
For $n\ge 5$, parity-odd contributions and one-loop $n$-edged polygons occur in
the higher orders in $\eps$~\cite{Bern:1996ja}. For example, 
an irreducible massless pentagon occurs in the parity-odd part
of the one-loop five-point amplitude, which to all orders in $\eps$ is~\cite{Bern:2006vw}
\beq
m_5^{(1)} = - \frac{1}{4} \sum_{\rm cyclic} s_{12} s_{23} I_4^{1m}(1,2,3,45,\eps) - 
\frac{\eps}{2} \eps_{1234} I_5^{6-2\eps}(\eps)\, ,\label{eq:pent}
\eeq
where $m_5^{(1)}$ denotes the one-loop coefficient, \emph{i.e.}, the one-loop amplitude rescaled by the tree-level amplitude, and where the cyclicity is over $i=1,\ldots,5$, $I_4^{1m}(1,2,3,45,\eps)$ is the one-mass box 
with the massive leg of virtuality $s_{45}$, $I_5^{6-2\eps}(\eps)$
is a pentagon evaluated in $6-2\eps$ dimensions, and the parity-odd factor is
$\eps_{1234}= {\rm tr}[\gamma_5\!\!\not\!k_1\!\!\not\!\!k_2\!\not\!\!k_3\!\!\not\!k_4]$.
A massless hexagon occurs in the parity-even 
part~\cite{Bern:2008ap,Bern:1996ja}, as well as a pentagon in the parity-odd 
part~\cite{Cachazo:2008hp,Bern:1996ja}, of the one-loop six-point amplitude at $\ord(\eps)$.
The hexagon may be reduced to one-mass pentagons.
Thus, it seems fair to guess that the analytic form of the remainder function $R_6^{(2)}$
may be somewhat linked to the properties of the special functions occurring in the hexagon
of the one-loop six-point amplitude. 

Because very little is known in the literature about the analytic properties of pentagons and hexagons
to higher orders in $\eps$, in this paper we perform a first study of the pentagon occurring in
the  parity-odd part of the one-loop five-point amplitude (\ref{eq:pent}). A massless pentagon can
in general be reduced to quadruple sums dependent on the four independent ratios of the kinematic
invariants. In order to simplify matters, we choose a particular kinematic realm, the {\it
multi-Regge} kinematics, where the quadruple sums reduce to double sums, which we proceed to study
and expand to $\ord(\eps^2)$.

Our paper is organised as follows.   As a prelude to our main objective which is the evaluation of
the scalar massless pentagon integral in $D=6-2\eps$ dimensions in the multi-Regge limit, we first
give a short review of multiple hypergeometric functions in Section~2.  In general one may
encounter a multiple hypergeometric function in several different representations - as a multiple
sum,  an Euler integral, a Mellin-Barnes representation or a Laplace integral. We then discuss in
Section~3 different representations of Feynman integrals and show how they naturally  match
onto the various representations for hypergeometric functions.   For example,  in the Negative
Dimension approach (NDIM)~\cite{Halliday:1987an,Ricotta:1989ia} based on Schwinger parameters, one naturally finds the hypergeometric functions in
the form of multiple sums.   On the other hand,  starting from Feynman parameters and utilising the
Mellin-Barnes identity one naturally finds the hypergeometric functions as multiple Mellin-Barnes
integrals.  The one-loop pentagon integral and multi-Regge kinematics are defined in Section~\ref{sec:pentagon}.  In
Euclidean kinematics,  there are three distinct regions which we call I, II(a) and II(b).   Regions
II(a) and II(b) are related by symmetry.   Regions I and II(a) are related by analytic
continuation.    We use two distinct methods to compute the massless pentagon integral in
$D=6-2\eps$ dimensions in each of these regions and show (a) that the solutions in each region are
the same for the two approaches and (b) that the solutions in region I and II(a) are related by
analytic continuation.       In Section~\ref{sec:PentNDIM}, we use NDIM to write
the integral as multiple hypergeometric sums.   In the multi-Regge kinematics these sums collapse to
hypergeometric sums of two variables.   It is straightforward to make a Laurent expansion around
$\eps =0$ and find that hypergeometric sums can be written in terms of transcendental double sums
which we denote by ${\cal M}$ functions.   In Section~\ref{sec:PentMB}, we revisit the problem starting from the
Mellin-Barnes representation of the pentagon integral.   In the multi-Regge kinematics we obtain
two Mellin-Barnes integrals.    We show that by summing the residues, one recovers exactly the same
hypergeometric sums as in NDIM.   Alternatively, by performing the Mellin-Barnes integrals sequentially one can write the pentagon integral in terms of Goncharov's
polylogarithms. In Section~\ref{sec:ACphys} we perform the analytic continuation of the pentagon to the physical region where all $s$-type invariants are positive. Appendices with more details concerning the generalised hypergeometric functions,
nested harmonic sums, ${\cal M}$ functions and Goncharov's multiple polylogarithms we encounter in
the calculation are enclosed.   Further appendices contain some of the technical details omitted
from Sections~\ref{sec:PentNDIM} and~\ref{sec:PentMB} of the main text.

\section{Short review of hypergeometric functions}

In this section we briefly review the different representations of hypergeometric functions. We focus on Gauss' hypergeometric function,
\beq\label{eq:2F1}
{_2F_1}(a,b,c;x) = \infsum{n} {\ph{a}{n}\ph{b}{n}\over \ph{c}{n}}\, {x^n\over n!},
\eeq
and for generalized hypergeometric functions we refer to Appendix~\ref{app:HypGeo} and to the literature~\cite{Exton, Prudnikov, Erdelyi}. \Eqn{eq:2F1} defines the series representation of the hypergeometric function. The series is absolutely convergent inside the unit disc $|x|<1$. The hypergeometric \emph{function} however is defined over the complex plane, so we need to analytically continue the series~(\ref{eq:2F1}) outside the unit disc according to the prescription $x\to 1/x$. Furthermore, in physical applications, one often encounters special cases where the indices $a$, $b$ and $c$ are of the form $\alpha+\beta\eps$, $\alpha,\beta$ being integers\footnote{In some applications, also half-integer values can be found. In the following we restrict the discussion to integer values.}, and it is desired to have the hypergeometric function in the form of a Laurent expansion up to a given order in $\eps$. We review in the following several other representations of the hypergeometric function, sometimes more suitable for specific problems.

Let us first turn to the \emph{Euler integral representation},
\beq\label{eq:2F1Euler}
{_2F_1}(a,b,c;x) = {\Gamma(c)\over\Gamma(b)\Gamma(c-b)}\,\int_0^1\rd t\, t^{b-1}\,(1-t)^{c-b-1}\,(1-xt)^{-a}.
\eeq
This identity follows immediately from the fact that for $|x|<1$, we can insert the series expansion of $(1-xt)^{-a}$ in the integrand,
\beq\bsp
{\Gamma(c)\over\Gamma(b)\Gamma(c-b)}&\,\int_0^1\rd t\, t^{b-1}\,(1-t)^{c-b-1}\,(1-xt)^{-a}\\
=&\, \infsum{n}{\ph{a}{n}\over n!}\,x^n\,{\Gamma(c)\over\Gamma(b)\Gamma(c-b)}\,\int_0^1\rd t\, t^{b+n-1}\,(1-t)^{c-b-1}\\
= &\,\infsum{n}{\ph{a}{n}\over n!}\,x^n\,{\Gamma(c)\over\Gamma(b)\Gamma(c-b)}\,{\Gamma(b+n)\Gamma(c-b)\over\Gamma(c+n)}\\
=&\,{_2F_1}(a,b,c;x),
\esp
\eeq
and the identity follows by analytic continuation for $|x|>1$. Note however that the Euler integral representation is meaningless if $x$ is real and greater than 1 and $a$ is a positive integer, since in that case the integral~(\ref{eq:2F1Euler}) is divergent. The Euler integral has the nice property that in the situation where all indices are of the form $\alpha+\beta\eps$, the expansion in $\eps$ can be easily performed using integration-by-parts identities and the Laporta algorithm~\cite{Aglietti:2008fe}. Indeed, since $t$ and $(1-t)$ vanish at the integration limits, we can write,
\beq
\int_0^1\rd t\,{\partial\over\partial t}\,\Big(t^{b-1}\,(1-t)^{c-b-1}\,(1-xt)^{-a}\Big) =0,
\eeq
and carrying out the derivative on the integrand generates a set of recursive relations for the hypergeometric function. Using the Laporta algorithm we can solve the recursion and express every integral of Euler type in terms of a small set of master integrals. We can write down a set of differential equations for the master integrals that can be solved order by order in $\eps$ in terms of harmonic polylogarithms (See Appendix~\ref{app:hpl} for a review of generalized polylogarithms). Note that in the case of Gauss' hypergeometric function this procedure might look like an overkill, since we could as well expand the Pochhammer symbols in the series~(\ref{eq:2F1})  and sum the resulting series in terms of nested sums. However, for more general hypergeometric functions, this naive approach might lead to series for which the sum is not necessarily known.

A third way of representing a hypergeometric function is in terms of a Mellin-Barnes integral,
\beq\label{eq:2F1mb}
{_2F_1}(a,b,c;x) = {\Gamma(c)\over\Gamma(a)\Gamma(b)}\,\mbint\rd z\,\Gamma(-z)\,{\Gamma(a+z)\Gamma(b+z)\over \Gamma(c+z)}\,(-x)^z,
\eeq
where the contour is chosen according to the usual prescription, \ie the contour should separate the poles in $\Gamma(-z)$ from the poles in $\Gamma(\ldots+z)$. The identity~(\ref{eq:2F1mb}) follows simply from the fact that for $|x|<1$ we can close the integration contour to the left and sum up the residues of $\Gamma(-z)$, which immediately results in the series~(\ref{eq:2F1}). The Mellin-Barnes representation is however much more general, because the integral~(\ref{eq:2F1mb}) is valid even for $|x|>1$, in which case it suffices to close the contour to the right. In this sense, the integral~(\ref{eq:2F1mb}) really represents the hypergeometric \emph{function}, because it is valid in all the regions\footnote{In \Ref{Appell}, Appell and Kamp\'e de F\'eriet propose for this reason to \emph{define} the hypergeometric function through the Mellin-Barnes integral~(\ref{eq:2F1mb}).}. This property gives us at the same time an effective way to perform the analytic continuation of the hypergeometric function outside the unit disc. Closing the integration contour to the right and taking residues at $z=-a-n$ and $z=-b-n$, $n\in \mathbb{N}$, we find
\beq\bsp
{_2F_1}(a,b,c;x) =&\, {\Gamma(c)\Gamma(b-a)\over \Gamma(b)\Gamma(c-a)}\,(-z)^{-a}\,{_2F_1}(a,1+a-c,1+a-b;{1/z}) \\
+&\,{\Gamma(c)\Gamma(a-b)\over \Gamma(a)\Gamma(c-b)}\,(-z)^{-b}\,{_2F_1}(b,1+b-c,1+b-a;1/z),
\esp\eeq
where we used the formula
\beq
\Gamma(a-n) = (-1)^n\,{\Gamma(a)\over \ph{1-a}{n}}.
\eeq

Finally, let us also mention the \emph{Laplace integral representation},
\beq\label{eq:2F1Laplace}
{_2F_1}(a,b,c;x) = {\Gamma(c)\over\Gamma(b)}\,\int_0^\infty\rd t_1\,\rd t_2\,e^{-t_1-t_2}\,t_1^{b-1}\,t_2^{c-1}\,(1-t_1\,t_2\,z)^{-a},
\eeq
which follows directly from the integral representation of the $\Gamma$ function,
\beq
\Gamma(z)=\int_0^\infty\rd t\,e^{-t}\,t^{z-1}.
\eeq
Since however in the following we do not use this representation explicitly, we do not comment on it further, but only quote it for completeness.

The lesson to learn from this is that depending on what one wants to do, a given representation of the same hypergeometric function might be more convenient than another.


\section{Feynman integrals}
In this section we review the different representations for $n$-point scalar one-loop Feynman integrals in $D=D_0-2\eps$ dimensions, $D_0$ being a positive integer,
\beq\label{eq:FeynmanIntegral}
I_n^D\Big(\{\nu_i\}; \{Q^2_i\}; \{M_i\}\Big) = e^{\eug\eps}\,\int{\rd^Dk\over i\pi^{D/2}}\, \prod_{i=1}^n\,{1\over D_i^{\nu_i}},
\eeq
where the external momenta $k_i$ are incoming such that $\sum_{i=1}^nk_i=0$ and the propagators have the form
\beq\bsp
D_1=&\,k^2-M_1^2+i0,\\
D_i=&\, \left(k+\sum_{j=1}^{i-1}k_j\right)^2-M_i^2+i0, \qquad i=2,\ldots,n.
\esp\eeq
The external momentum scales are the Mandelstam variables $Q_i^2$, and we work in the Euclidean region, $Q_i^2 < 0$.

Feynman integrals can often be expressed in terms of (generalized) hypergeometric functions. In the previous section we showed that different representations of hypergeometric functions are useful to derive different properties. In the following we will argue that the different parametrizations used to evaluate Feynman integrals, Schwinger and Feynman parameters, series and Mellin-Barnes representations, are the equivalents to the four representations of the representations of the hypergeometric function, and switching from one parametrization to another might allow one to obtain valuable information about the Feynman integral. In the rest of this section we briefly review the different parametrizations, and in the rest of this work we give two examples how representations for Feynman integrals can be combined when computing Feynman integrals.

\subsection{The Schwinger parametrization}
The Schwinger parametrization\footnote{Also known as $\alpha$ parametrization.} is based on the identity
\beq\label{eq:Schwinger}
{1\over D_i^{\nu_i}} = {(-1)^{\nu_i}\over \Gamma(\nu_i)}\,\int_0^\infty\rd \alpha_i\,\alpha_i^{\nu_i-1}\,e^{\alpha_i\,D_i},
\eeq
Note that we explicitly derived \Eqn{eq:Schwinger} in Euclidean space, where we used the fact that in the Euclidean region $D_i<0$ in order to get a convergent integral. The corresponding relation in Minkowski space is similar, up to some factors of $i$ in the exponent.
Inserting \Eqn{eq:Schwinger} into the loop integral~(\ref{eq:FeynmanIntegral}), we obtain,
\beq\label{eq:Schwinger2}
I_n^D\Big(\{\nu_i\}; \{Q^2_i\}; \{M_i\}\Big) = \int\cD \alpha\,\int{\rd^Dk\over i\pi^{D/2}}\, \exp\Big(\sum_{i=1}^n\alpha_i\,D_i\Big),
\eeq
where we introduced the shorthand
\beq
\int\cD \alpha =e^{\eug\eps}\,\prod_{i=1}^n\,{(-1)^{\nu_i}\over \Gamma(\nu_i)}\,\int_0^\infty\,\rd \alpha_i\,\alpha_i^{\nu_i-1},
\eeq
and performing the Gaussian integral in \Eqn{eq:Schwinger2} leads to
\beq\label{eq:Schwinger4}
I_n^D\Big(\{\nu_i\}; \{Q^2_i\}; \{M_i\}\Big) =\int\cD \alpha\,{1\over \cP^{D/2}}\, \exp(\cQ/\cP)\, \exp(-\cM).
\eeq
The quantities $\cP$, $\cQ$ and $\cM$ are polynomials in the Schwinger parameters $\alpha_i$, the internal masses $M_i^2$ and the momentum scales $Q_i^2$,
\beq\bsp\label{eq:Schwinger3}
\cP=\,& \sum_{i=1}^n\alpha_i,\\
\cQ=\,& \sum_{i=1}^{n-1}\sum_{j=i+1}^n\alpha_i\alpha_j\,\left(\sum_{l=i}^{j-1}k_l\right)^2,\\
\cM=\,& \sum_{i=1}^n\alpha_i\,M_i^2.
\esp\eeq 
Note that the polynomials can directly be read off from the Feynman diagram in terms of trees and two-trees~\cite{Bogoliubov,Itzykson,SmirnovBook2, SmirnovBook}.

The Schwinger parametrization is the starting point of the NDIM method, which we describe in the next section. Note however at this point the formal similarity between the Schwinger parametrization~(\ref{eq:Schwinger3}) and the Laplace integral representation of the hypergeometric function, \Eqn{eq:2F1Laplace}.

\subsection{The Negative Dimension approach}

The crucial point in the NDIM approach is that the
Gaussian integral (\ref{eq:Schwinger2}) is an analytic function of the
space-time dimension.  Hence it is possible to consider $D < 0$ and to
make the definition \cite{Halliday:1987an,Ricotta:1989ia}
\begin{equation}
\int \frac{d^Dk}{i\pi^{D/2}} \, (k^2)^n = n! \;\delta_{n+\halfD,\, 0}
\label{eq:def}
\end{equation}
for positive values of $n$.

For the one-loop integrals we are interested in here, we view
Eqs.~(\ref{eq:Schwinger2}) and~(\ref{eq:Schwinger4}) as existing in negative
dimensions.  
Making the same series expansion of the exponential as above,
Eq.~(\ref{eq:Schwinger2}) becomes
\begin{eqnarray}
\Id &=& \Dx
\sum_{n_1,\ldots,n_n=0}^{\infty}
\int 
\frac{d^Dk}{i\pi^{D/2}}
\prod_{i=1}^{n} \frac{(x_iD_i)^{n_i}}{n_i!}\nonumber \\
&=& 
\Dx
\sum_{n_1,\ldots,n_n=0}^{\infty}
I_n^D\Big(-n_1,\ldots,-n_n;\{Q_i^2\},\{M_i^2\}\Big)
\prod_{i=1}^n\frac{ x_i^{n_i}}{n_i!},\phantom{aaaa}
\label{eq:LHS}
\end{eqnarray}
where the $n_i$ are positive integers.  The target loop integral is an
infinite sum of (integrals over the Schwinger parameters of) 
loop integrals with negative powers of the propagators.

Likewise, we expand the exponentials in Eq.~(\ref{eq:Schwinger4})
\begin{equation}
\label{eq:lhs_expans}
\Id =\Dx
\sum_{n=0}^{\infty}
\frac{\Q^n \P^{-n-\halfD}}{n!} 
\sum_{m=0}^{\infty}
\frac{(-\M)^m}{m!}, 
\end{equation}
and introduce the integers $q_1, \ldots, q_q$, $p_1, \ldots, p_n$
and $m_1,\ldots,m_n$  
to make 
multinomial expansions of $\Q$, $\P$ and $\M$ respectively
\begin{eqnarray}
\label{eq:mult_expans}
\Q^n &=& \sum_{q_{1},\ldots,q_{q}=0}^{\infty}  \frac{\Q_1^{q_1}}{q_1!}
\ldots \frac{\Q_q^{q_q}}{q_q!}\, (q_1+\ldots+q_q)! ,\nonumber \\
\P^{-n-\halfD} &=& \sum_{p_1,\ldots,p_n=0}^{\infty}  \frac{\alpha_1^{p_1}}{p_1!}
\ldots \frac{\alpha_n^{p_n}}{p_n!} \, (p_1+\ldots+p_n)!  ,\\
(-\M)^m &=& \sum_{m_1,\ldots,m_n=0}^{\infty}  \frac{(-\alpha_1M_1^2)^{m_1}}{m_1!}
\ldots \frac{(-\alpha_nM_n^2)^{m_n}}{m_n!} \, (m_1+\ldots+m_n)!,\nonumber
\end{eqnarray}
subject to the constraints 
\begin{eqnarray}
\label{eq:constraints}
\sum_{i=1}^q q_i &=& n,  \nonumber \\
\sum_{i=1}^n p_i &=& -n-\halfD ,\nonumber \\
\sum_{i=1}^n m_i &=& m,\nonumber \\
p_1+\ldots + p_n +q_1 +\ldots + q_q &=& -\frac{D}{2}.
\end{eqnarray}

Altogether, Eqs.~(\ref{eq:lhs_expans}) and~(\ref{eq:mult_expans}) give
\begin{eqnarray}
\label{eq:RHS}
\lefteqn{\Id =} \nonumber \\
&&\Dx
\sum_{{p_1,\ldots,p_n =0 \atop {q_1,\ldots,q_q =0 \atop
m_1,\ldots,m_n=0 }}}^{\infty}
\frac{\Q_1^{q_1}\ldots \Q_q^{q_q}}{q_1!\ldots q_q!}
\frac{\alpha_1^{p_1}\ldots
\alpha_n^{p_n}}{ p_1!\ldots p_n!}
\frac{(-\alpha_1M_1^2)^{m_1}}{m_1!}
\ldots \frac{(-\alpha_nM_n^2)^{m_n}}{m_n!}
\, (p_1+\ldots+p_n)!, \nonumber \\
\end{eqnarray}
with the constraints expressed by Eq.~(\ref{eq:constraints}).

We recall that each of the $\Q_i$ is a bilinear in the Schwinger parameters, so
that the target loop integral is now an infinite sum of powers of the 
scales of the process (with each of the $M_i^2$ and the $Q_i^2$ raised to a
different summation variable) integrated over the Schwinger parameters.

Equations~(\ref{eq:LHS}) and~(\ref{eq:RHS}) are two different expressions
for the same quantity: $I_n^D$.   Matching up powers of the Schwinger parameters,  
we obtain an expression for the 
loop integral with negative powers of the propagators in negative
dimensions
\begin{eqnarray}
\label{eq:sum}
\Id   
&\equiv&
e^{\gamma_E\eps}\,\sum_{{p_1,\ldots,p_n =0 \atop {q_1,\ldots,q_q =0 \atop
m_1,\ldots,m_n=0 }}}^{\infty}
(Q_1^2)^{q_1}\ldots(Q_q^2)^{q_q}\,
(-M_1^2)^{m_1}\ldots (-M_n^2)^{m_n}\nonumber \\
&\times&\left(
\prod_{i=1}^{n} \frac{\Gamma(1-\nu_i)}{\Gamma(1+m_i)\Gamma(1+p_i)}
\right)
\left(
\prod_{i=1}^{q} \frac{1}{\Gamma(1+q_i)}
\right)
\Gamma\left(1+\sum_{k=1}^n p_k\right), \nonumber \\
\end{eqnarray}
subject to the constraints Eq.~(\ref{eq:constraints}).
This is the main result of the negative dimension approach. 
The loop integral is written directly as an infinite sum.  Given
that $\Q$ can be read off directly from the Feynman graph, so can the precise
form of Eq.~(\ref{eq:sum}) as well as the system of constraints. Of course,
strictly speaking we have assumed that both $\nu_i$ and $D/2$ are negative
integers and we must be careful in interpreting this result in the physically
interesting domain where the $\nu_i$ and $D$ are all positive. Furthermore,  of the many possible solutions, only those that converge in the appropriate kinematic region should be retained.  

\subsection{The Feynman parametrization}

The Feynman parametrization is based on the identity
\beq
\prod_{i=1}^n\,{1\over D_i^{\nu_i}}={\Gamma(\nu)\over \Gamma(\nu_1)\ldots\Gamma(\nu_n)}\,\prod_{i=1}^n\,\int_0^1\rd x_i\, x_i^{\nu_i-1}\,{\delta(1-x_1\ldots-x_n)\over (x_1D_1+\ldots+x_nD_n)^\nu},
\eeq
with $\nu=\sum_{i=1}^n\nu_i$. Inserting this relation into \Eqn{eq:FeynmanIntegral}, we obtain the following representation for the Feynman integral,
\beq
I_n^D\Big(\{\nu_i\}; \{Q^2_i\}; \{M_i\}\Big) = \Gamma(\nu)\,\int\cD x\, \int{\rd^Dk\over i\pi^{D/2}}\,{1\over (k^2-2\cQ\cdot k+\cJ)^\nu},
\eeq
where we introduced the shorthand
\beq
\int\cD x=e^{\gamma_E\eps}\,\prod_{i=1}^n\,\int_0^1\rd x_i\, {x_i^{\nu_i-1}\over \Gamma(\nu_i)}\,\delta\left(1-\sum_{i=1}^nx_i\right).
\eeq
The coefficients $\cQ$ and $\cJ$ are given by
\beq\bsp
\cQ=&\,-\sum_{i=2}^nx_i\,\sum_{j=1}^{i-1}k_j,\\
\cJ=&\, \sum_{i=1}^nx_i\,\left(\sum_{j=1}^{i-1}k_j\right)^2-\sum_{i=1}^nx_i\,M_i^2.
\esp\eeq
Performing the loop integration, we can reduce the Feynman integral to
\beq\label{eq:FeynPar}
I_n^D\Big(\{\nu_i\}; \{Q^2_i\}; \{M_i\}\Big) = (-1)^\nu\,\Gamma(\nu-D/2)\,\int\cD x\, {1\over F^{\nu-D/2}},
\eeq
where the $F$-polynomial is defined by
\beq
F = \cQ^2-\cJ.
\eeq
We have reduced the Feynman integral to an integral of a rational function over the unit cube in $n$ dimensions. Note the formal equivalence of \Eqn{eq:FeynPar} with the Euler integral representation of the hypergeometric function, \Eqn{eq:2F1Euler}.

\subsection{The Mellin-Barnes representation}

The Mellin-Barnes techniques rely on the following identity,
\beq\label{eq:MBdefinition}
{1\over (A+B)^\lam} = {1\over \Gamma(\lam)}\,{1\over 2\pi i}\,\mbint \rd z\, \Gamma(-z)\,\Gamma(\lam+z)\,{B^z\over A^{\lam+z}}.
\eeq
The contour in \Eqn{eq:MBdefinition} is chosen in the standard way, \ie it should separate the poles in $\Gamma(-z)$ from the poles in $\Gamma(\lam+z)$. We can apply \Eqn{eq:MBdefinition} to the $F$-polynomial in \Eqn{eq:FeynPar}, and break it up into monomials in the Feynman parameters $x_i$. The integration over the Feynman parameters can now be easily performed in terms of $\Gamma$ functions,
\beq
\int_0^1\prod_{i=1}^n \rd x_i\,x_i^{a_i-1} \,\delta(1-x_1\ldots-x_n)= {\Gamma(a_1)\ldots\Gamma(a_n)\over \Gamma(a_1+\ldots+a_n)}.
\eeq
In this way we have eliminated all the Feynman parameter integrals in terms of Mellin-Barnes integrals, and we obtain a representation equivalent to the Mellin-Barnes representation of the hypergeometric function, \Eqn{eq:2F1mb}.


\section{The massless scalar pentagon in multi-Regge kinematics}
\label{sec:pentagon}

\begin{center}
\begin{figure}[!t]
\begin{center}
\includegraphics[scale=0.6]{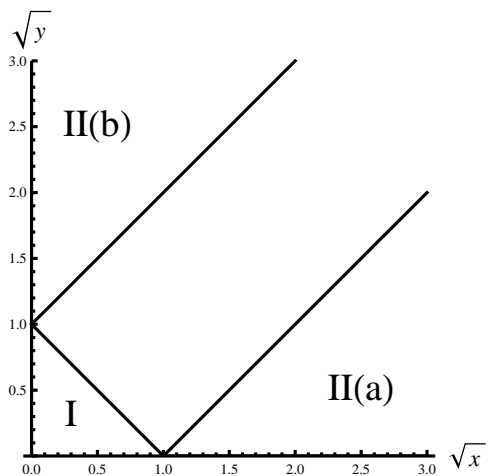}
\end{center}
\caption{\label{fig:psbound} The three regions contributing to the scalar massless pentagon in Euclidean kinematics.}
\end{figure}
\end{center}

In the rest of this work we compute the scalar massless pentagon in $D=6-2\eps$ dimensions in multi-Regge kinematics. The scalar pentagon corresponds to the integral
\beq
I_5^D(\nu_1,\nu_2, \nu_3, \nu_4, \nu_5;{Q_i^2})=e^{\gamma_E\eps}\,\int\frac{\rd^Dk}{i\pi^{D/2}}\,\frac{1}{D_1^{\nu_1}D_2^{\nu_2}D_3^{\nu_3}D_4^{\nu_4}D_5^{\nu_5}},
\eeq
where the external momenta $k_i$ are lightlike, $k^2_i = 0$, and are incoming 
so that $\sum^5_{i=1} k^\mu_i = 0$.
The external momentum scales are the Mandelstam variables $Q ^2_i = s_{12} , s_{23} , s_{34} , s_{45} , s_{15}$ with $s_{ij}= (k_i+ k_j)^2$, and we work in the Euclidean region, $s_{ij}<0$.
Let us introduce the shorthands
  \beq\bsp
  s=s_{12},\quad s_1=&s_{45},\quad s_2=s_{34},\\
     t_1=s_{51},&\quad  t_2=s_{23}.
     \esp\eeq      
We are concerned with the one-loop pentagon in the very peculiar kinematics,
\beq
-s \gg -s_1,\, -s_2 \gg -t_1,\, -t_2.
\eeq
The hierarchy of scales is such that
\beq
s_1s_2\sim s t_1 \sim st_2.
\eeq
Equivalently, we can define the limit by the scaling
  \beq\label{scaling}
  s\rightarrow\,s,\qquad s_1\rightarrow\lambda\,s_1,\qquad s_2\rightarrow\lambda\,s_2,\qquad t_1\rightarrow \lambda^2\,t_1,\qquad t_2\rightarrow \lambda^2\,t_2,
  \eeq
  where $\lambda\rightarrow0$ .
The Euclidean region is itself divided into three other regions, in which the pentagon is represented by different analytic expressions. For later convenience let us introduce the following definitions,
\beq\label{xydef}
x_1={st_1\over s_1s_2} = {t_1\over \kappa}{\rm ~~and~~} x_2 = {st_2\over s_1s_2} = {t_2\over\kappa},
\eeq
where we introduced the transverse momentum scale
\beq\label{eq:kappadef}
-\kappa = {(-s_1)\,(-s_2)\over (-s)}.
\eeq
Note that $\kappa$ behaves as a $t$-type invariant under the scaling~\Eqn{scaling}.
In terms of these quantities the Euclidean region can be divided into three regions
\ben
\item Region I, where $\sqrt{x_1}+\sqrt{x_2} <1$.
\item Region II(a), where $-\sqrt{x_1}+\sqrt{x_2} >1$.
\item Region II(b), where $\sqrt{x_1}-\sqrt{x_2} >1$.
\een
A graphical representation of these three regions in the $(x_1,x_2)$ plane can be found in Fig.~\ref{fig:psbound}. Note that Region I is symmetric in $x_1$ and $x_2$, whereas Regions II(a) and II(b) exchange their roles under an exchange of $x_1$ and $x_2$. It is easy to see that Regions II(a) and II(b) can be furthermore characterized by
\ben
\item Region II(a): $(-t_1)<(-t_2)$.
\item Region II(b): $(-t_1)>(-t_2).$
\een
Note that the region where $k_4$ is soft, $s_1,s_2\rightarrow 0$, corresponds to $x_1,x_2\rightarrow +\infty$ in the $(x_1,x_2)$-plane.

\section{The pentagon integral from NDIM}
\label{sec:PentNDIM}

\subsection{General considerations} 
Solving the system of constraints from negative dimensions, we identify 125 quadruple series contributing to the massless scalar pentagon in general kinematics. Each series has the form of a multiple generalized hypergeometric series.  For example,
\beq\bsp\label{eq:PentND}
I&^{\{n_1,n_2,n_3,n_4\}} = (-s)^{\nu_{45}-{D\over 2}}\,(-t_2)^{\nu_{51}-{D\over 2}}\,(-s_2)^{-\nu_{345}+{D\over 2}}\,(-s_1)^{\nu_{45}-{D\over 2}}\,(-t_1)^{-\nu_{512}+{D\over 2}}\\
&\times (-1)^{{D\over 2}}\,e^{\gamma_E\eps}\,{\Gamma(\nu_1)\Gamma(\nu_2)\Gamma(\nu_3)\Gamma(\nu_4)\Gamma(\nu_5)\over \Gamma(\nu_{345}-{D\over 2})\Gamma(\nu_{451}-{D\over 2})\Gamma(\nu_{512}-{D\over 2})\Gamma({D\over 2}-\nu_{45})\Gamma({D\over 2}-\nu_{51})}\\
&\times F\Bigg(D-\nu,{D\over 2}-\nu_{45},{D\over 2}-\nu_{51}, 1+{D\over 2}-\nu_{345}, 1+{D\over 2}-\nu_{451}, 1+{D\over 2}-\nu_{512};\, x_1,x_2,x_3,x_4\Bigg).
\esp\eeq
The arguments of the hypergeometric functions are ratios of scales, \eg
\beq
x_1={s_2\over s},{\rm~~}x_2=-{s_1s_2\over st_2},{\rm~~} x_3={s_1t_1\over st_2},{\rm~~}x_4= {t_1\over t_2},
\eeq
and we introduced the definitions $\nu_{123}=\nu_1+\nu_2+\nu_3$, \etc
For convenience we have introduced the shorthand for quadruple sums,
\beq\bsp
F(a,b,c,&d,e,f;x_1,x_2,x_3,x_4)\\
=\,&\infsum{n_1,n_2,n_3,n_4}{\ph{a}{n_1+n_2+n_3+n_4}\,\ph{b}{n_1+n_2+n_3}\,\ph{c}{n_2+n_3+n_4}\over \ph{d}{n_1+n_2}\,\ph{e}{n_2+n_3}\,\ph{f}{n_3+n_4}}\, {x_1^{n_1}\over n_1!}\, {x_2^{n_2}\over n_2!}\, {x_3^{n_3}\over n_3!}\, {x_4^{n_4}\over n_4!}.
\esp\eeq

The hierarchy of scales in multi-Regge kinematics eliminates many of the 125 solutions for the pentagon integral. The procedure for reducing the number of solutions is as follows,
\ben
\item Any solution containing a summation that contains ratios of a ``large'' scale divided by a ``small'' scale, such as
\beq
\left({s\over s_1}\right)^n,
\eeq
cannot converge and is therefore discarded. This reduces the number of solutions from 125 to 22. 
\item Solutions with a prefactor that are less singular than
\beq
{1\over s_1s_2}, \qquad {1\over st_1},\qquad {1\over st_2},
\eeq
 when $D=6-2\eps$ and $\nu_i=1$ are discarded. This reduces the number of solutions from 22 to 20.
 \item Any sum that contains ratios of a ``small'' scale divided by a ``large'' scale such as 
\beq
\left({s_1\over s}\right)^n,
\eeq  
gives its leading contribution when the summation variable $n$ is zero. This leads to sums with fewer than four summations.
\item The remaining solutions contain only double sums of ratios of the three scales of \Eqn{eq:PentND}, defined in \Eqn{xydef}. 
\een

\subsection{The solution in Region II}

As an example, we list here the solutions of the system of constraints in Region II(a),
\beq
I^{(IIa)}(s,s_1,s_2,t_1,t_2) =r_\Gamma\,e^{\gamma_E\eps}\,{(-\kappa)^{-\eps}\over st_2}\,\cI^{(IIa)}_{\rm ND}(s,s_1,s_2,t_1,t_2),
\eeq
with
\beq
\cI^{(IIa)}_{\rm ND}(s,s_1,s_2,t_1,t_2) = \sum_{i=1}^6\,I^{(IIa)}_i(s,s_1,s_2,t_1,t_2),
\eeq
and
\beq\bsp
I^{(IIa)}_1&(s,s_1,s_2,t_1,t_2) =\, 
  - {1\over\eps^3}\,y_2^{-\eps}\, \Gamma (1-2\epsilon )\, \Gamma (1+\epsilon )^2\, F_4\Big(1-2\eps, 1-\eps, 1-\eps, 1-\eps; -y_1, y_2\Big),\\
  I^{(IIa)}_2&(s,s_1,s_2,t_1,t_2)=\, 
  {1\over \eps^3}\,\Gamma(1+\eps)\,\Gamma(1-\eps)\, F_4\Big(1, 1-\eps, 1-\eps, 1+\eps; -y_1, y_2\Big),\\
  I^{(IIa)}_3&(s,s_1,s_2,t_1,t_2)=\,  
   -\frac{  \Gamma (1-2\eps )\, \Gamma (-\delta )\,
   \Gamma (-\epsilon -\delta +1)\, \Gamma
  (\epsilon +\delta ) }{\eps\,\Gamma(1+\eps)\,\Gamma(1-\eps)\,  \Gamma (1-2 \epsilon -\delta)}\\
  &\times
  (-s_1)^{-\delta }\,
  y_1^\eps\,y_2^{-\eps-\delta}\,F^{2,1}_{0,2}\left(\begin{array}{cc|cccc|} 1-\delta& 1-\eps-\delta & 1&-&-&-\\ -&-& 1+\eps&1-\eps-\delta&1-\delta&- \end{array}\, -y_1, y_2\right),\\
   I^{(IIa)}_4&(s,s_1,s_2,t_1,t_2)=\,    
  -\frac{\Gamma (-\delta )\, \Gamma (1-2\epsilon )\,
  \Gamma (-\epsilon -\delta )\, \Gamma (\epsilon +\delta )\,
  }{\Gamma(1+\eps)\,\Gamma(1-\eps)\, \Gamma (1-2 \epsilon -\delta)}\\
  &\times (-s_1)^{-\delta} \,y_1^{\eps+\delta}\,y_2^{-\eps-\delta}\,F_4\Big(1,1-\eps,1+\eps+\delta,1-\eps-\delta; -y_1,y_2\Big),\\
      I^{(IIa)}_5&(s,s_1,s_2,t_1,t_2)=\,    
  \frac{  \Gamma (\delta )\, \Gamma
  (1-2\epsilon) \,
  \Gamma (-\epsilon -\delta )}{\eps\,\Gamma(1-\eps)\,\Gamma
  (\delta +1) \, \Gamma (-2 \epsilon
  -\delta +1)}\\
  &\times(-s_1)^{-\delta }\,
  y_1^{\epsilon}\,
  F^{2,1}_{0,2}\left(\begin{array}{cc|cccc|} 1& 1+\eps& 1&-&-&-\\ -&-& 1+\eps&1+\eps+\delta&1-\delta&- \end{array}\, -y_1, y_2\right),\\
   I^{(IIa)}_6&(s,s_1,s_2,t_1,t_2)=\,   
  -\frac{  \Gamma (-\delta )\, \Gamma (\delta +1)\, \Gamma (-\epsilon
  -\delta )^2\, \Gamma (\epsilon +\delta +1)\,\Gamma(1-2\eps)}{\Gamma(1-\eps)^2\,\Gamma(1+\eps)\Gamma (-2 \epsilon -\delta
  +1)}\\
  &\times (-s_1)^{-\delta}\, y_1^{\eps+\delta}\,F_4\Big(1+\delta,1+\eps+\delta,1+\eps+\delta,1+\eps+\delta;-y_1, y_2\Big),
\esp   \eeq
where we defined
\beq\label{eq:y12def}
y_1={1\over x_2}={\kappa\over t_2} {\rm ~~and~~} y_2={x_1\over x_2}={t_1\over t_2},
\eeq
and we factored out the usual loop factor
\beq
r_\Gamma={\Gamma(1+\eps)\,\Gamma(1-\eps)^2\over\Gamma(1-2\eps)}.
\eeq
The regulator $\delta$ is needed in order to prevent divergences in the prefactors in the limit $\nu_i=1$ and $D=6-2\eps$. The singularities in $\delta$ are spurious and cancel out in the sum over all six contributions, \eg,
\beq\bsp
I^{(IIa)}_3+I^{(IIa)}_4=&\, -{1\over\eps^2}\,y_1^\eps\,y_2^{-\eps}\,\Bigg\{\big[\ln y_1 + \psi(1-\eps) - \psi(-\eps)\big]\,F_4\big(1,1-\eps,1+\eps,1-\eps;-y_1, y_2\big)\\
&\,+{\partial\over \partial \delta}\,F^{2,1}_{0,2}\left(\begin{array}{cc|cccc|} 1+\delta & 1+\delta-\eps& 1&- & -& -\\
-&-&1+\delta&1-\eps& 1+\eps+\delta&-\end{array}\, -y_1, y_2\right)_{|\delta=0}\Bigg\},\\
I^{(IIa)}_5+I^{(IIa)}_6=&\,{1\over\eps^2}\, y_1^{\eps}\,\Bigg\{\big[\ln y_1 + \psi(1+\eps) - \psi(-\eps)\big]\,F_4\big(1,1+\eps,1+\eps,1+\eps;-y_1, y_2\big)\\
&\,+{\partial\over \partial \delta}\,F^{2,1}_{0,2}\left(\begin{array}{cc|cccc|} 1+\delta & 1+\delta+\eps& 1&- & -& -\\
-&-&1+\delta&1+\eps& 1+\eps+\delta&-\end{array}\, -y_1, y_2\right)_{|\delta=0}\Bigg\}.
\esp\eeq
The final result for the massless scalar pentagon in multi-Regge kinematics to all orders in $\eps$ in Region II(a) is then simply given by the sum
\beq\bsp\label{eq:PentNDRegIIa}
\cI^{(IIa)}_{\rm ND}&(s,s_1,s_2,t_1,t_2)\\
=&\, -{1\over\eps^3}\,y_2^{-\eps}\, \Gamma (1-2\epsilon )\, \Gamma (1+\epsilon )^2\, F_4\Big(1-2\eps, 1-\eps, 1-\eps, 1-\eps;-y_1, y_2\Big)\\
  +&\,
  {1\over \eps^3}\,
  \Gamma (1+\epsilon ) \,\Gamma (1-\epsilon ) \, F_4\Big(1, 1-\eps, 1-\eps, 1+\eps; -y_1, y_2\Big)\\
   -&\,{1\over \eps^2}\,y_1^\eps\,y_2^{-\eps}\,\Bigg\{\big[\ln y_1 + \psi(1-\eps) - \psi(-\eps)\big]\,F_4\big(1,1-\eps,1+\eps,1-\eps;-y_1, y_2\big)\\
&\qquad + {\partial\over \partial \delta}\,F^{2,1}_{0,2}\left(\begin{array}{cc|cccc|} 1+\delta & 1+\delta-\eps& 1&- & -& -\\
-&-&1+\delta&1-\eps& 1+\eps+\delta&-\end{array}\, -y_1, y_2\right)_{|\delta=0}\Bigg\}\\
+&\,{1\over \eps^2}\,y_1^{\eps}\,\Bigg\{\big[\ln y_1 + \psi(1+\eps) - \psi(-\eps)\big]\,F_4\big(1,1+\eps,1+\eps,1+\eps;-y_1, y_2\big)\\
&\qquad+{\partial\over \partial \delta}\,F^{2,1}_{0,2}\left(\begin{array}{cc|cccc|} 1+\delta & 1+\delta+\eps& 1&- & -& -\\
-&-&1+\delta&1+\eps& 1+\eps+\delta&-\end{array}\, -y_1, y_2\right)_{|\delta=0}\Bigg\}.
\esp\eeq
Note that the only functional dependence of $\cI^{(IIa)}_{\rm ND}$ is in the ratio of scales $y_1$ and $y_2$, \ie, in the transverse momentum scales $t_1$, $t_2$ and $\kappa$,
\beq
\cI^{(IIa)}_{\rm ND}(s,s_1,s_2,t_1,t_2) = \cI^{(IIa)}_{\rm ND}(\kappa, t_1,t_2).
\eeq
The solution in Region II(b) is related to the Region II(a) by analytic continuation according to the prescription $t_1/t_2\rightarrow t_2/t_1$, or equivalently $y_2 \rightarrow 1/y_2$. From the symmetry of the multi-Regge limit in $t_1$ and $t_2$ it is easy to see that we must have
\beq\label{eq:RegIIaIIb}
\cI^{(IIb)}_{\rm ND}(\kappa,t_1,t_2) = {t_2\over t_1}\,\cI^{(IIa)}_{\rm ND}(\kappa,t_2,t_1).
\eeq
In Appendix~\ref{app:AnalContPent} we explicitly show that \Eqn{eq:PentNDRegIIa} enjoys this property. 

Using the reduction formulas for the Appell function given in Appendix~\ref{app:HypGeo}, we could reexpress all the $F_4$ in \Eqn{eq:PentNDRegIIa} in terms of Gauss' hypergeometric function, 
\beq\bsp
{F_4}\Big(1-2\eps,1-\eps,&1-\eps,1-\eps,{-x\over (1-x)(1-y)},{-y\over (1-x)(1-y)}\Big)\\
=&\,(1-x)^{1-\eps}\,(1-y)^{1-\eps}\,{_2F_1}(1-2\eps,1-\eps,1-\eps;xy)\\
=&\, {(1-x)^{1-\eps}\,(1-y)^{1-\eps}\over (1-x y)^{1-2\eps}},\\
{F_4}\Big(1,1-\eps,&1+\eps,1-\eps,{-x\over (1-x)(1-y)},{-y\over (1-x)(1-y)}\Big)\\
=&\,(1-x)\,(1-y)\,F_1(1,2\eps,1-\eps,1+\eps;x y).
\esp\eeq
The Appell $F_1$ function appearing in this reduction can be easily expanded into a Laurent series in $\eps$ using {\tt XSummer}~\cite{Moch:2005uc}.
Note however that we do not know the corresponding reduction formulas for the Kamp\'e de F\'eriet function appearing in \Eqn{eq:PentNDRegIIa}. For this reason, we do not apply the reduction formulas of the Appell $F_4$ functions, but we proceed and perform the $\eps$ expansion directly on the series representation of the hypergeometric functions appearing in \Eqn{eq:PentNDRegIIa}.
Since all the hypergeometric functions in \Eqn{eq:PentNDRegIIa} are finite for $\eps=0$, we can safely expand the Pochhammer symbols into a power series under the summation sign,
\beq\bsp\label{eq:phexp}
(1+\eps)_n &\,= n!\,\left(1+\eps\,Z_1(n)+\eps^2\,Z_{11}(n)+\eps^3\,Z_{111}(n)+\ord(\eps^4)\right),\\
{1\over(1+\eps)_n} &\,= {1\over n!}\,\left(1-\eps\,S_1(n)-\eps^2\,S_{11}(n)-\eps^3\,S_{111}(n)+\ord(\eps^4)\right),
\esp\eeq
where $S$ and $Z$ denote nested harmonic sums and Euler-Zagier sums, defined recursively by~~\cite{Vermaseren:1998uu}
\beq\bsp\label{eq:SZSums}
&S_i(n) = Z_i(n) = H_n^{(i)} = \sum_{k=1}^n\,{1\over k^i},\\
& S_{i\vec \jmath}(n) = \sum_{k=1}^n\,{S_{\vec \jmath}(k)\over k^i} {\rm ~~and~~} Z_{i\vec \jmath}(n) = \sum_{k=1}^n\,{Z_{\vec \jmath}(k-1)\over k^i}.
\esp\eeq
Using the algorithm described in Appendix~\ref{app:Ssums}, we can express all the Euler-Zagier sums in terms of harmonic sums. Finally, using the algebra properties of the $S$-sums, we can reduce all the products of harmonic sums to linear combinations of the latter. Inserting \Eqn{eq:phexp} into the series representation for hypergeometric functions we obtain the desired $\eps$ expansions, \eg,
\beq\bsp
F_4(1,&\,1+\eps,1+\eps,1+\eps;x_1,x_2) = \cM(0,0,0;x_1,x_2) \\
+&\, \eps\Big[\cM(0,0,1;x_1,x_2)-\cM(1,0,0;x_1,x_2)-\cM(0,1,0;x_1,x_2)\Big]\\
+&\,\eps^2\Big[\cM((1,1),0,0;x_1,x_2)+\cM(0,(1,1),0;x_1,x_2)+\cM(0,0,(1,1);x_1,x_2)\\
&\,\phantom{\eps^2\Big[}+\cM(1,1,0;x_1,x_2)- \cM(1,0,1;x_1,x_2)- \cM(0,1,1;x_1,x_2)- \cM(0,0,2;x_1,x_2)\Big]\\
+&\,\ord(\eps^3).
\esp\eeq
The $\cM$ functions appearing in this expansion are transcendental functions defined by the double series
\beq\label{eq:Mdef}
\cM(\vec \imath, \vec \jmath, \vec k;x_1,x_2) = \sum_{n_1=0}^\infty\sum_{n_2=0}^\infty\,\binom{n_1+n_2}{n_1}^2\,S_{\vec \imath}(n_1)\,S_{\vec \jmath}(n_2)\,S_{\vec k}(n_1+n_2)\,x_1^{n_1}\,x_2^{n_2}.
\eeq
Note that due to the appearance  of the binomial squared term in \Eqn{eq:Mdef}, we cannot reduce the double sums in general to known function using the standard techniques~\cite{Moch:2005uc,Moch:2001zr}. We can however sum the series in some particular cases in which we can relate the $\cM$-function to the expansion of a known hypergeometric function. This issue will be addressed in Appendix~\ref{app:Mfunc}.

Since \Eqn{eq:PentNDRegIIa} only involves Appell functions and Kamp\'e de F\'eriet functions with indices $1+c_i\eps$, it can be easily expanded in terms of $\cM$ functions. The first two orders read,
\beq
\cI^{(IIa)}_{\rm ND}(\kappa, t_1,t_2)= i^{(IIa)}_0(y_1,y_2)+\eps\,i^{(IIa)}_1(y_1,y_2)+\ord(\eps^2), \label{eq:i2a}
\eeq
with
\beq\bsp\label{eq:i02a}
i^{(IIa)}_0&(y_1,y_2)=
(-8 \ln y_1-4 \ln y_2) \cM\big(0,0,(1,1);-y_1, y_2\big)-4 \ln y_2 \cM\big((1,1),0,0;-y_1, y_2\big)\\
+&\,18 \cM\big(0,0,(1,2);-y_1, y_2\big)+18 \cM\big(0,0,(2,1);-y_1, y_2\big)-24 \cM\big(0,0,(1,1,1);-y_1, y_2\big)\\
+&\,8 \cM\big(0,1,(1,1);-y_1, y_2\big)+16 \cM\big(1,0,(1,1);-y_1, y_2\big)-8 \cM\big((1,1),0,1;-y_1, y_2\big)\\
+&\,8 \cM\big((1,1),1,0;-y_1, y_2\big)-\cM\big(0,0,0;-y_1, y_2\big) \Big(\frac{\pi ^2 \ln y_1}{3}+\frac{\ln^2 y_1 \ln y_2}{2}+\frac{\pi ^2 \ln y_2}{2}-2 \zeta_3\Big)\\
-&\,\cM\big(0,0,1;-y_1, y_2\big) \Big(2 \ln y_1 \ln y_2+\ln^2 y_1+\frac{5 \pi ^2}{3}\Big)+(6 \ln y_1+3 \ln y_2) \cM\big(0,0,2;-y_1, y_2\big)\\
+&\,\Big(2 \ln y_1 \ln y_2+\frac{2 \pi ^2}{3}\Big) \cM\big(1,0,0;-y_1, y_2\big)+(4 \ln y_1+4 \ln y_2) \cM\big(1,0,1;-y_1, y_2\big)\\
+&\,4 \ln y_1 \cM\big(0,1,1;-y_1, y_2\big)-4 \ln y_1 \cM\big(1,1,0;-y_1, y_2\big)+\Big(\ln^2 y_1+\pi ^2\Big) \cM\big(0,1,0;-y_1, y_2\big)\\
+&\,\ln y_2 \cM\big(2,0,0;-y_1, y_2\big)-12 \cM\big(0,0,3;-y_1, y_2\big)-6 \cM\big(0,1,2;-y_1, y_2\big)\\
-&\,12 \cM\big(1,0,2;-y_1, y_2\big)-8 \cM\big(1,1,1;-y_1, y_2\big)+2 \cM\big(2,0,1;-y_1, y_2\big)\\
-&\,2 \cM\big(2,1,0;-y_1, y_2\big),
\esp\eeq
\beq\bsp\label{eq:i12a}
i^{(IIa)}_1&(y_1,y_2)=
\cM\big(0,0,(1,1);-y_1, y_2\big) \Big(4 \ln y_1 \ln y_2-4 \ln^2 y_1+2 \ln^2 y_2+4 \pi ^2\Big)\\
+&\,(2 \ln^2 y_2-4 \ln y_1 \ln y_2) \cM\big((1,1),0,0;-y_1, y_2\big)+(8 \ln y_1-12 \ln y_2) \cM\big(1,0,(1,1);-y_1, y_2\big)\\
+&\,(4 \ln y_2-8 \ln y_1) \cM\big((1,1),0,1;-y_1, y_2\big)+(8 \ln y_1-4 \ln y_2) \cM\big((1,1),1,0;-y_1, y_2\big)\\
-&\,15 \ln y_2 \cM\big(0,0,(1,2);-y_1, y_2\big)-15 \ln y_2 \cM\big(0,0,(2,1);-y_1, y_2\big)\\
+&\,20 \ln y_2 \cM\big(0,0,(1,1,1);-y_1, y_2\big)-4 \ln y_2 \cM\big(0,1,(1,1);-y_1, y_2\big)\\
-&\,\ln y_2 \cM\big((1,2),0,0;-y_1, y_2\big)-\ln y_2 \cM\big((2,1),0,0;-y_1, y_2\big)
\esp\eeq
\beq\bsp\nonumber
\phantom{i^{(IIa)}}+&\,4 \ln y_2 \cM\big((1,1,1),0,0;-y_1, y_2\big)+32 \cM\big(0,0,(1,3);-y_1, y_2\big)+36 \cM\big(0,0,(2,2);-y_1, y_2\big)\\
+&\,32 \cM\big(0,0,(3,1);-y_1, y_2\big)-48 \cM\big(0,0,(1,1,2);-y_1, y_2\big)-48 \cM\big(0,0,(1,2,1);-y_1, y_2\big)\\
-&\,48 \cM\big(0,0,(2,1,1);-y_1, y_2\big)+64 \cM\big(0,0,(1,1,1,1);-y_1, y_2\big)+12 \cM\big(0,1,(1,2);-y_1, y_2\big)\\
&\,+12 \cM\big(0,1,(2,1);-y_1, y_2\big)-16 \cM\big(0,1,(1,1,1);-y_1, y_2\big)+18 \cM\big(1,0,(1,2);-y_1, y_2\big)\\
+&\,18 \cM\big(1,0,(2,1);-y_1, y_2\big)-24 \cM\big(1,0,(1,1,1);-y_1, y_2\big)+8 \cM\big(1,1,(1,1);-y_1, y_2\big)\\
-&\,2 \cM\big((1,2),0,1;-y_1, y_2\big)+2 \cM\big((1,2),1,0;-y_1, y_2\big)-2 \cM\big((2,1),0,1;-y_1, y_2\big)\\
+&\,2 \cM\big((2,1),1,0;-y_1, y_2\big)+8 \cM\big((1,1,1),0,1;-y_1, y_2\big)-8 \cM\big((1,1,1),1,0;-y_1, y_2\big)\\
+&\,\cM\big(0,0,1;-y_1, y_2\big) \Big(\ln y_1 \ln^2 y_2-\frac{\pi ^2 \ln y_1}{3}-\frac{\ln^2 y_1 \ln y_2}{2}-\frac{2 \ln^3 y_1}{3}+\frac{3 \pi ^2 \ln y_2}{2}-6 \zeta_3\Big)\\
+&\,\cM\big(0,1,0;-y_1, y_2\big) \Big(\frac{\pi ^2 \ln y_1}{3}-\frac{\ln^2 y_1 \ln y_2}{2}+\frac{2 \ln^3 y_1}{3}-\frac{\pi ^2 \ln y_2}{2}+2 \zeta_3\Big)\\
+&\,\cM\big(1,0,0;-y_1, y_2\big) \Big(-\ln y_1 \,\ln^2 y_2+\frac{\pi ^2 \ln y_1}{3}+\frac{3 \ln^2 y_1 \ln y_2}{2}-\frac{\pi ^2 \ln y_2}{2}+2 \zeta_3\Big)\\
+&\,\cM\big(0,0,2;-y_1, y_2\big) \Big(-3 \ln y_1 \ln y_2+3 \ln^2 y_1-\frac{3 \ln^2 y_2}{2}-3 \pi ^2\Big)\\
+&\,\cM\big(0,1,1;-y_1, y_2\big) \Big(-2 \ln y_1 \ln y_2+2 \ln^2 y_1-\frac{4 \pi ^2}{3}\Big)\\
+&\,\cM\big(1,1,0;-y_1, y_2\big) \Big(2 \ln y_1 \ln y_2-3 \ln^2 y_1+\frac{\pi ^2}{3}\Big)+(\ln y_2-2 \ln y_1) \cM\big(2,1,0;-y_1, y_2\big)\\
+&\,\cM\big(2,0,0;-y_1, y_2\big) \Big(\ln y_1 \ln y_2-\frac{\ln^2 y_2}{2}\Big)+(9 \ln y_2-6 \ln y_1) \cM\big(1,0,2;-y_1, y_2\big)\\
+&\,(4 \ln y_2-4 \ln y_1) \cM\big(1,1,1;-y_1, y_2\big)+(2 \ln y_1-\ln y_2) \cM\big(2,0,1;-y_1, y_2\big)\\
+&\,\cM\big(0,0,0;-y_1, y_2\big) \Big(\frac{\ln^2 y_1 \ln^2 y_2}{4}-\frac{\pi ^2 \ln^2 y_1}{6}-\frac{\ln^3 y_1 \ln y_2}{3}-2 \ln y_2 \zeta_3+\frac{\pi ^2 \ln^2 y_2}{4}+\frac{2 \pi ^4}{15}\Big)\\
+&\,\Big(3 \ln^2 y_1-2 \ln^2 y_2-\pi ^2\Big) \cM\big(1,0,1;-y_1, y_2\big)+10 \ln y_2 \cM\big(0,0,3;-y_1, y_2\big)\\
+&\,3 \ln y_2 \cM\big(0,1,2;-y_1, y_2\big)-20 \cM\big(0,0,4;-y_1, y_2\big)-8 \cM\big(0,1,3;-y_1, y_2\big)\\
-&\,12 \cM\big(1,0,3;-y_1, y_2\big)-6 \cM\big(1,1,2;-y_1, y_2\big).
\esp\eeq
Note that $i^{(IIa)}_0$ and $i^{(IIa)}_1$ are of uniform transcendental weight $3$ and $4$, as expected. 

\subsection{The solution in Region I}

The solutions in Region I read
\beq
I^{(I)}(s,s_1,s_2,t_1,t_2)= r_\Gamma\,e^{\gamma_E\eps}\,{(-\kappa)^{-\eps}\over s_1s_2}\,\cI^{(I)}_{\rm ND}(s,s_1,s_2,t_1,t_2),
\eeq
with
\beq
\cI^{(I)}_{\rm ND}(s,s_1,s_2,t_1,t_2) = \sum_{i=1}^6 I^{(I)}_i(s,s_1,s_2,t_1,t_2),
\eeq
and
\beq\bsp\label{eq:SolRegI}
I^{(I)}_1&(s,s_1,s_2,t_1,t_2)= 
-{1\over \eps^3}\,x_1^{-\eps}\,x_2^{-\eps}\, \Gamma (1-2\epsilon )\, \Gamma (1+\epsilon )^2\\
&\times\, F_4(1-2\eps,1-\eps,1-\eps,1-\eps;-x_1,-x_2),\\
I^{(I)}_2&(s,s_1,s_2,t_1,t_2)= 
  {1\over\eps^3}\,\Gamma(1+\eps)\,\Gamma(1-\eps)\, F_4(1,1+\eps,1+\eps,1+\eps;-x_1,-x_2),\\
   I^{(I)}_3&(s,s_1,s_2,t_1,t_2)=   
     \left(-s_2\right)^{\delta}\, \left(-s_1\right)^{-\delta}\,\left(-t_1\right)^{-\delta}\, \left(-t_2\right)^{\delta
  }\,x_1^{-\eps}\,\frac{ \Gamma (-\delta )\,  \Gamma (\epsilon +\delta )}{\eps\,\Gamma(1+\eps)}\\
  &\times\infsum{n_1}\infsum{n_2} { (1-\delta )_{n_2}\, (-\delta )_{n_1-n_2}\, (\delta +1)_{n_2}\, (1-\epsilon )_{n_2}\, (-\epsilon
  )_{n_1-n_2} \over  (-\epsilon -\delta +1)_{n_1}\, n_1!\,n_2!}\left(\frac{x_1}{x_2}\right)^{n_1} \left(-x_2\right)^{n_2},
\\
   I^{(I)}_4&(s,s_1,s_2,t_1,t_2)=      
  -(-s)^{-\delta }\, \left(-s_2\right)^{2 \delta}\, \left(-t_1\right)^{-\delta}\,x_1^{-\eps}\\
   &\times
 \frac{ \Gamma (1-2 \delta )\, \Gamma (\delta )
    \Gamma (-\epsilon -\delta +1)\, \Gamma (\delta -\epsilon )\, \Gamma (\epsilon +\delta )}{\Gamma (1-\delta ) \,\Gamma (\delta +1)\, \Gamma (1+\epsilon )\,\Gamma(1-\eps)^2}\\
    &\times F^{3,0}_{1,2}\left(\begin{array}{ccc|cc|}
  1 &  1-\eps-\delta &1-2 \delta & -&- \\
  -&-&1-\delta & 1-\eps-\delta&1+\eps-\delta\end{array}\, -x_1, -x_2\right),\\
     I^{(I)}_5&(s,s_1,s_2,t_1,t_2)= 
  -(-s)^{\delta }\, \left(-s_1\right)^{-2 \delta}\,\left(-t_2\right)^{\delta }\,x_2^{-\eps}\\
  &\times\frac{ \Gamma (-\delta )\, \Gamma (2 \delta +1)\,
   \Gamma (-\epsilon -\delta )\, \Gamma (\epsilon -\delta )\, \Gamma (-\epsilon +\delta +1)}{\Gamma (1-\delta ) \,\Gamma (\delta +1)\, \Gamma (1+ \epsilon )\,\Gamma(1-\eps)^2}\\
       &\times F^{3,0}_{1,2}\left(\begin{array}{ccc|cc|}
  1 &  1-\eps+\delta &1+2 \delta & -&- \\
  -&-&1+\delta & 1+\eps+\delta&1-\eps+\delta\end{array}\, -x_1, -x_2\right),\\
     I^{(I)}_6&(s,s_1,s_2,t_1,t_2)=    
  -\left(-s_2\right)^{\delta }\, \left(-s_1\right)^{-\delta }\,x_2^{-\eps}\, \frac{ \Gamma (\delta )\, \Gamma (-\epsilon -\delta )}{\epsilon \,\Gamma(1-\eps) }\\
  &\times\infsum{n_1}\infsum{n_2} { (1-\delta )_{n_2} (\delta )_{n_1-n_2} (\delta +1)_{n_2} (1-\epsilon
  )_{n_2} (\epsilon )_{n_1-n_2}\over (\epsilon +\delta +1)_{n_1}   n_1! n_2! } \left(\frac{x_1}{x_2}\right)^{n_1} \left(-x_2\right)^{n_2}.
  \esp\eeq
  The regulator $\delta$ is introduced to prevent divergences in the $\Gamma$ functions in the prefactor. The cancellation of the spurious $\delta$-poles is not as straightforward as in Region II(a), due the appearance of a new type of hypergeometric series. Since however this new series involves only Pochhammer symbols of the form $\ph{.}{n_1-n_2}$, $\ph{.}{n_1}$ and $\ph{.}{n_2}$, it can be reduced to Kamp\'e de F\'eriet functions using the techniques described in Appendix~\ref{app:HypGeo}. We find
\beq\bsp
\infsum{n_1}&\infsum{n_2} { (1-\delta )_{n_2}\, (-\delta )_{n_1-n_2}\, (\delta +1)_{n_2}\, (1-\epsilon )_{n_2}\, (-\epsilon
  )_{n_1-n_2} \over  (-\epsilon -\delta +1)_{n_1}\, n_1!\,n_2!}\left(\frac{x_1}{x_2}\right)^{n_1} \left(-x_2\right)^{n_2}\\
&\,= {\delta\,\eps\over 1-\eps-\delta}\,{x_1\over x_2}\,F^{0,3}_{2,0}\left(\begin{array}{cc|cccccc|}
-&-&1+\delta & 1 & 1-\delta & 1-\delta &1-\eps&1-\eps\\
2&2-\eps-\delta&-&-&-&-&-&-\end{array}\,-x_1,{x_1\over x_2}\right)\\
&\,+ F^{3,1}_{1,2}\left(\begin{array}{ccc|cccc|}
1-\delta&1+\delta&1-\eps & - & 1& -&-\\
1&-&-&-&1+\delta&1-\delta-\eps&1+\eps\end{array}\,-x_1,-x_2\right),\esp\eeq
\beq\bsp
\infsum{n_1}&\infsum{n_2} { (1-\delta )_{n_2} (\delta )_{n_1-n_2} (\delta +1)_{n_2} (1-\epsilon
  )_{n_2} (\epsilon )_{n_1-n_2}\over (\epsilon +\delta +1)_{n_1}   n_1! n_2! } \left(\frac{x_1}{x_2}\right)^{n_1} \left(-x_2\right)^{n_2}\\
  &\,={\delta\,\eps\over 1+\eps+\delta}\,{x_1\over x_2}\,F^{0,3}_{2,0}\left(\begin{array}{cc|cccccc|}
-&-&1-\delta & 1 & 1+\delta & 1+\delta &1-\eps&1+\eps\\
2&2+\eps+\delta&-&-&-&-&-&-\end{array}\,-x_1,{x_1\over x_2}\right)\\
&\,+ F^{3,1}_{1,2}\left(\begin{array}{ccc|cccc|}
1-\delta&1+\delta&1-\eps & - & 1& -&-\\
1&-&-&-&1-\delta&1+\delta+\eps&1-\eps\end{array}\,-x_1,-x_2\right).
\esp\eeq
After inserting these expressions into \Eqn{eq:SolRegI}, we expand the solution into a Laurent series in $\delta$. The poles in $\delta$ cancel mutually between $I^{(I)}_3$ and $I^{(I)}_4$ and $I^{(I)}_5$ and $I^{(I)}_6$. After some algebra, we find the following expression for the solution in Region I,
\beq\bsp\label{eq:SolRegIKF}
\cI^{(I)}_{\rm ND}&(s,s_1,s_2,t_1,t_2)\\
=\,& -\frac{1}{\eps^3}\,x_1^{-\eps}\,x_2^{-\eps}\,\Gamma(1-2\eps)\, \Gamma (1+\epsilon )^2\, F_4(1-2\eps,1-\eps,1-\eps,1-\eps;-x_1,-x_2)\\
 +\,&   {1\over\eps^3}\,
  \Gamma (1-\epsilon ) \Gamma (1+\epsilon)\, F_4(1,1+\eps,1+\eps,1+\eps;-x_1,-x_2)\\
     -\,&{1\over\eps^2}\,x_1^{-\eps}\,\Bigg\{\big[\ln x_2 + \psi(1-\eps)-\psi(-\eps)\big]\,F_4(1,1-\eps,1-\eps,1+\eps;-x_1,-x_2)
\\
 &\quad +{\partial\over\partial\delta} F^{2,1}_{0,2}\left(\begin{array}{cc|cccc|}
 1+\delta & 1-\eps+\delta &-&-&-&1\\
 -&-&1-\eps&1+\eps+\delta&-&1+\delta\end{array}\,-x_1,-x_2\right)_{|\delta=0}\\
 &\quad +{\eps\over 1-\eps}\,{x_1\over x_2}\, F^{0,3}_{2,0}\left(\begin{array}{cc|cccccc|}
 - & - &1&1&1&1&1-\eps&1-\eps\\
 2&2-\eps&-&-&-&-&-&-\end{array}\,-x_1,{x_1\over x_2}\right)\Bigg\}\\
 -\,&{1\over \eps^2}\,x_2^{-\eps}\,\Bigg\{\big[\ln x_2 + \psi(1-\eps)-\psi(\eps)\big]\,F_4(1,1-\eps,1+\eps,1-\eps;-x_1,-x_2)\\
 &\quad +{\partial\over\partial\delta} F^{2,1}_{0,2}\left(\begin{array}{cc|cccc|}
 1+\delta & 1-\eps+\delta &-&-&-&1\\
 -&-&1+\eps&1-\eps+\delta&-&1+\delta\end{array}\,-x_1,-x_2\right)_{|\delta=0}\\
 &\quad -{\eps\over 1+\eps}\,{x_1\over x_2}\, F^{0,3}_{2,0}\left(\begin{array}{cc|cccccc|}
 - & - &1&1&1&1&1-\eps&1+\eps\\
 2&2+\eps&-&-&-&-&-&-\end{array}\,-x_1,{x_1\over x_2}\right)\Bigg\}.
 \esp\eeq
 Note that the right-hand side of \Eqn{eq:SolRegIKF} only depends on the dimensionless quantities $x_1$ and $x_2$, \ie, on the transverse momentum scales $\kappa$, $t_1$ and $t_2$,
 \beq
 \cI^{(I)}_{\rm ND}(s,s_1,s_2,t_1,t_2) = \cI^{(I)}_{\rm ND}(\kappa,t_1,t_2).
 \eeq
Furthermore, we know that the pentagon in Region I must fulfill the symmetry relation
\beq
\cI^{(I)}_{\rm ND}(\kappa,t_1,t_2) = \cI^{(I)}_{\rm ND}(\kappa,t_2,t_1).
\eeq
The solution given in \Eqn{eq:SolRegIKF} however apparently breaks this symmetry, due to the appearance of the ratio $x_1/x_2$. We show in Appendix~\ref{app:AnalContPent} that \Eqn{eq:SolRegIKF} indeed has the correct symmetry properties, which becomes explicit only after a proper analytic continuation has been performed. We also show that the solution in Region I, \Eqn{eq:SolRegIKF}, can be obtained from the solution in Region II(a) by performing analytic continuation according to the prescription $y_1\rightarrow 1/y_1$.


\section{The pentagon integral from Mellin-Barnes integrals}
\label{sec:PentMB}

In Ref.~\cite{Bern:2006ew} a Mellin-Barnes representation for the pentagon  was given,
\beq\bsp
I_5^D(\nu_1,&\nu_2, \nu_3, \nu_4, \nu_5;{Q_i^2})=\frac{(-1)^{N_\nu}\,e^{\gamma_E\eps}}{ \Gamma \left(D-N_{\nu }\right)}\\
&\times\frac{1}{(2\pi i)^4}\int_{-i\infty}^{+i\infty}\prod_{i=1}^4\rd z_i\,\Gamma(-z_i)\,(-s)^{\frac{D}{2}-N_{\nu }} \left(\frac{s_1}{s}\right)^{z_4}
  \left(\frac{s_2}{s}\right)^{z_1} \left(\frac{t_1}{s}\right)^{z_2} \left(\frac{t_2}{s}\right)^{z_3} \\
  &\quad\times\Gamma \left(\nu _5+z_1+z_2\right) \Gamma
  \left(\frac{D}{2}-N_{\nu }+\nu _1-z_1-z_2-z_3\right)  \Gamma \left(\nu _2+z_2+z_3\right) \\
  &\quad\times\Gamma
  \left(\frac{D}{2}-N_{\nu }+\nu _3-z_2-z_3-z_4\right) \Gamma \left(\nu _4+z_3+z_4\right) \\
  &\quad\times\Gamma
  \left(-\frac{D}{2}+N_{\nu }+z_1+z_2+z_3+z_4\right),
  \esp\eeq
where $N_{\nu }=\sum \nu_i$.
In particular, if we put $\nu_i=1$ and $D=6-2\epsilon$, then we get
\beq\bsp\label{MBpent}
I_5^D(1,&1, 1, 1, 1;{Q_i^2})=\frac{-\,e^{\gamma_E\eps}\,(-s)^{-2-\epsilon}}{\Gamma (1-2 \epsilon )}\\
&\times\frac{1}{(2\pi i)^4}\int_{-i\infty}^{+i\infty}\prod_{i=1}^4\rd z_i\,\Gamma(-z_i)\, \left(\frac{s_1}{s}\right)^{z_4}
  \left(\frac{s_2}{s}\right)^{z_1} \left(\frac{t_1}{s}\right)^{z_2} \left(\frac{t_2}{s}\right)^{z_3}\\
  &\quad\times \Gamma \left(z_1+z_2+1\right) \Gamma
  \left(-\epsilon -z_1-z_2-z_3-1\right)
  \Gamma \left(z_2+z_3+1\right) \\
  &\quad\times\Gamma \left(-\epsilon
  -z_2-z_3-z_4-1\right) \Gamma
  \left(z_3+z_4+1\right) \Gamma \left(\epsilon
  +z_1+z_2+z_3+z_4+2\right).
  \esp\eeq
In the following we extract the leading behavior of this integral in the limit defined by the scaling~(\ref{scaling}), and we show that this leading behavior is described in all the regions by a twofold Mellin-Barnes integral, which can be evaluated in terms of multiple polylogarithms.

\subsection{Evaluation of the Mellin-Barnes integral in Region I}
  Performing this rescaling~(\ref{scaling}) in the Mellin-Barnes representation~(\ref{MBpent}) and after the change of variable
$ z_4= z-z_1-z_2-z_3$,
 we find
 \beq\bsp
 I_5^D(1,&1, 1, 1, 1;{Q_i^2})=\frac{-\,e^{\gamma_E\eps}\,(-s)^{-\epsilon -2}}{\Gamma (1-2 \epsilon )}\\
 &\times\,{1\over (2\pi i)^4}\,\mbint\rd z \,\rd z_1\,\rd z_2\,\rd z_3\,
  \left(\frac{s_1}{s}\right)^{z-z_1-z_2-z_3} \left(\frac{s_2}{s}\right)^{z_1-z_2-z_3} \left(\frac{t_1}{s}\right)^{z_2}
  \left(\frac{t_2}{s}\right)^{z_3} \lambda^z \\
  &\qquad\times\Gamma \left(-\epsilon -z_1-1\right) \Gamma \left(-\epsilon -z+z_1-1\right) \Gamma \left(z-z_1-z_2+1\right)
  \Gamma \left(-z_2\right) \Gamma \left(z_1-z_3+1\right)\\
  &\qquad\times \Gamma \left(\epsilon +z-z_2-z_3+2\right) \Gamma \left(-z_3\right) \Gamma \left(z_2+z_3+1\right)\\
  &\qquad\times \Gamma
  \left(-z_1+z_2+z_3\right) \Gamma \left(-z+z_1+z_2+z_3\right).
  \esp\eeq
To obtain the leading behavior in our limit $\lambda\to 0$ let us follow the
strategy formulated, e.g., in Chap.~4 of~\cite{SmirnovBook2, SmirnovBook}.
  We think of the integration over $z$ as the last one, and we analyze how poles in $\Gamma(\ldots-z)$ with leading behavior $\lambda^{-2}$ might arise. There is only one possibility, coming from the product $\Gamma \left(-\epsilon -z+z_1-1\right)\Gamma \left(-\epsilon -z_1-1\right)$. Taking the residues at $z_1=-1-\eps+n_1$, $n_1\in \mathbb{N}$, we find 
  \beq
  \lambda^z\Gamma \left(-\epsilon -z+z_1-1\right)\rightarrow\lambda^z\Gamma \left(-2\epsilon-2 -z+n_1\right).
  \eeq
  If we now take the residues at $z=-2-2\eps+n_1+n_2$, $n_2\in \mathbb{N}$, we find 
  \beq
  \lambda^z\Gamma \left(-2\epsilon-2 -z+n_1\right)\rightarrow \lambda^{-2-2\eps+n_1+n_2}.
  \eeq
  Since we are only interested in the leading behavior in $\lambda^{-2}$, we only keep the terms in $n_1=n_2=0$. Hence, we find a twofold Mellin-Barnes representation for the pentagon in multi-Regge kinematics,
   \beq\bsp\label{MBMRpent}
 I_5^D(1,&1, 1, 1, 1;{Q_i^2})=r_\Gamma\,\,e^{\gamma_E\eps}\,\frac{(-\kappa)^{-\epsilon}}{s_1s_2}\,
 \cI^{(I)}_{\rm MB}(\kappa,t_1,t_2)
  \esp\eeq
  with
      \beq\bsp\label{eq:RegIMGI}
 \cI^{(I)}_{\rm MB}&(\kappa,t_1,t_2)={-1\over\Gamma(1+\eps)\,\Gamma(1-\eps)^2}\,{1\over (2\pi i)^2}\,\mbint\,\rd z_1\,\rd z_2\,
 x_1^{z_1} \,x_2^{z_2} \Gamma \left(-\epsilon -z_1\right) \Gamma \left(-z_1\right) \\
  &\qquad\times\Gamma
  \left(-\epsilon -z_2\right)\, \Gamma \left(-\epsilon -z_1-z_2\right) \,\Gamma \left(-z_2\right) \Gamma \left(z_1+z_2+1\right) \Gamma \left(\epsilon
  +z_1+z_2+1\right)^2,
  \esp\eeq
  where $x_1$ and $x_2$ are defined in \Eqn{xydef}.
Note that this expression is symmetric in $x_1$ and $x_2$, as expected in Region I. We checked that if we close the integration contours to the right, and take residues, we reproduce exactly the expression of the pentagon obtained from NDIM, \Eqn{eq:SolRegIKF}.

We now evaluate the Mellin-Barnes representation~(\ref{MBMRpent}) and we derive an Euler integral representation for the pentagon in multi-Regge kinematics. Let us concentrate only on the Mellin-Barnes integral. We start with the change of variable $z_2=z-z_1$ and we find
   \beq\bsp
 \cI^{(I)}_{\rm MB}(\kappa,t_1,t_2)&= {-1\over\Gamma(1+\eps)\,\Gamma(1-\eps)^2}\,{1\over (2\pi i)^2}\,\mbint\,\rd z_1\,\rd z\,x_1^{z_1} x_2^{z-z_1} \Gamma (-\epsilon -z) \Gamma (z+1) \\
  &\qquad\times  \Gamma
  (\epsilon +z+1)^2\,\Gamma \left(-\epsilon -z_1\right)\,\Gamma
  \left(-z_1\right) \Gamma \left(z_1-z\right) \Gamma
  \left(-\epsilon -z+z_1\right).
\esp\eeq
We now replace the Mellin-Barnes integral over $z_1$ by an Euler integral by using the transformation formula,
\beq\bsp\label{eq:MBtoEuler}
{1\over 2\pi i}\,&\mbint\,\rd z_1\,\Gamma(-z_1)\,\Gamma(c-z_1)\,\Gamma(b+z_1)\,\Gamma(a+z_1)\,X^{z_1}\\
&\, = \Gamma(a)\,\Gamma(b+c)\,\int_0^1\,\rd v\,v^{b-1}\,(1-v)^{a+c-1}\,(1-(1-X)v)^{-a},
\esp\eeq
and we find
\beq\bsp
 \cI^{(I)}_{\rm MB}(\kappa,t_1,t_2)=&{-1\over\Gamma(1+\eps)\,\Gamma(1-\eps)^2}\,{1\over 2\pi i}\,\mbint\,\rd z\,\int_0^1\rd v\, \left(1-v
  \left(1-\frac{x_1}{x_2}\right)\right)^{\epsilon +z} x_2^z\\
  &\qquad\times
  (1-v)^{-2 \epsilon -z-1}\, v^{-z-1}\,\Gamma (-\epsilon -z)^3 \Gamma (z+1) \Gamma (\epsilon +z+1)^2.
  \esp\eeq
  To continue, we exchange the Euler and the Mellin-Barnes integration. Note that we must be careful when doing this, because some of the poles in $z$ could be generated by the Euler integration. We checked numerically that in the present case the exchange of the two integrations is allowed and produces the same answer.
We now close the $z$-contour to the right and take residues at $z=n-\eps, n\in\mathbb{N}$.
         \beq\bsp
 \cI^{(I)}_{\rm MB}&(\kappa,t_1,t_2)=-{x_2^{-\eps}\over 2\Gamma(1+\eps)\,\Gamma(1-\eps)}\,\int_0^1\rd v\,\sum_{n=0}^\infty\,
{(1-\eps)_n\over n!}\,(-x_2)^n\\
&\times v^{\eps-1-n}\,(1-v)^{-\eps-1-n}\,\Big(1-v\Big(1-{x_1\over x_2}\Big)\Big)^n\\
  &\times
  \Bigg\{ \ln ^2(1-v)  + \ln ^2v+ \ln ^2\left(1-v
  \left(1-\frac{x_1}{x_2}\right)\right)+\ln ^2x_2 + \psi (n+1)^2+ \psi (-\epsilon +n+1)^2 
  \\&\qquad+\pi^2
  +2 \ln (1-v) \ln v -2\ln (1-v) \ln \left(1-v
  \left(1-\frac{x_1}{x_2}\right)\right)
     \\&\qquad-2 \ln v \ln \left(1-v
  \left(1-\frac{x_1}{x_2}\right)\right)
  -2 \ln (1-v) \ln x_2-2 \ln v \ln x_2
     \\&\qquad
     +2 \ln \left(1-v
  \left(1-\frac{x_1}{x_2}\right)\right) \ln x_2
  +2 \ln (1-v)\, \psi (n+1)+2 \ln v\, \psi (n+1) 
     \\&\qquad-2 \ln \left(1-v
  \left(1-\frac{x_1}{x_2}\right)\right)\, \psi (n+1)
    -2 \ln x_2\, \psi (n+1)
-2 \ln (1-v)\, \psi (-\epsilon +n+1)
   \\&\qquad
  -2 \ln v \,\psi (-\epsilon +n+1)+2 \ln \left(1-v
  \left(1-\frac{x_1}{x_2}\right)\right)\, \psi (-\epsilon +n+1)
        \\&\qquad
 +2\ln x_2\, \psi (-\epsilon
  +n+1)
-2\psi (n+1) \psi (-\epsilon +n+1)
  - \psi ^{(1)}(n+1)
           \\&\qquad+ \psi ^{(1)}(-\epsilon +n+1)\Bigg\}.
  \esp\eeq
To continue, we perform the $\eps$-expansion under the integration sign\footnote{Note that we must be careful when we do this, since the integrals might contain poles coming from the terms $1/v(1-v)$. As we will see in the following however these poles are spurious and cancel out.}
\beq
\cI^{(I)}_{\rm MB}(\kappa,t_1,t_2)= \cI^{(I)}_0(x_1,x_2)+\eps\,\cI^{(I)}_1(x_1,x_2)+\ord(\eps^2). \label{eq:ix1x2}
\eeq
We find
\beq\bsp\label{cI0}
\cI^{(I)}_0(x_1,x_2)&\, = \int_0^1\,\rd v\,{i^{(0)}(x_1,x_2,v)\over v^2-x_1 v+x_2 v-v-x_2},
  \esp\eeq
and
\beq\bsp
\label{cI1}
\cI^{(I)}_1(x_1,x_2)&\, = \int_0^1\,\rd v\,{i^{(1)}(x_1,x_2,v)\over v^2+(-x_1+x_2-1)v-x_2}.
  \esp\eeq
where $i^{(0)}$ and $i^{(1)}$ are functions depending on (poly)logarithms of weight 2 and 3 respectively in $x_1$, $x_2$ and $v$ (See Appendix~\ref{app:integrands} for the explicit expressions). Note that this implies that $\cI^{(I)}_0(x_1,x_2)$ and $\cI^{(I)}_1(x_1,x_2)$ will have uniform weight 3 and 4 respectively, as expected. Furthermore note that the poles in $v=0$ and $v=1$ have cancelled out. However, we still need to be careful with the quadratic polynomial in the denominator of the integrand, since it might vanish in the integration region. We analyze this situation in the rest of this section.

We know already that the phase space boundaries in Region I require 
\beq
\sqrt{x_1}+\sqrt{x_2}<1.
\eeq
This subspace of the square $[0,1]\times[0,1]$ is at the same time the domain of the integral $\cI(x_1,x_2;\eps)$.
We can divide this domain further into
\begin{enumerate}
\item Region I(a): $x_1<x_2$.
\item Region I(b): $x_2<x_1$.
\end{enumerate}

We now turn to the quadratic denominator in Eqs.~(\ref{cI0}) and~(\ref{cI1}). The roots of this quadratic polynomial are
\beq\bsp\label{l1l2}
\lambda_1\equiv\lambda_1(x_1,x_2)&\,={1\over 2}\,\left(1+x_1-x_2-\sqrt{\lambda_K}\right),\\
\lambda_2\equiv\lambda_1(x_1,x_2)&\,={1\over 2}\,\left(1+x_1-x_2+\sqrt{\lambda_K}\right),
\esp\eeq
where $\lambda_K$ denotes the K\"allen function
\beq\label{kaellen}
\lambda_K\equiv \lambda_K(x_1,x_2)= \lambda(x_1,x_2,-1)=1+x_1^2+x_2^2+2x_1+2x_2-2x_1x_2.
\eeq
First, let us note that $\lambda_K(x_1,x_2)>0$, $\forall \,(x_1,x_2)\in[0,1]\times[0,1]$, and hence the square root in Eq.~(\ref{l1l2}) is well defined in the Region I. Second, it is easy to show that on the square $[0,1]\times[0,1]$ we have,
\beq\label{l1l2PS}
-1<\lambda_1(x_1,x_2)<0\qquad \textrm{ and }\qquad 1<\lambda_2(x_1,x_2)<2.
\eeq
For later convenience, let us note at this point the following useful identities
\beq\bsp\label{l1l2relations}
\lambda_1\lambda_2&\,=-x_2,\\
\lambda_1+\lambda_2&\,=1+x_1-x_2,\\
 \lambda_1-\lambda_2&\,=-\sqrt{\lambda_K},\\
\left(1-\frac{1}{\lambda_1}\right)\left(1-\frac{1}{\lambda_2}\right)&=\,\frac{x_1}{x_2}.
\esp\eeq
From Eq.~(\ref{l1l2PS}) it follows now immediately that the quadratic denominators in Eqs.~(\ref{cI0}) and~(\ref{cI1}) do not vanish in the whole integration  range $[0,1]$, and hence all the integrals in Eqs.~(\ref{cI0}) and~(\ref{cI1}) are convergent. Using partial fractioning and the relations~(\ref{l1l2relations}) we can write
\beq\bsp\label{cI02}
\cI^{(I)}_0(x_1,x_2)&\, = {1\over \sqrt{\lambda_K}}\,\int_0^1\,\rd v\,{i^{(0)}(x_1,x_2,v)\over v-\lambda_2}-{1\over \sqrt{\lambda_K}}\,\int_0^1\,\rd v\,{i^{(0)}(x_1,x_2,v)\over v-\lambda_1},
  \esp\eeq
and
\beq\bsp
\label{cI12}
\cI^{(I)}_1(x_1,x_2)&\, = {1\over \sqrt{\lambda_K}}\,\int_0^1\,\rd v\,{i^{(1)}(x_1,x_2,v)\over v-\lambda_2}-{1\over \sqrt{\lambda_K}}\,\int_0^1\,\rd v\,{i^{(1)}(x_1,x_2,v)\over v-\lambda_1}.
  \esp\eeq
Let us introduce at this point the function
\beq
\lambda_3\equiv\lambda_3(x_1,x_2)=\frac{x_2}{x_2-x_1}=\frac{\lambda_1\lambda_2}{\lambda_1+\lambda_2-1}.
\eeq
This function appears in $i^{(0)}$ and $i^{(1)}$ through the logarithm
\beq
\ln\left(1-v\left(1-\frac{x_1}{x_2}\right)\right)=\ln\left(1-\frac{v}{\lambda_3}\right)=\int_0^v{\rd t\over t-\lambda_3}.
\eeq
It is easy to see that this logarithm is well defined in Region I, since
\begin{enumerate}
\item
In Region I(a), $x_1<x_2$, and hence $\lambda_3>1$.
\item
In Region I(b), $x_2<x_1$, and hence $\lambda_3<0$.
\end{enumerate}
Note however that $\lambda_3$ diverges on the diagonal $x_1=x_2$. This is not a contradiction since
\beq
\lim_{x_2\rightarrow x_1}\int_0^v{\rd t\over t-\lambda_3(x_1,x_2)}=\lim_{\lambda_3\rightarrow \infty}\int_0^v{\rd t\over t-\lambda_3}=0,
\eeq
in agreement with
\beq
\lim_{x_2\rightarrow x_1}\ln\left(1-v\left(1-\frac{x_1}{x_2}\right)\right)=0.
\eeq

We now evaluate the integrals $\cI^{(I)}_0$ and $\cI^{(I)}_1$ explicitly. To this effect, let us introduce some generalized multiple polylogarithms defined by
\beq
G(a,\vec w;z)=\int_0^z\rd t\,f(a,t)\,G(\vec w;t),
\eeq
where 
\beq
f(a,t)={1\over t-a}.
\eeq
If all indices are zero we define
\beq
G(\vec 0_n;z)=\int_1^z{\rd t\over t} \,G(\vec 0_{n-1};t)={1\over n!}\,\ln^n z.
\eeq
In particular cases the $G$-functions reduce to ordinary logarithms and polylogarithms,
\beq
G(\vec a_n;z)=\frac{1}{n!}\,\ln^n\left(1-\frac{z}{a}\right),\quad G(\vec 0_{n-1},a;z) = -\textrm{Li}_n\left(\frac{z}{a}\right).
\eeq
Note that these definitions are straightforward generalizations of the harmonic polylogarithms, and hence these functions inherit all the properties of the \emph{HPL's}. In particular they fulfill a shuffle algebra
\beq
G(\vec w_1;z)\,G(\vec w_2;z)=\sum_{\vec w=\vec w_1\uplus \vec w_2}\,G(\vec w;z).
\eeq
If the weight vector $\vec w$ has a parametric dependence on a second variable, then we obtain multidimensional harmonic polylogarithms.
Finally, let us introduce the following set of functions, which will be useful to write down the answer for the pentagon
\beq
M(\vec w)\equiv G(\vec w;1).
\eeq
The $M$-functions defined in this way are in fact Goncharov's multiple polylogarithm (up to a sign).
For a more detailed discussions of these functions, and their relations to Goncharov's multiple polylogarithm, see Appendix~\ref{app:hpl}.
It is clear from the definition that these functions form a shuffle algebra
\beq
M(\vec w_1)\,M(\vec w_2)=\sum_{\vec w=\vec w_1\uplus \vec w_2}\,M(\vec w).
\eeq

Using these functions we can easily integrate $\cI^{(I)}_0$ and $\cI^{(I)}_1$. We illustrate this procedure explicitly for the integral
\beq
\int_0^1{\rd v\over v-\lambda_1}\,\ln \left(v
  \left(\frac{x_1}{x_2}-1\right)+1\right)\, \ln (1-v).
  \eeq
  First we can express all logarithms in terms of the $G$-functions we defined:
  \beq\bsp
  \ln \left(v
  \left(\frac{x_1}{x_2}-1\right)+1\right)\, \ln (1-v) &\,= \ln \left(1-\frac{v}{\lambda_3}\right)\, \ln (1-v)\\
  &\,=G(\lambda_3;v)\,G(1;v)\\
  &\,=G(\lambda_3,1;v)+G(1,\lambda_3;v).
  \esp\eeq
  Then we get
\beq\bsp
\int_0^1{\rd v\over v-\lambda_1}&\,\ln \left(v
  \left(\frac{x_1}{x_2}-1\right)+1\right)\, \ln (1-v) \\
  &=\,\int_0^1\,\rd v\,{G(\lambda_3,1;v)\over v-\lambda_1}+\int_0^1\,\rd v\,{G(1,\lambda_3;v)\over v-\lambda_1}\\
  &=G(\lambda_1,\lambda_3,1;1)+G(\lambda_1,1,\lambda_3;1)\\
  &=M(\lambda_1,\lambda_3,1)+M(\lambda_1,1,\lambda_3).
\esp   \eeq
All other integrals can be performed in exactly the same way, and we can hence express $\cI^{(I)}_0$ as a combination of $M$-functions. We find
\btxtsloppy
\parbox{130mm}{\raggedright\(
\cI^{(I)}_0(x_1,x_2)=\)}
 \raggedleft\refstepcounter{equation}(\theequation)\label{eq:cI0}\\
 \raggedright
\({1\over\sqrt{\lambda_K}}\,\Bigg\{\Big(\frac{1}{2} \ln ^2x_2+\frac{\pi ^2}{2}\Big) M\big(\lambda _1\big)+\Big(-\frac{1}{2} \ln ^2x_2-\frac{\pi ^2}{2}\Big) M\big(\lambda _2\big)-\ln x_2 M\big(\lambda _1,0\big)-\ln x_2 M\big(\lambda _1,1\big)+\ln x_2 M\big(\lambda _1,\lambda _3\big)+\ln x_2 M\big(\lambda _2,0\big)+\ln x_2 M\big(\lambda _2,1\big)-\ln x_2 M\big(\lambda _2,\lambda _3\big)+M\big(\lambda _1,0,0\big)+M\big(\lambda _1,0,1\big)-M\big(\lambda _1,0,\lambda _3\big)+M\big(\lambda _1,1,0\big)+M\big(\lambda _1,1,1\big)-M\big(\lambda _1,1,\lambda _3\big)-M\big(\lambda _1,\lambda _3,0\big)-M\big(\lambda _1,\lambda _3,1\big)+M\big(\lambda _1,\lambda _3,\lambda _3\big)-M\big(\lambda _2,0,0\big)-M\big(\lambda _2,0,1\big)+M\big(\lambda _2,0,\lambda _3\big)-M\big(\lambda _2,1,0\big)-M\big(\lambda _2,1,1\big)+M\big(\lambda _2,1,\lambda _3\big)+M\big(\lambda _2,\lambda _3,0\big)+M\big(\lambda _2,\lambda _3,1\big)-M\big(\lambda _2,\lambda _3,\lambda _3\big)\Bigg\}.\)
\etxtsloppy
Note that this expression is of uniform weight 3, as expected.

The integration of $\cI^{(I)}_1$ can be done in a similar way as for $\cI^{(I)}_0$. However, there is a slight complication. The function $i^{(1)}$ contains polylogarithms of the form 
\beq
\mathrm{Li}_n\left({v(x_1-x_2)+x_2\over v(v-1)}\right).
\eeq
In order to perform the integration in terms of $G$-functions, we have to express these functions in terms of objects of the form $G(\ldots;v)$. In Appendix~\ref{app:LiRed} we show that the following identities hold:
\btxtsloppy
\parbox{130mm}{\raggedright\(
\mathrm{Li}_2\left({v(x_1-x_2)+x_2\over v(v-1)}\right)=\)}
 \raggedleft\refstepcounter{equation}(\theequation)\label{eq:Li2RegI}\\
 \raggedright\(
-\frac{1}{2} \ln ^2x_1+\ln x_2 \ln
  x_1-\ln
  ^2x_2-G(0,0;v)-G(0,1;v)+G\left(0,\lambda
  _1;v\right)
  +G\left(0,\lambda
  _2;v\right)-G(1,0;v)-G(1,1;v)+G\left(1,\lambda
  _1;v\right)+G\left(1,\lambda _2;v\right)+G\left(\lambda
  _3,0;v\right)+G\left(\lambda _3,1;v\right)
  -G\left(\lambda _3,\lambda
  _1;v\right)-G\left(\lambda _3,\lambda _2;v\right)+G(0;v) \ln
  x_2+G(1;v) \ln x_2-G\left(\lambda
  _3;v\right) \ln x_2-M\left(0,\lambda
  _1\right)
  -M\left(0,\lambda _2\right)+M\left(\lambda
  _1,1\right)+M\left(\lambda _2,1\right)-M\left(\lambda
  _3,0\right)-M\left(\lambda _3,1\right)+M\left(\lambda _3,\lambda
  _1\right)+M\left(\lambda _3,\lambda _2\right)-\frac{\pi ^2}{6}.
\)\etxtsloppy

\btxtsloppy
\parbox{130mm}{\raggedright\(
\mathrm{Li}_3\left({v(x_1-x_2)+x_2\over v(v-1)}\right)=\)}
 \raggedleft\refstepcounter{equation}(\theequation)\label{eq:Li3RegI}\\
 \raggedright \(
-\Bigg(\frac{1}{6} \ln ^3x_1-\frac{1}{2} G(0;v) \ln
  ^2x_1-\frac{1}{2} G(1;v) \ln
  ^2x_1+\frac{1}{2} G\left(\lambda _3;v\right) \ln
  ^2x_1-\frac{1}{2} M\left(\lambda _3\right) \ln
  ^2x_1+G(0;v) \ln x_2 \ln
  x_1+G(1;v) \ln x_2 \ln
  x_1-G\left(\lambda _3;v\right) \ln x_2 \ln
  x_1+\ln x_2 M\left(\lambda _3\right) \ln
  x_1+\frac{1}{6} \pi ^2 \ln x_1-G(0;v) \ln
  ^2x_2-G(1;v) \ln ^2x_2+G\left(\lambda
  _3;v\right) \ln ^2x_2-\frac{1}{6} \pi ^2
  G(0;v)-\frac{1}{6} \pi ^2 G(1;v)+\frac{1}{6} \pi ^2 G\left(\lambda
  _3;v\right)-G(0,0,0;v)-G(0,0,1;v)+G\left(0,0,\lambda
  _1;v\right)+G\left(0,0,\lambda
  _2;v\right)-G(0,1,0;v)-G(0,1,1;v)+G\left(0,1,\lambda
  _1;v\right)+G\left(0,1,\lambda _2;v\right)+G\left(0,\lambda
  _3,0;v\right)+G\left(0,\lambda _3,1;v\right)-G\left(0,\lambda
  _3,\lambda _1;v\right)-G\left(0,\lambda _3,\lambda
  _2;v\right)-G(1,0,0;v)-G(1,0,1;v)+G\left(1,0,\lambda
  _1;v\right)+G\left(1,0,\lambda
  _2;v\right)-G(1,1,0;v)-G(1,1,1;v)+G\left(1,1,\lambda
  _1;v\right)+G\left(1,1,\lambda _2;v\right)+G\left(1,\lambda
  _3,0;v\right)+G\left(1,\lambda _3,1;v\right)-G\left(1,\lambda
  _3,\lambda _1;v\right)-G\left(1,\lambda _3,\lambda
  _2;v\right)+G\left(\lambda _3,0,0;v\right)+G\left(\lambda
  _3,0,1;v\right)-G\left(\lambda _3,0,\lambda _1;v\right)-G\left(\lambda
  _3,0,\lambda _2;v\right)+G\left(\lambda _3,1,0;v\right)+G\left(\lambda
  _3,1,1;v\right)-G\left(\lambda _3,1,\lambda _1;v\right)-G\left(\lambda
  _3,1,\lambda _2;v\right)-G\left(\lambda _3,\lambda
  _3,0;v\right)-G\left(\lambda _3,\lambda _3,1;v\right)+G\left(\lambda
  _3,\lambda _3,\lambda _1;v\right)+G\left(\lambda _3,\lambda _3,\lambda
  _2;v\right)+G(0,0;v) \ln x_2+G(0,1;v) \ln
  x_2-G\left(0,\lambda _3;v\right) \ln
  x_2+G(1,0;v) \ln x_2+G(1,1;v) \ln
  x_2-G\left(1,\lambda _3;v\right) \ln
  x_2-G\left(\lambda _3,0;v\right) \ln
  x_2-G\left(\lambda _3,1;v\right) \ln
  x_2+G\left(\lambda _3,\lambda _3;v\right) \ln
  x_2-\ln ^2x_2 M\left(\lambda
  _3\right)-\frac{1}{6} \pi ^2 M\left(\lambda _3\right)-G(0;v)
  M\left(0,\lambda _1\right)-G(1;v) M\left(0,\lambda
  _1\right)+G\left(\lambda _3;v\right) M\left(0,\lambda
  _1\right)-M\left(\lambda _3\right) M\left(0,\lambda _1\right)-G(0;v)
  M\left(0,\lambda _2\right)-G(1;v) M\left(0,\lambda
  _2\right)+G\left(\lambda _3;v\right) M\left(0,\lambda
  _2\right)-M\left(\lambda _3\right) M\left(0,\lambda _2\right)+\ln
  x_2 M\left(0,\lambda _3\right)+G(0;v) M\left(\lambda
  _1,1\right)+G(1;v) M\left(\lambda _1,1\right)-G\left(\lambda
  _3;v\right) M\left(\lambda _1,1\right)+M\left(\lambda _3\right)
  M\left(\lambda _1,1\right)+G(0;v) M\left(\lambda _2,1\right)+G(1;v)
  M\left(\lambda _2,1\right)-G\left(\lambda _3;v\right) M\left(\lambda
  _2,1\right)+M\left(\lambda _3\right) M\left(\lambda _2,1\right)-G(0;v)
  M\left(\lambda _3,0\right)-G(1;v) M\left(\lambda
  _3,0\right)+G\left(\lambda _3;v\right) M\left(\lambda _3,0\right)+\ln
  x_2 M\left(\lambda _3,0\right)-M\left(\lambda _3\right)
  M\left(\lambda _3,0\right)-G(0;v) M\left(\lambda _3,1\right)-G(1;v)
  M\left(\lambda _3,1\right)+G\left(\lambda _3;v\right) M\left(\lambda
  _3,1\right)-M\left(\lambda _3\right) M\left(\lambda _3,1\right)+G(0;v)
  M\left(\lambda _3,\lambda _1\right)+G(1;v) M\left(\lambda _3,\lambda
  _1\right)-G\left(\lambda _3;v\right) M\left(\lambda _3,\lambda
  _1\right)+M\left(\lambda _3\right) M\left(\lambda _3,\lambda
  _1\right)+G(0;v) M\left(\lambda _3,\lambda _2\right)+G(1;v)
  M\left(\lambda _3,\lambda _2\right)-G\left(\lambda _3;v\right)
  M\left(\lambda _3,\lambda _2\right)+M\left(\lambda _3\right)
  M\left(\lambda _3,\lambda _2\right)-\ln x_2
  M\left(\lambda _3,\lambda _3\right)-M\left(0,0,\lambda
  _1\right)-M\left(0,0,\lambda _2\right)+M\left(0,\lambda
  _1,1\right)+M\left(0,\lambda _2,1\right)-M\left(0,\lambda
  _3,0\right)-M\left(0,\lambda _3,1\right)+M\left(0,\lambda _3,\lambda
  _1\right)+M\left(0,\lambda _3,\lambda _2\right)-M\left(\lambda
  _1,1,1\right)-M\left(\lambda _2,1,1\right)-M\left(\lambda
  _3,0,0\right)+M\left(\lambda _3,0,\lambda _1\right)+M\left(\lambda
  _3,0,\lambda _2\right)+M\left(\lambda _3,1,1\right)-M\left(\lambda
  _3,\lambda _1,1\right)-M\left(\lambda _3,\lambda
  _2,1\right)+M\left(\lambda _3,\lambda _3,0\right)+M\left(\lambda
  _3,\lambda _3,1\right)-M\left(\lambda _3,\lambda _3,\lambda
  _1\right)-M\left(\lambda _3,\lambda _3,\lambda _2\right)\Bigg).
\)\etxtsloppy
Using these identities we can express $i^{(1)}$ completely in terms of $G$ and $M$-functions, and perform the integration in exactly the same way as for $\cI^{(I)}_0$. The result is
\btxtsloppy
\parbox{130mm}{\raggedright\(
\cI^{(I)}_1(x_1,x_2)=\)}
 \raggedleft\refstepcounter{equation}(\theequation)\label{eq:cI1}\\
 \raggedright 
 \(
{1\over\sqrt{\lambda_K}}\,\Bigg\{
\Big(\ln ^2x_2+\frac{\pi ^2}{6}\Big) M\big(\lambda _1,0\big)-\frac{5}{6} \pi ^2 M\big(\lambda _1,1\big)+\Big(\frac{1}{2} \ln ^2x_2+\frac{\pi ^2}{2}\Big) M\big(\lambda _1,\lambda _1\big)+\)\\
\(\Big(\frac{1}{2} \ln ^2x_2+\frac{\pi ^2}{2}\Big) M\big(\lambda _1,\lambda _2\big)+
\Big(-\frac{1}{2} \ln ^2x_1\,+\ln x_2 \ln x_1\,-2 \ln ^2x_2-\frac{\pi ^2}{3}\Big) M\big(\lambda _1,\lambda _3\big)+\)\\
\(\Big(-\ln ^2x_2-\frac{\pi ^2}{6}\Big) M\big(\lambda _2,0\big)+\frac{5}{6} \pi ^2 M\big(\lambda _2,1\big)+\Big(-\frac{1}{2} \ln ^2x_2-\frac{\pi ^2}{2}\Big) M\big(\lambda _2,\lambda _1\big)+\)\\
\(\Big(-\frac{1}{2} \ln ^2x_2-\frac{\pi ^2}{2}\Big) M\big(\lambda _2,\lambda _2\big)+\Big(\frac{1}{2} \ln ^2x_1\,-\ln x_2 \ln x_1\,+2 \ln ^2x_2+\frac{\pi ^2}{3}\Big) M\big(\lambda _2,\lambda _3\big)+\)\\
\(\Big(-\frac{1}{2} \ln ^2x_1\,+\ln x_2 \ln x_1\,-\ln ^2x_2-\frac{\pi ^2}{6}\Big) M\big(\lambda _3,\lambda _1\big)+\)\\
\(\Big(\frac{1}{2} \ln ^2x_1\,-\ln x_2 \ln x_1\,+\ln ^2x_2+\frac{\pi ^2}{6}\Big) M\big(\lambda _3,\lambda _2\big)-2 \ln x_2 M\big(0,\lambda _1,\lambda _1\big)+\ln x_2 M\big(0,\lambda _1,\lambda _3\big)+2 \ln x_2 M\big(0,\lambda _2,\lambda _2\big)-\ln x_2 M\big(0,\lambda _2,\lambda _3\big)+\ln x_2 M\big(0,\lambda _3,\lambda _1\big)-\ln x_2 M\big(0,\lambda _3,\lambda _2\big)-2 \ln x_2 M\big(\lambda _1,0,0\big)-\ln x_2 M\big(\lambda _1,0,\lambda _1\big)-\ln x_2 M\big(\lambda _1,0,\lambda _2\big)+2 \ln x_2 M\big(\lambda _1,0,\lambda _3\big)+2 \ln x_2 M\big(\lambda _1,1,1\big)+\ln x_2 M\big(\lambda _1,1,\lambda _1\big)-\ln x_2 M\big(\lambda _1,1,\lambda _2\big)-\ln x_2 M\big(\lambda _1,1,\lambda _3\big)-\ln x_2 M\big(\lambda _1,\lambda _1,0\big)+\ln x_2 M\big(\lambda _1,\lambda _1,1\big)+\ln x_2 M\big(\lambda _1,\lambda _1,\lambda _3\big)-\ln x_2 M\big(\lambda _1,\lambda _2,0\big)-\ln x_2 M\big(\lambda _1,\lambda _2,1\big)+\ln x_2 M\big(\lambda _1,\lambda _2,\lambda _3\big)+2 \ln x_2 M\big(\lambda _1,\lambda _3,0\big)-\ln x_2 M\big(\lambda _1,\lambda _3,1\big)+\ln x_2 M\big(\lambda _1,\lambda _3,\lambda _1\big)+\ln x_2 M\big(\lambda _1,\lambda _3,\lambda _2\big)-2 \ln x_2 M\big(\lambda _1,\lambda _3,\lambda _3\big)+2 \ln x_2 M\big(\lambda _2,0,0\big)+\ln x_2 M\big(\lambda _2,0,\lambda _1\big)+\ln x_2 M\big(\lambda _2,0,\lambda _2\big)-2 \ln x_2 M\big(\lambda _2,0,\lambda _3\big)-2 \ln x_2 M\big(\lambda _2,1,1\big)+\ln x_2 M\big(\lambda _2,1,\lambda _1\big)-\ln x_2 M\big(\lambda _2,1,\lambda _2\big)+\ln x_2 M\big(\lambda _2,1,\lambda _3\big)+\ln x_2 M\big(\lambda _2,\lambda _1,0\big)+\ln x_2 M\big(\lambda _2,\lambda _1,1\big)-\ln x_2 M\big(\lambda _2,\lambda _1,\lambda _3\big)+\ln x_2 M\big(\lambda _2,\lambda _2,0\big)-\ln x_2 M\big(\lambda _2,\lambda _2,1\big)-\ln x_2 M\big(\lambda _2,\lambda _2,\lambda _3\big)-2 \ln x_2 M\big(\lambda _2,\lambda _3,0\big)+\ln x_2 M\big(\lambda _2,\lambda _3,1\big)-\ln x_2 M\big(\lambda _2,\lambda _3,\lambda _1\big)-\ln x_2 M\big(\lambda _2,\lambda _3,\lambda _2\big)+2 \ln x_2 M\big(\lambda _2,\lambda _3,\lambda _3\big)-\ln x_2 M\big(\lambda _3,1,\lambda _1\big)+\ln x_2 M\big(\lambda _3,1,\lambda _2\big)-\ln x_2 M\big(\lambda _3,\lambda _1,1\big)+2 \ln x_2 M\big(\lambda _3,\lambda _1,\lambda _1\big)-\ln x_2 M\big(\lambda _3,\lambda _1,\lambda _3\big)+\ln x_2 M\big(\lambda _3,\lambda _2,1\big)-2 \ln x_2 M\big(\lambda _3,\lambda _2,\lambda _2\big)+\ln x_2 M\big(\lambda _3,\lambda _2,\lambda _3\big)-\ln x_2 M\big(\lambda _3,\lambda _3,\lambda _1\big)+\ln x_2 M\big(\lambda _3,\lambda _3,\lambda _2\big)-2 M\big(0,0,\lambda _1,\lambda _1\big)+2 M\big(0,0,\lambda _2,\lambda _2\big)-M\big(0,\lambda _1,0,\lambda _1\big)-M\big(0,\lambda _1,0,\lambda _2\big)+M\big(0,\lambda _1,1,\lambda _1\big)-M\big(0,\lambda _1,1,\lambda _2\big)+2 M\big(0,\lambda _1,\lambda _1,1\big)-2 M\big(0,\lambda _1,\lambda _1,\lambda _3\big)-M\big(0,\lambda _1,\lambda _3,0\big)-M\big(0,\lambda _1,\lambda _3,1\big)-M\big(0,\lambda _1,\lambda _3,\lambda _1\big)+M\big(0,\lambda _1,\lambda _3,\lambda _2\big)+M\big(0,\lambda _2,0,\lambda _1\big)+M\big(0,\lambda _2,0,\lambda _2\big)+M\big(0,\lambda _2,1,\lambda _1\big)-M\big(0,\lambda _2,1,\lambda _2\big)-2 M\big(0,\lambda _2,\lambda _2,1\big)+2 M\big(0,\lambda _2,\lambda _2,\lambda _3\big)+M\big(0,\lambda _2,\lambda _3,0\big)+M\big(0,\lambda _2,\lambda _3,1\big)-M\big(0,\lambda _2,\lambda _3,\lambda _1\big)+M\big(0,\lambda _2,\lambda _3,\lambda _2\big)-M\big(0,\lambda _3,0,\lambda _1\big)+M\big(0,\lambda _3,0,\lambda _2\big)-M\big(0,\lambda _3,1,\lambda _1\big)+M\big(0,\lambda _3,1,\lambda _2\big)-M\big(0,\lambda _3,\lambda _1,0\big)-M\big(0,\lambda _3,\lambda _1,1\big)+M\big(0,\lambda _3,\lambda _2,0\big)+M\big(0,\lambda _3,\lambda _2,1\big)+2 M\big(\lambda _1,0,0,0\big)-M\big(\lambda _1,0,0,\lambda _1\big)-M\big(\lambda _1,0,0,\lambda _2\big)-M\big(\lambda _1,0,0,\lambda _3\big)-2 M\big(\lambda _1,0,1,1\big)+M\big(\lambda _1,0,1,\lambda _3\big)+M\big(\lambda _1,0,\lambda _1,1\big)-M\big(\lambda _1,0,\lambda _1,\lambda _3\big)+M\big(\lambda _1,0,\lambda _2,1\big)-M\big(\lambda _1,0,\lambda _2,\lambda _3\big)-2 M\big(\lambda _1,0,\lambda _3,0\big)-2 M\big(\lambda _1,1,0,1\big)+M\big(\lambda _1,1,0,\lambda _3\big)-2 M\big(\lambda _1,1,1,0\big)-4 M\big(\lambda _1,1,1,1\big)-M\big(\lambda _1,1,1,\lambda _1\big)+M\big(\lambda _1,1,1,\lambda _2\big)+3 M\big(\lambda _1,1,1,\lambda _3\big)-M\big(\lambda _1,1,\lambda _1,1\big)+M\big(\lambda _1,1,\lambda _1,\lambda _3\big)+M\big(\lambda _1,1,\lambda _2,1\big)-M\big(\lambda _1,1,\lambda _2,\lambda _3\big)+M\big(\lambda _1,1,\lambda _3,0\big)+3 M\big(\lambda _1,1,\lambda _3,1\big)+M\big(\lambda _1,1,\lambda _3,\lambda _1\big)-M\big(\lambda _1,1,\lambda _3,\lambda _2\big)-2 M\big(\lambda _1,1,\lambda _3,\lambda _3\big)+M\big(\lambda _1,\lambda _1,0,0\big)+M\big(\lambda _1,\lambda _1,0,1\big)-M\big(\lambda _1,\lambda _1,0,\lambda _3\big)+M\big(\lambda _1,\lambda _1,1,0\big)-M\big(\lambda _1,\lambda _1,1,1\big)+M\big(\lambda _1,\lambda _1,1,\lambda _3\big)-M\big(\lambda _1,\lambda _1,\lambda _3,0\big)+M\big(\lambda _1,\lambda _1,\lambda _3,1\big)+M\big(\lambda _1,\lambda _1,\lambda _3,\lambda _3\big)+M\big(\lambda _1,\lambda _2,0,0\big)+M\big(\lambda _1,\lambda _2,0,1\big)-M\big(\lambda _1,\lambda _2,0,\lambda _3\big)+M\big(\lambda _1,\lambda _2,1,0\big)+M\big(\lambda _1,\lambda _2,1,1\big)-M\big(\lambda _1,\lambda _2,1,\lambda _3\big)-M\big(\lambda _1,\lambda _2,\lambda _3,0\big)-M\big(\lambda _1,\lambda _2,\lambda _3,1\big)+M\big(\lambda _1,\lambda _2,\lambda _3,\lambda _3\big)-3 M\big(\lambda _1,\lambda _3,0,0\big)+3 M\big(\lambda _1,\lambda _3,1,1\big)+M\big(\lambda _1,\lambda _3,1,\lambda _1\big)-M\big(\lambda _1,\lambda _3,1,\lambda _2\big)-2 M\big(\lambda _1,\lambda _3,1,\lambda _3\big)+M\big(\lambda _1,\lambda _3,\lambda _1,1\big)+M\big(\lambda _1,\lambda _3,\lambda _1,\lambda _3\big)-M\big(\lambda _1,\lambda _3,\lambda _2,1\big)+M\big(\lambda _1,\lambda _3,\lambda _2,\lambda _3\big)-2 M\big(\lambda _1,\lambda _3,\lambda _3,1\big)+M\big(\lambda _1,\lambda _3,\lambda _3,\lambda _1\big)+M\big(\lambda _1,\lambda _3,\lambda _3,\lambda _2\big)-2 M\big(\lambda _2,0,0,0\big)+M\big(\lambda _2,0,0,\lambda _1\big)+M\big(\lambda _2,0,0,\lambda _2\big)+M\big(\lambda _2,0,0,\lambda _3\big)+2 M\big(\lambda _2,0,1,1\big)-M\big(\lambda _2,0,1,\lambda _3\big)-M\big(\lambda _2,0,\lambda _1,1\big)+M\big(\lambda _2,0,\lambda _1,\lambda _3\big)-M\big(\lambda _2,0,\lambda _2,1\big)+M\big(\lambda _2,0,\lambda _2,\lambda _3\big)+2 M\big(\lambda _2,0,\lambda _3,0\big)+2 M\big(\lambda _2,1,0,1\big)-M\big(\lambda _2,1,0,\lambda _3\big)+2 M\big(\lambda _2,1,1,0\big)+4 M\big(\lambda _2,1,1,1\big)-M\big(\lambda _2,1,1,\lambda _1\big)+M\big(\lambda _2,1,1,\lambda _2\big)-3 M\big(\lambda _2,1,1,\lambda _3\big)-M\big(\lambda _2,1,\lambda _1,1\big)+M\big(\lambda _2,1,\lambda _1,\lambda _3\big)+M\big(\lambda _2,1,\lambda _2,1\big)-M\big(\lambda _2,1,\lambda _2,\lambda _3\big)-M\big(\lambda _2,1,\lambda _3,0\big)-3 M\big(\lambda _2,1,\lambda _3,1\big)+M\big(\lambda _2,1,\lambda _3,\lambda _1\big)-M\big(\lambda _2,1,\lambda _3,\lambda _2\big)+2 M\big(\lambda _2,1,\lambda _3,\lambda _3\big)-M\big(\lambda _2,\lambda _1,0,0\big)-M\big(\lambda _2,\lambda _1,0,1\big)+M\big(\lambda _2,\lambda _1,0,\lambda _3\big)-M\big(\lambda _2,\lambda _1,1,0\big)-M\big(\lambda _2,\lambda _1,1,1\big)+M\big(\lambda _2,\lambda _1,1,\lambda _3\big)+M\big(\lambda _2,\lambda _1,\lambda _3,0\big)+M\big(\lambda _2,\lambda _1,\lambda _3,1\big)-M\big(\lambda _2,\lambda _1,\lambda _3,\lambda _3\big)-M\big(\lambda _2,\lambda _2,0,0\big)-M\big(\lambda _2,\lambda _2,0,1\big)+M\big(\lambda _2,\lambda _2,0,\lambda _3\big)-M\big(\lambda _2,\lambda _2,1,0\big)+M\big(\lambda _2,\lambda _2,1,1\big)-M\big(\lambda _2,\lambda _2,1,\lambda _3\big)+M\big(\lambda _2,\lambda _2,\lambda _3,0\big)-M\big(\lambda _2,\lambda _2,\lambda _3,1\big)-M\big(\lambda _2,\lambda _2,\lambda _3,\lambda _3\big)+3 M\big(\lambda _2,\lambda _3,0,0\big)-3 M\big(\lambda _2,\lambda _3,1,1\big)+M\big(\lambda _2,\lambda _3,1,\lambda _1\big)-M\big(\lambda _2,\lambda _3,1,\lambda _2\big)+2 M\big(\lambda _2,\lambda _3,1,\lambda _3\big)+M\big(\lambda _2,\lambda _3,\lambda _1,1\big)-M\big(\lambda _2,\lambda _3,\lambda _1,\lambda _3\big)-M\big(\lambda _2,\lambda _3,\lambda _2,1\big)-M\big(\lambda _2,\lambda _3,\lambda _2,\lambda _3\big)+2 M\big(\lambda _2,\lambda _3,\lambda _3,1\big)-M\big(\lambda _2,\lambda _3,\lambda _3,\lambda _1\big)-M\big(\lambda _2,\lambda _3,\lambda _3,\lambda _2\big)-M\big(\lambda _3,0,0,\lambda _1\big)+M\big(\lambda _3,0,0,\lambda _2\big)-M\big(\lambda _3,0,\lambda _1,0\big)-M\big(\lambda _3,0,\lambda _1,\lambda _3\big)+M\big(\lambda _3,0,\lambda _2,0\big)+M\big(\lambda _3,0,\lambda _2,\lambda _3\big)-M\big(\lambda _3,0,\lambda _3,\lambda _1\big)+M\big(\lambda _3,0,\lambda _3,\lambda _2\big)+M\big(\lambda _3,1,1,\lambda _1\big)-M\big(\lambda _3,1,1,\lambda _2\big)+M\big(\lambda _3,1,\lambda _1,1\big)-M\big(\lambda _3,1,\lambda _1,\lambda _3\big)-M\big(\lambda _3,1,\lambda _2,1\big)+M\big(\lambda _3,1,\lambda _2,\lambda _3\big)-M\big(\lambda _3,1,\lambda _3,\lambda _1\big)+M\big(\lambda _3,1,\lambda _3,\lambda _2\big)-M\big(\lambda _3,\lambda _1,0,0\big)-M\big(\lambda _3,\lambda _1,0,\lambda _3\big)+M\big(\lambda _3,\lambda _1,1,1\big)-M\big(\lambda _3,\lambda _1,1,\lambda _3\big)+2 M\big(\lambda _3,\lambda _1,\lambda _1,\lambda _3\big)-M\big(\lambda _3,\lambda _1,\lambda _3,0\big)-M\big(\lambda _3,\lambda _1,\lambda _3,1\big)+2 M\big(\lambda _3,\lambda _1,\lambda _3,\lambda _1\big)+M\big(\lambda _3,\lambda _2,0,0\big)+M\big(\lambda _3,\lambda _2,0,\lambda _3\big)-M\big(\lambda _3,\lambda _2,1,1\big)+M\big(\lambda _3,\lambda _2,1,\lambda _3\big)-2 M\big(\lambda _3,\lambda _2,\lambda _2,\lambda _3\big)+M\big(\lambda _3,\lambda _2,\lambda _3,0\big)+M\big(\lambda _3,\lambda _2,\lambda _3,1\big)-2 M\big(\lambda _3,\lambda _2,\lambda _3,\lambda _2\big)-M\big(\lambda _3,\lambda _3,0,\lambda _1\big)+M\big(\lambda _3,\lambda _3,0,\lambda _2\big)-M\big(\lambda _3,\lambda _3,1,\lambda _1\big)+M\big(\lambda _3,\lambda _3,1,\lambda _2\big)-M\big(\lambda _3,\lambda _3,\lambda _1,0\big)-M\big(\lambda _3,\lambda _3,\lambda _1,1\big)+2 M\big(\lambda _3,\lambda _3,\lambda _1,\lambda _1\big)+M\big(\lambda _3,\lambda _3,\lambda _2,0\big)+M\big(\lambda _3,\lambda _3,\lambda _2,1\big)-2 M\big(\lambda _3,\lambda _3,\lambda _2,\lambda _2\big)+\)\\
\(M\big(\lambda _2\big) \Big(-\frac{1}{6} \ln ^3x_1\,+\frac{1}{2} \ln x_2 \ln ^2x_1\,-\ln ^2x_2 \ln x_1\,-\frac{1}{6} \pi ^2 \ln x_1\,+\ln ^3x_2+\frac{1}{3} \pi ^2 \ln x_2-\zeta_3\Big)+M\big(\lambda _1\big) \Big(\frac{1}{6} \ln ^3x_1\,-\frac{1}{2} \ln x_2 \ln ^2x_1\,+\ln ^2x_2 \ln x_1\,+\frac{1}{6} \pi ^2 \ln x_1\,-\ln ^3x_2-\frac{1}{3} \pi ^2 \ln x_2+\zeta_3\Big)\Bigg\}.
\)
\etxtsloppy
Note that $\cI^{(I)}_1$ is of uniform weight 4, as expected.


\subsection{Evaluation of the Mellin-Barnes integral in Region II}

In this section we evaluate the pentagon in Region II. Since the Regions II(a) and II(b) are related simply by $x_1\leftrightarrow x_2$, we only concentrate on Region II(a). The procedure is very similar to Region I, \ie we start by deriving a twofold Mellin-Barnes representation for the pentagon in this region, which we then reduce to an Euler-type integral, and finally we express the result in terms of Goncharov's multiple polylogarithms.
Performing this rescaling~(\ref{scaling}) in the Mellin-Barnes representation~(\ref{MBpent}) and after the change of variable
  \beq\bsp
 z_3\,&\rightarrow z'-z_1-z_2,\\
 z_4\,&\rightarrow -z-2z'+z_1,
 \esp\eeq
 we find
 \beq\bsp
 I_5^D(1,&1, 1, 1, 1;{Q_i^2})=\frac{-\,e^{\gamma_E\eps}\,(-s)^{-\epsilon -2}}{\Gamma (1-2 \epsilon )}\\
 &\times\,{1\over (2\pi i)^4}\,\mbint\rd z \,\rd z'\,\rd z_1\,\rd z_2\,
  \left(\frac{s_1}{s}\right)^{-z-2z'+z_1} \left(\frac{s_2}{s}\right)^{z_1}
  \left(\frac{t_1}{s}\right)^{z_2} \left(\frac{t_2}{s}\right)^{-z_1-z_2+z'}
  \lambda ^{-z} \\
  &\qquad\times\Gamma \left(-z_1\right) \Gamma \left(-z_2\right) \Gamma
  \left(z_1+z_2+1\right) \Gamma \left(-\epsilon -z'-1\right) \Gamma \left(\epsilon
  -z+z_1-z'+2\right) \\
  &\qquad\times\Gamma \left(-z-z_2-z'+1\right) \Gamma \left(z_1+z_2-z'\right)
  \Gamma \left(-\epsilon +z+z'-1\right) \Gamma \left(-z_1+z'+1\right)\\
  &\qquad\times \Gamma
  \left(z+z_1+2 z_2+2 \left(-z_1-z_2+z'\right)\right).
  \esp\eeq
  We think of the integration over $z$ as the last one, and we analyze how poles in $\Gamma(\ldots+z)$ with leading behavior $\lambda^{-2}$ might arise. There is only one possibility, coming from the product $\Gamma \left(-\epsilon -z'-1\right)\Gamma \left(-\epsilon +z+z'-1\right)$. Taking the residues at $z'=-1-\eps+n'$, $n'\in \mathbb{N}$, we find 
  \beq
  \lambda^{-z}\Gamma \left(-\epsilon +z+z'-1\right)\rightarrow\lambda^{-z}\Gamma \left(-2\epsilon-2 +z+n'\right).
  \eeq
Taking the residues at $z=2+2\eps-n-v'$, $n\in \mathbb{N}$, we find 
  \beq
  \lambda^z\Gamma \left(-2\epsilon-2 +z+n'\right)\rightarrow \lambda^{-2-2\eps+n+n'}.
  \eeq
  Since we are only interested in the leading behavior in $\lambda^{-2}$, we only keep the terms in $n=n'=0$. Hence, we find a twofold Mellin-Barnes representation for the pentagon in multi-Regge kinematics,
   \beq\bsp\label{MBMRpentRegII}
 I_5^D(1,&1, 1, 1, 1;{Q_i^2})=r_\Gamma\,\,e^{\gamma_E\eps}\,\frac{(-\kappa)^{-\epsilon}}{st_2}\,
 \cI^{(IIa)}_{\rm MB}(\kappa,t_1,t_2)
  \esp\eeq
  with
   \beq\bsp
 \cI^{(IIa)}_{\rm MB}&(\kappa,t_1,t_2)={-y_1^\eps\over\Gamma(1+\eps)\,\Gamma(1-\eps)^2}\,{1\over (2\pi i)^2}\,\mbint\,\rd z_1\,\rd z_2\,
  y_1^{z_1} y_2^{z_2} \,\Gamma \left(-\epsilon
  -z_1\right) \Gamma \left(-z_1\right)^2\\
  &\qquad\times \Gamma \left(z_1+1\right) \Gamma
  \left(-\epsilon -z_2\right) \Gamma \left(-z_2\right) \Gamma \left(z_1+z_2+1\right)
  \Gamma \left(\epsilon +z_1+z_2+1\right),
  \esp\eeq
  where $y_1$ and $y_2$ are defined in \Eqn{eq:y12def}. We checked that if we close the integration contours to the right, and take residues, we reproduce exactly the expression of the pentagon obtained from NDIM, \Eqn{eq:PentNDRegIIa}.

We now evaluate the Mellin-Barnes representation~(\ref{MBMRpentRegII}) and we derive an Euler integral representation for the pentagon in multi-Regge kinematics. Let us concentrate only on the Mellin-Barnes integral.
We can now use the identity~(\ref{eq:MBtoEuler}), and we checked again numerically that we can exchange the Euler and the Mellin-Barnes integration. We find
         \beq\bsp
  \cI^{(IIa)}_{\rm MB}&(\kappa,t_1,t_2)={-y_1^\eps\over\Gamma(1+\eps)\,\Gamma(1-\eps)^2}\,{1\over 2\pi i}\,\int_0^1\rd v\,\mbint\,\rd z_1\,(1-v)^{z_1-\epsilon } v^{\epsilon +z_1} y_1^{z_1}
   \\
  &\qquad\times\left(1-v \left(1-y_2\right)\right)^{-z_1-1} \Gamma \left(-\epsilon -z_1\right)
  \Gamma \left(-z_1\right)^2 \Gamma \left(z_1+1\right)^3.
  \esp\eeq
  Finally, we close the $z_1$-contour to the right and take residues at $z_1=n_1-\eps, n_1\in\mathbb{N}$. As in the case of $\cI^{(I)}_{\rm MB}$, we can sum up the series of residues and expand the integrand in a power series in $\eps$,
\beq
 \cI^{(IIa)}_{\rm MB}(\kappa,t_1,t_2)=\cI^{(IIa)}_0(y_1,y_2)+\eps\,\cI^{(IIa)}_1(y_1,y_2)+\ord(\eps^2). \label{eq:jy1y2}
\eeq
We find
\beq\bsp\label{cJ0}
\cI^{(IIa)}_0(y_1,y_2)&\, =\int_0^1\,\rd v\,{j^{(0)}(y_1,y_2,v)\over (y_1v^2-y_1 v-y_2v+v-1)},
  \esp\eeq
and
\beq\bsp
\label{cJ1}
\cI^{(IIa)}_1(y_1,y_2)&\, = \int_0^1\,\rd v\,{j^{(1)}(y_1,y_2,v)\over (y_1v^2-y_1 v-y_2v+v-1)}.
  \esp\eeq
where $j^{(0)}$ and $j^{(1)}$ are functions depending on (poly)logarithms of weight 2 and 3 respectively in $y_1$, $y_2$ and $v$ (See Appendix~\ref{app:integrands} for the explicit expressions). Note that this implies that $\cI^{(IIa)}_0(y_1,y_2)$ and $\cI^{(IIa)}_1(y_1,y_2)$ will have uniform weight 3 and 4 respectively, as expected. Furthermore note that the poles in $v=0$ and $v=1$ have cancelled out. We need however still to be careful with the quadratic polynomial in the denominator of the integrand, since it might vanish in the integration region. We will analyze this situation in the rest of this section.

We know already that the phase space boundaries in Region II(a) require 
\beq
-\sqrt{x_1}+\sqrt{x_2}>1.
\eeq
Since this implies $x_2>1$ and $x_2>x_1$, we get from Eq.~(\ref{eq:y12def}) that $0<y_1,y_2<1$.
We now turn to the quadratic denominator in Eqs.~(\ref{cJ0}) and~(\ref{cJ1}). The roots of this quadratic polynomial are
\beq\bsp\label{J1J2}
\lambda'_1\equiv\lambda'_1(y_1,y_2)&\,={1\over 2y_1}\,\left(-1+y_1+y_2-\sqrt{\lambda'_K}\right),\\
\lambda'_2\equiv\lambda'_1(y_1,y_2)&\,={1\over 2y_1}\,\left(-1+y_1+y_2+\sqrt{\lambda'_K}\right),
\esp\eeq
where $\lambda'_K$ denotes the K\"allen function
\beq\label{kaellenII}
\lambda'_K\equiv \lambda'_K(y_1,y_2)= \lambda(-y_1,y_2,1)=1+y_1^2+y_2^2+2y_1-2y_2+2y_1y_2.
\eeq
First, let us note that $\lambda'_K(y_1,y_2)>0$ everywhere in Region II(a), and hence the square root in Eq.~(\ref{J1J2}) is well-defined. Second, it is easy to show that  we have,
\beq\label{J1J2PS}
-1<\lambda'_1(y_1,y_2)<0\qquad \textrm{ and }\qquad 1<\lambda'_2(y_1,y_2)<2.
\eeq
For later convenience, let us note at this point the following useful identities
\beq\bsp\label{J1J2relations}
\lambda'_1\lambda'_2&\,={-1\over y_2},\\
\lambda'_1+\lambda'_2&\,={-1+y_1-y_2\over y_1},\\
 \lambda'_1-\lambda'_2&\,=-{\sqrt{\lambda'_K}\over y_1},\\
\left(1-\frac{1}{\lambda'_1}\right)\left(1-\frac{1}{\lambda'_2}\right)&=y_2.
\esp\eeq
From Eq.~(\ref{J1J2PS}) it follows now immediately that the quadratic denominators in Eqs.~(\ref{cJ0}) and~(\ref{cJ1}) do not vanish in the whole integration $v\in[0,1]$, and hence all the integrals in Eqs.~(\ref{cJ0}) and~(\ref{cJ1}) are convergent. Using partial fractioning and the relations~(\ref{J1J2relations}) we can write
\beq\bsp\label{cJ02}
\cI^{(IIa)}_0(y_1,y_2)&\, = {-1\over\sqrt{\lambda'_K}}\,\int_0^1\,\rd v\,{j^{(0)}(y_1,y_2,v)\over v-\lambda_2}+{1\over\sqrt{\lambda'_K}}\int_0^1\,\rd v\,{j^{(0)}(y_1,y_2,v)\over v-\lambda_1},
  \esp\eeq
and
\beq\bsp
\label{cJ12}
\cI^{(IIa)}_1(y_1,y_2)&\, = {-1\over\sqrt{\lambda'_K}}\,\int_0^1\,\rd v\,{j^{(1)}(y_1,y_2,v)\over v-\lambda_2}+{1\over\sqrt{\lambda'_K}}\int_0^1\,\rd v\,{j^{(1)}(y_1,y_2,v)\over v-\lambda_1}.
  \esp\eeq
Let us conclude this section by introducing the function
\beq
\lambda'_3\equiv\lambda'_3(y_1,y_2)={1\over 1-y_2}= {1\over 1-t_1/t_2}=\lambda_3.
\eeq
This function appears in $j^{(0)}$ and $j^{(1)}$ through the logarithm
\beq
\ln\left(1-v\left(1-y_2\right)\right)=\ln\left(1-\frac{v}{\lambda'_3}\right)=\int_0^v{\rd t\over t-\lambda'_3}.
\eeq
We now evaluate the integrals $\cI^{(IIa)}_0$ and $\cI^{(IIa)}_1$ explicitly.  For $\cI^{(IIa)}_0$, we find
\btxtsloppy
\parbox{130mm}{\raggedright\(
\cI^{(IIa)}_0(y_1,y_2)=\)}
 \raggedleft\refstepcounter{equation}(\theequation)\label{eq:cJ0}\\
 \raggedright \(
{1\over\sqrt{\lambda'_K}}\,
\Bigg\{\Big(-\frac{1}{2} \ln ^2y_1\,-\frac{\pi ^2}{2}\Big) M\big(\lambda' _1\big)+\Big(\frac{1}{2} \ln ^2y_1\,+\frac{\pi ^2}{2}\Big) M\big(\lambda' _2\big)-\ln y_1\, M\big(\lambda' _1,0\big)-\ln y_1\, M\big(\lambda' _1,1\big)+\ln y_1\, M\big(\lambda' _1,\lambda' _3\big)+\ln y_1\, M\big(\lambda' _2,0\big)+\ln y_1\, M\big(\lambda' _2,1\big)-\ln y_1\, M\big(\lambda' _2,\lambda' _3\big)-M\big(\lambda' _1,0,0\big)-M\big(\lambda' _1,0,1\big)+M\big(\lambda' _1,0,\lambda' _3\big)-M\big(\lambda' _1,1,0\big)-M\big(\lambda' _1,1,1\big)+M\big(\lambda' _1,1,\lambda' _3\big)+M\big(\lambda' _1,\lambda' _3,0\big)+M\big(\lambda' _1,\lambda' _3,1\big)-M\big(\lambda' _1,\lambda' _3,\lambda' _3\big)+M\big(\lambda' _2,0,0\big)+M\big(\lambda' _2,0,1\big)-M\big(\lambda' _2,0,\lambda' _3\big)+M\big(\lambda' _2,1,0\big)+M\big(\lambda' _2,1,1\big)-M\big(\lambda' _2,1,\lambda' _3\big)-M\big(\lambda' _2,\lambda' _3,0\big)-M\big(\lambda' _2,\lambda' _3,1\big)+M\big(\lambda' _2,\lambda' _3,\lambda' _3\big)\Bigg\}.
\)\etxtsloppy
Note that this expression is of uniform weight 3, as expected.

The integration of $\cI^{(IIa)}_1$ can be done in a similar way as for $\cI^{(IIa)}_0$. However, there is again a slight complication. The function $j^{(1)}$ contains polylogarithms of the form 
\beq
\mathrm{Li}_n\left({y_2 v-v+1\over v(v-1)y_1}\right).
\eeq
In order to perform the integration in terms of $G$-functions, we have to express these functions in terms of objects of the form $G(\ldots;v)$. In Appendix~\ref{app:LiRedII} we show that the following identities hold:
\btxtsloppy
\parbox{130mm}{\raggedright\(
\mathrm{Li}_2\left({y_2 v-v+1\over v(v-1)y_1}\right)=\)}
 \raggedleft\refstepcounter{equation}(\theequation)\label{eq:Li2RegII}\\
 \raggedright \(
-\frac{1}{2} \ln
  ^2\left(\frac{y_2}{y_1}\right)-G(0,0;v)-G(0,1;v)-G\left(0,\lambda
  _1',1\right)+G\left(0,\lambda _1';v\right)-G\left(0,\lambda
  _2',1\right)+G\left(0,\lambda
  _2';v\right)-G(1,0;v)-G(1,1;v)+G\left(1,\lambda
  _1';v\right)+G\left(1,\lambda _2';v\right)+G\left(\lambda
  _1',1,1\right)+G\left(\lambda _2',1,1\right)-G\left(\lambda
  _3',0,1\right)+G\left(\lambda _3',0;v\right)-G\left(\lambda
  _3',1,1\right)+G\left(\lambda _3',1;v\right)+G\left(\lambda
  _3',\lambda _1',1\right)-G\left(\lambda _3',\lambda
  _1';v\right)+G\left(\lambda _3',\lambda _2',1\right)-G\left(\lambda
  _3',\lambda _2';v\right)-G(0;v) \ln y_1-G(1;v) \ln
  y_1+G\left(\lambda _3';v\right) \ln
  y_1-\ln y_1 \ln y_2-\frac{\pi
  ^2}{6}.
\)\etxtsloppy

\btxtsloppy
\parbox{130mm}{\raggedright\(
\mathrm{Li}_3\left({y_2 v-v+1\over v(v-1)y_1}\right)=\)}
 \raggedleft\refstepcounter{equation}(\theequation)\label{eq:Li3RegII}\\
 \raggedright\(
\frac{1}{6} \ln ^3\left(\frac{y_2}{y_1}\right)+\frac{1}{2} G(0,1) \ln
  ^2\left(\frac{y_2}{y_1}\right)-\frac{1}{2} G(0;v) \ln
  ^2\left(\frac{y_2}{y_1}\right)-\frac{1}{2} G(1;v) \ln
  ^2\left(\frac{y_2}{y_1}\right)-\frac{1}{2} G\left(\lambda _3',1\right)
  \ln ^2\left(\frac{y_2}{y_1}\right)+\frac{1}{2} G\left(\lambda
  _3';v\right) \ln ^2\left(\frac{y_2}{y_1}\right)+\frac{1}{6} \pi ^2
  \ln \left(\frac{y_2}{y_1}\right)+\frac{1}{6} \pi ^2
  G(0,1)-\frac{1}{6} \pi ^2 G(0;v)-\frac{1}{6} \pi ^2 G(1;v)-\frac{1}{6}
  \pi ^2 G\left(\lambda _3',1\right)+\frac{1}{6} \pi ^2 G\left(\lambda
  _3';v\right)+G(0,1) G\left(0,\lambda _1',1\right)-G(0;v)
  G\left(0,\lambda _1',1\right)-G(1;v) G\left(0,\lambda
  _1',1\right)-G\left(\lambda _3',1\right) G\left(0,\lambda
  _1',1\right)+G\left(\lambda _3';v\right) G\left(0,\lambda
  _1',1\right)+G(0,1) G\left(0,\lambda _2',1\right)-G(0;v)
  G\left(0,\lambda _2',1\right)-G(1;v) G\left(0,\lambda
  _2',1\right)-G\left(\lambda _3',1\right) G\left(0,\lambda
  _2',1\right)+G\left(\lambda _3';v\right) G\left(0,\lambda
  _2',1\right)-G(0,1) G\left(\lambda _1',1,1\right)+G(0;v)
  G\left(\lambda _1',1,1\right)+G(1;v) G\left(\lambda
  _1',1,1\right)+G\left(\lambda _3',1\right) G\left(\lambda
  _1',1,1\right)-G\left(\lambda _3';v\right) G\left(\lambda
  _1',1,1\right)-G(0,1) G\left(\lambda _2',1,1\right)+G(0;v)
  G\left(\lambda _2',1,1\right)+G(1;v) G\left(\lambda
  _2',1,1\right)+G\left(\lambda _3',1\right) G\left(\lambda
  _2',1,1\right)-G\left(\lambda _3';v\right) G\left(\lambda
  _2',1,1\right)+G(0,1) G\left(\lambda _3',0,1\right)-G(0;v)
  G\left(\lambda _3',0,1\right)-G(1;v) G\left(\lambda
  _3',0,1\right)-G\left(\lambda _3',1\right) G\left(\lambda
  _3',0,1\right)+G\left(\lambda _3';v\right) G\left(\lambda
  _3',0,1\right)+G(0,1) G\left(\lambda _3',1,1\right)-G(0;v)
  G\left(\lambda _3',1,1\right)-G(1;v) G\left(\lambda
  _3',1,1\right)-G\left(\lambda _3',1\right) G\left(\lambda
  _3',1,1\right)+G\left(\lambda _3';v\right) G\left(\lambda
  _3',1,1\right)-G(0,1) G\left(\lambda _3',\lambda _1',1\right)+G(0;v)
  G\left(\lambda _3',\lambda _1',1\right)+G(1;v) G\left(\lambda
  _3',\lambda _1',1\right)+G\left(\lambda _3',1\right) G\left(\lambda
  _3',\lambda _1',1\right)-G\left(\lambda _3';v\right) G\left(\lambda
  _3',\lambda _1',1\right)-G(0,1) G\left(\lambda _3',\lambda
  _2',1\right)+G(0;v) G\left(\lambda _3',\lambda _2',1\right)+G(1;v)
  G\left(\lambda _3',\lambda _2',1\right)+G\left(\lambda _3',1\right)
  G\left(\lambda _3',\lambda _2',1\right)-G\left(\lambda _3';v\right)
  G\left(\lambda _3',\lambda
  _2',1\right)+G(0,0,0,1)-G(0,0,0;v)-G(0,0,1;v)-G\left(0,0,\lambda
  _1',1\right)+G\left(0,0,\lambda _1';v\right)-G\left(0,0,\lambda
  _2',1\right)+G\left(0,0,\lambda
  _2';v\right)-G(0,1,0;v)-G(0,1,1;v)+G\left(0,1,\lambda
  _1';v\right)+G\left(0,1,\lambda _2';v\right)+G\left(0,\lambda
  _1',1,1\right)+G\left(0,\lambda _2',1,1\right)-G\left(0,\lambda
  _3',0,1\right)+G\left(0,\lambda _3',0;v\right)-G\left(0,\lambda
  _3',1,1\right)+G\left(0,\lambda _3',1;v\right)+G\left(0,\lambda
  _3',\lambda _1',1\right)-G\left(0,\lambda _3',\lambda
  _1';v\right)+G\left(0,\lambda _3',\lambda
  _2',1\right)-G\left(0,\lambda _3',\lambda
  _2';v\right)-G(1,0,0;v)-G(1,0,1;v)+G\left(1,0,\lambda
  _1';v\right)+G\left(1,0,\lambda
  _2';v\right)-G(1,1,0;v)-G(1,1,1;v)+G\left(1,1,\lambda
  _1';v\right)+G\left(1,1,\lambda _2';v\right)+G\left(1,\lambda
  _3',0;v\right)+G\left(1,\lambda _3',1;v\right)-G\left(1,\lambda
  _3',\lambda _1';v\right)-G\left(1,\lambda _3',\lambda
  _2';v\right)-G\left(\lambda _1',1,1,1\right)-G\left(\lambda
  _2',1,1,1\right)-G\left(\lambda _3',0,0,1\right)+G\left(\lambda
  _3',0,0;v\right)+G\left(\lambda _3',0,1;v\right)+G\left(\lambda
  _3',0,\lambda _1',1\right)-G\left(\lambda _3',0,\lambda
  _1';v\right)+G\left(\lambda _3',0,\lambda _2',1\right)-G\left(\lambda
  _3',0,\lambda _2';v\right)+G\left(\lambda
  _3',1,0;v\right)+G\left(\lambda _3',1,1,1\right)+G\left(\lambda
  _3',1,1;v\right)-G\left(\lambda _3',1,\lambda
  _1';v\right)-G\left(\lambda _3',1,\lambda _2';v\right)-G\left(\lambda
  _3',\lambda _1',1,1\right)-G\left(\lambda _3',\lambda
  _2',1,1\right)+G\left(\lambda _3',\lambda
  _3',0,1\right)-G\left(\lambda _3',\lambda
  _3',0;v\right)+G\left(\lambda _3',\lambda
  _3',1,1\right)-G\left(\lambda _3',\lambda
  _3',1;v\right)-G\left(\lambda _3',\lambda _3',\lambda
  _1',1\right)+G\left(\lambda _3',\lambda _3',\lambda
  _1';v\right)-G\left(\lambda _3',\lambda _3',\lambda
  _2',1\right)+G\left(\lambda _3',\lambda _3',\lambda
  _2';v\right)+G(0,0,1) \ln \left(y_1\right)-G(0,0;v) \ln
  \left(y_1\right)-G(0,1;v) \ln \left(y_1\right)-G\left(0,\lambda
  _3',1\right) \ln \left(y_1\right)+G\left(0,\lambda _3';v\right) \ln
  \left(y_1\right)-G(1,0;v) \ln \left(y_1\right)-G(1,1;v) \ln
  \left(y_1\right)+G\left(1,\lambda _3';v\right) \ln
  \left(y_1\right)-G\left(\lambda _3',0,1\right) \ln
  \left(y_1\right)+G\left(\lambda _3',0;v\right) \ln
  \left(y_1\right)+G\left(\lambda _3',1;v\right) \ln
  \left(y_1\right)+G\left(\lambda _3',\lambda _3',1\right) \ln
  \left(y_1\right)-G\left(\lambda _3',\lambda _3';v\right) \ln
  \left(y_1\right)+G(0,1) \ln \left(y_1\right) \ln
  \left(y_2\right)-G(0;v) \ln \left(y_1\right) \ln
  \left(y_2\right)-G(1;v) \ln \left(y_1\right) \ln
  \left(y_2\right)-G\left(\lambda _3',1\right) \ln \left(y_1\right)
  \ln \left(y_2\right)+G\left(\lambda _3';v\right) \ln
  \left(y_1\right) \ln \left(y_2\right).
\)\etxtsloppy
Using these identities we can express $j^{(1)}$ completely in terms of $G$ and $M$-functions, and perform the integration in exactly the same way as for $\cI^{(IIa)}_0$. The result is
\btxtsloppy
\parbox{130mm}{\raggedright\(
\cI^{(IIa)}_1(y_1,y_2)=\)}
 \raggedleft\refstepcounter{equation}(\theequation)\label{eq:cJ1}\\
 \raggedright\(
 {1\over\sqrt{\lambda'_K}}\,\Bigg\{
\Big(-2 \ln ^2y_1\,-\ln ^2y_2-\frac{\pi ^2}{2}\Big) M\big(\lambda' _1,0\big)+\Big(-\ln ^2y_1\,-\ln ^2y_2+\frac{\pi ^2}{2}\Big) M\big(\lambda' _1,1\big)+\)\\
\(\Big(-\frac{1}{2} \ln ^2y_1\,-\frac{\pi ^2}{2}\Big) M\big(\lambda' _1,\lambda' _1\big)+\Big(-\frac{1}{2} \ln ^2y_1\,-\frac{\pi ^2}{2}\Big) M\big(\lambda' _1,\lambda' _2\big)+\)\\
\(\Big(\frac{3}{2} \ln ^2y_1\,+\frac{1}{2} \ln ^2y_2+\frac{\pi ^2}{3}\Big) M\big(\lambda' _1,\lambda' _3\big)+\Big(2 \ln ^2y_1\,+\ln ^2y_2+\frac{\pi ^2}{2}\Big) M\big(\lambda' _2,0\big)+\)\\
\(\Big(\ln ^2y_1\,+\ln ^2y_2-\frac{\pi ^2}{2}\Big) M\big(\lambda' _2,1\big)+\Big(\frac{1}{2} \ln ^2y_1\,+\frac{\pi ^2}{2}\Big) M\big(\lambda' _2,\lambda' _1\big)+\Big(\frac{1}{2} \ln ^2y_1\,+\frac{\pi ^2}{2}\Big) M\big(\lambda' _2,\lambda' _2\big)+\Big(-\frac{3}{2} \ln ^2y_1\,-\frac{1}{2} \ln ^2y_2-\frac{\pi ^2}{3}\Big) M\big(\lambda' _2,\lambda' _3\big)+\Big(-\frac{1}{2} \ln ^2y_1\,-\frac{1}{2} \ln ^2y_2-\frac{\pi ^2}{6}\Big) M\big(\lambda' _3,\lambda' _1\big)+\)\\
\(\Big(\frac{1}{2} \ln ^2y_1\,+\frac{1}{2} \ln ^2y_2+\frac{\pi ^2}{6}\Big) M\big(\lambda' _3,\lambda' _2\big)-2 \ln y_1\, M\big(0,\lambda' _1,\lambda' _1\big)-\ln y_1\, M\big(0,\lambda' _1,\lambda' _3\big)+2 \ln y_1\, M\big(0,\lambda' _2,\lambda' _2\big)+\ln y_1\, M\big(0,\lambda' _2,\lambda' _3\big)-\ln y_1\, M\big(0,\lambda' _3,\lambda' _1\big)+\ln y_1\, M\big(0,\lambda' _3,\lambda' _2\big)-4 \ln y_1\, M\big(\lambda' _1,0,0\big)-2 \ln y_1\, M\big(\lambda' _1,0,1\big)-\ln y_1\, M\big(\lambda' _1,0,\lambda' _1\big)-\ln y_1\, M\big(\lambda' _1,0,\lambda' _2\big)+2 \ln y_1\, M\big(\lambda' _1,0,\lambda' _3\big)-2 \ln y_1\, M\big(\lambda' _1,1,0\big)+\ln y_1\, M\big(\lambda' _1,1,\lambda' _1\big)-\ln y_1\, M\big(\lambda' _1,1,\lambda' _2\big)+\ln y_1\, M\big(\lambda' _1,1,\lambda' _3\big)-\ln y_1\, M\big(\lambda' _1,\lambda' _1,0\big)+\ln y_1\, M\big(\lambda' _1,\lambda' _1,1\big)+\ln y_1\, M\big(\lambda' _1,\lambda' _1,\lambda' _3\big)-\ln y_1\, M\big(\lambda' _1,\lambda' _2,0\big)-\ln y_1\, M\big(\lambda' _1,\lambda' _2,1\big)+\ln y_1\, M\big(\lambda' _1,\lambda' _2,\lambda' _3\big)+2 \ln y_1\, M\big(\lambda' _1,\lambda' _3,0\big)+\ln y_1\, M\big(\lambda' _1,\lambda' _3,1\big)+\ln y_1\, M\big(\lambda' _1,\lambda' _3,\lambda' _1\big)+\ln y_1\, M\big(\lambda' _1,\lambda' _3,\lambda' _2\big)-2 \ln y_1\, M\big(\lambda' _1,\lambda' _3,\lambda' _3\big)+4 \ln y_1\, M\big(\lambda' _2,0,0\big)+2 \ln y_1\, M\big(\lambda' _2,0,1\big)+\ln y_1\, M\big(\lambda' _2,0,\lambda' _1\big)+\ln y_1\, M\big(\lambda' _2,0,\lambda' _2\big)-2 \ln y_1\, M\big(\lambda' _2,0,\lambda' _3\big)+2 \ln y_1\, M\big(\lambda' _2,1,0\big)+\ln y_1\, M\big(\lambda' _2,1,\lambda' _1\big)-\ln y_1\, M\big(\lambda' _2,1,\lambda' _2\big)-\ln y_1\, M\big(\lambda' _2,1,\lambda' _3\big)+\ln y_1\, M\big(\lambda' _2,\lambda' _1,0\big)+\ln y_1\, M\big(\lambda' _2,\lambda' _1,1\big)-\ln y_1\, M\big(\lambda' _2,\lambda' _1,\lambda' _3\big)+\ln y_1\, M\big(\lambda' _2,\lambda' _2,0\big)-\ln y_1\, M\big(\lambda' _2,\lambda' _2,1\big)-\ln y_1\, M\big(\lambda' _2,\lambda' _2,\lambda' _3\big)-2 \ln y_1\, M\big(\lambda' _2,\lambda' _3,0\big)-\ln y_1\, M\big(\lambda' _2,\lambda' _3,1\big)-\ln y_1\, M\big(\lambda' _2,\lambda' _3,\lambda' _1\big)-\ln y_1\, M\big(\lambda' _2,\lambda' _3,\lambda' _2\big)+2 \ln y_1\, M\big(\lambda' _2,\lambda' _3,\lambda' _3\big)-2 \ln y_1\, M\big(\lambda' _3,0,\lambda' _1\big)+2 \ln y_1\, M\big(\lambda' _3,0,\lambda' _2\big)-\ln y_1\, M\big(\lambda' _3,1,\lambda' _1\big)+\ln y_1\, M\big(\lambda' _3,1,\lambda' _2\big)-2 \ln y_1\, M\big(\lambda' _3,\lambda' _1,0\big)-\ln y_1\, M\big(\lambda' _3,\lambda' _1,1\big)+2 \ln y_1\, M\big(\lambda' _3,\lambda' _1,\lambda' _1\big)+\ln y_1\, M\big(\lambda' _3,\lambda' _1,\lambda' _3\big)+2 \ln y_1\, M\big(\lambda' _3,\lambda' _2,0\big)+\ln y_1\, M\big(\lambda' _3,\lambda' _2,1\big)-2 \ln y_1\, M\big(\lambda' _3,\lambda' _2,\lambda' _2\big)-\ln y_1\, M\big(\lambda' _3,\lambda' _2,\lambda' _3\big)+\ln y_1\, M\big(\lambda' _3,\lambda' _3,\lambda' _1\big)-\ln y_1\, M\big(\lambda' _3,\lambda' _3,\lambda' _2\big)-2 M\big(0,0,\lambda' _1,\lambda' _1\big)+2 M\big(0,0,\lambda' _2,\lambda' _2\big)-3 M\big(0,\lambda' _1,0,\lambda' _1\big)-3 M\big(0,\lambda' _1,0,\lambda' _2\big)-M\big(0,\lambda' _1,1,\lambda' _1\big)-3 M\big(0,\lambda' _1,1,\lambda' _2\big)-4 M\big(0,\lambda' _1,\lambda' _1,0\big)-2 M\big(0,\lambda' _1,\lambda' _1,1\big)+2 M\big(0,\lambda' _1,\lambda' _1,\lambda' _3\big)-M\big(0,\lambda' _1,\lambda' _3,0\big)-M\big(0,\lambda' _1,\lambda' _3,1\big)+M\big(0,\lambda' _1,\lambda' _3,\lambda' _1\big)+3 M\big(0,\lambda' _1,\lambda' _3,\lambda' _2\big)+3 M\big(0,\lambda' _2,0,\lambda' _1\big)+3 M\big(0,\lambda' _2,0,\lambda' _2\big)+3 M\big(0,\lambda' _2,1,\lambda' _1\big)+M\big(0,\lambda' _2,1,\lambda' _2\big)+4 M\big(0,\lambda' _2,\lambda' _2,0\big)+2 M\big(0,\lambda' _2,\lambda' _2,1\big)-2 M\big(0,\lambda' _2,\lambda' _2,\lambda' _3\big)+M\big(0,\lambda' _2,\lambda' _3,0\big)+M\big(0,\lambda' _2,\lambda' _3,1\big)-3 M\big(0,\lambda' _2,\lambda' _3,\lambda' _1\big)-M\big(0,\lambda' _2,\lambda' _3,\lambda' _2\big)-M\big(0,\lambda' _3,0,\lambda' _1\big)+M\big(0,\lambda' _3,0,\lambda' _2\big)-M\big(0,\lambda' _3,1,\lambda' _1\big)+M\big(0,\lambda' _3,1,\lambda' _2\big)-M\big(0,\lambda' _3,\lambda' _1,0\big)-M\big(0,\lambda' _3,\lambda' _1,1\big)+M\big(0,\lambda' _3,\lambda' _2,0\big)+M\big(0,\lambda' _3,\lambda' _2,1\big)-4 M\big(\lambda' _1,0,0,0\big)-2 M\big(\lambda' _1,0,0,1\big)-3 M\big(\lambda' _1,0,0,\lambda' _1\big)-3 M\big(\lambda' _1,0,0,\lambda' _2\big)+M\big(\lambda' _1,0,0,\lambda' _3\big)-2 M\big(\lambda' _1,0,1,0\big)-M\big(\lambda' _1,0,1,\lambda' _3\big)-2 M\big(\lambda' _1,0,\lambda' _1,0\big)+M\big(\lambda' _1,0,\lambda' _1,1\big)+M\big(\lambda' _1,0,\lambda' _1,\lambda' _3\big)-2 M\big(\lambda' _1,0,\lambda' _2,0\big)+M\big(\lambda' _1,0,\lambda' _2,1\big)+M\big(\lambda' _1,0,\lambda' _2,\lambda' _3\big)-2 M\big(\lambda' _1,0,\lambda' _3,1\big)+2 M\big(\lambda' _1,0,\lambda' _3,\lambda' _1\big)+2 M\big(\lambda' _1,0,\lambda' _3,\lambda' _2\big)-2 M\big(\lambda' _1,1,0,0\big)-M\big(\lambda' _1,1,0,\lambda' _3\big)+2 M\big(\lambda' _1,1,1,1\big)+M\big(\lambda' _1,1,1,\lambda' _1\big)+3 M\big(\lambda' _1,1,1,\lambda' _2\big)-3 M\big(\lambda' _1,1,1,\lambda' _3\big)+2 M\big(\lambda' _1,1,\lambda' _1,0\big)+3 M\big(\lambda' _1,1,\lambda' _1,1\big)-M\big(\lambda' _1,1,\lambda' _1,\lambda' _3\big)-2 M\big(\lambda' _1,1,\lambda' _2,0\big)+M\big(\lambda' _1,1,\lambda' _2,1\big)+M\big(\lambda' _1,1,\lambda' _2,\lambda' _3\big)-M\big(\lambda' _1,1,\lambda' _3,0\big)-3 M\big(\lambda' _1,1,\lambda' _3,1\big)+M\big(\lambda' _1,1,\lambda' _3,\lambda' _1\big)-M\big(\lambda' _1,1,\lambda' _3,\lambda' _2\big)+2 M\big(\lambda' _1,1,\lambda' _3,\lambda' _3\big)-M\big(\lambda' _1,\lambda' _1,0,0\big)+3 M\big(\lambda' _1,\lambda' _1,0,1\big)+M\big(\lambda' _1,\lambda' _1,0,\lambda' _3\big)+3 M\big(\lambda' _1,\lambda' _1,1,0\big)+5 M\big(\lambda' _1,\lambda' _1,1,1\big)-M\big(\lambda' _1,\lambda' _1,1,\lambda' _3\big)+M\big(\lambda' _1,\lambda' _1,\lambda' _3,0\big)-M\big(\lambda' _1,\lambda' _1,\lambda' _3,1\big)-M\big(\lambda' _1,\lambda' _1,\lambda' _3,\lambda' _3\big)-M\big(\lambda' _1,\lambda' _2,0,0\big)-M\big(\lambda' _1,\lambda' _2,0,1\big)+M\big(\lambda' _1,\lambda' _2,0,\lambda' _3\big)-M\big(\lambda' _1,\lambda' _2,1,0\big)-M\big(\lambda' _1,\lambda' _2,1,1\big)+M\big(\lambda' _1,\lambda' _2,1,\lambda' _3\big)+M\big(\lambda' _1,\lambda' _2,\lambda' _3,0\big)+M\big(\lambda' _1,\lambda' _2,\lambda' _3,1\big)-M\big(\lambda' _1,\lambda' _2,\lambda' _3,\lambda' _3\big)-M\big(\lambda' _1,\lambda' _3,0,0\big)-2 M\big(\lambda' _1,\lambda' _3,0,1\big)+2 M\big(\lambda' _1,\lambda' _3,0,\lambda' _1\big)+2 M\big(\lambda' _1,\lambda' _3,0,\lambda' _2\big)-2 M\big(\lambda' _1,\lambda' _3,1,0\big)-3 M\big(\lambda' _1,\lambda' _3,1,1\big)+M\big(\lambda' _1,\lambda' _3,1,\lambda' _1\big)-M\big(\lambda' _1,\lambda' _3,1,\lambda' _2\big)+2 M\big(\lambda' _1,\lambda' _3,1,\lambda' _3\big)+2 M\big(\lambda' _1,\lambda' _3,\lambda' _1,0\big)+M\big(\lambda' _1,\lambda' _3,\lambda' _1,1\big)-M\big(\lambda' _1,\lambda' _3,\lambda' _1,\lambda' _3\big)+2 M\big(\lambda' _1,\lambda' _3,\lambda' _2,0\big)-M\big(\lambda' _1,\lambda' _3,\lambda' _2,1\big)-M\big(\lambda' _1,\lambda' _3,\lambda' _2,\lambda' _3\big)+2 M\big(\lambda' _1,\lambda' _3,\lambda' _3,1\big)-M\big(\lambda' _1,\lambda' _3,\lambda' _3,\lambda' _1\big)-M\big(\lambda' _1,\lambda' _3,\lambda' _3,\lambda' _2\big)+4 M\big(\lambda' _2,0,0,0\big)+2 M\big(\lambda' _2,0,0,1\big)+3 M\big(\lambda' _2,0,0,\lambda' _1\big)+3 M\big(\lambda' _2,0,0,\lambda' _2\big)-M\big(\lambda' _2,0,0,\lambda' _3\big)+2 M\big(\lambda' _2,0,1,0\big)+M\big(\lambda' _2,0,1,\lambda' _3\big)+2 M\big(\lambda' _2,0,\lambda' _1,0\big)-M\big(\lambda' _2,0,\lambda' _1,1\big)-M\big(\lambda' _2,0,\lambda' _1,\lambda' _3\big)+2 M\big(\lambda' _2,0,\lambda' _2,0\big)-M\big(\lambda' _2,0,\lambda' _2,1\big)-M\big(\lambda' _2,0,\lambda' _2,\lambda' _3\big)+2 M\big(\lambda' _2,0,\lambda' _3,1\big)-2 M\big(\lambda' _2,0,\lambda' _3,\lambda' _1\big)-2 M\big(\lambda' _2,0,\lambda' _3,\lambda' _2\big)+2 M\big(\lambda' _2,1,0,0\big)+M\big(\lambda' _2,1,0,\lambda' _3\big)-2 M\big(\lambda' _2,1,1,1\big)-3 M\big(\lambda' _2,1,1,\lambda' _1\big)-M\big(\lambda' _2,1,1,\lambda' _2\big)+3 M\big(\lambda' _2,1,1,\lambda' _3\big)+2 M\big(\lambda' _2,1,\lambda' _1,0\big)-M\big(\lambda' _2,1,\lambda' _1,1\big)-M\big(\lambda' _2,1,\lambda' _1,\lambda' _3\big)-2 M\big(\lambda' _2,1,\lambda' _2,0\big)-3 M\big(\lambda' _2,1,\lambda' _2,1\big)+M\big(\lambda' _2,1,\lambda' _2,\lambda' _3\big)+M\big(\lambda' _2,1,\lambda' _3,0\big)+3 M\big(\lambda' _2,1,\lambda' _3,1\big)+M\big(\lambda' _2,1,\lambda' _3,\lambda' _1\big)-M\big(\lambda' _2,1,\lambda' _3,\lambda' _2\big)-2 M\big(\lambda' _2,1,\lambda' _3,\lambda' _3\big)+M\big(\lambda' _2,\lambda' _1,0,0\big)+M\big(\lambda' _2,\lambda' _1,0,1\big)-M\big(\lambda' _2,\lambda' _1,0,\lambda' _3\big)+M\big(\lambda' _2,\lambda' _1,1,0\big)+M\big(\lambda' _2,\lambda' _1,1,1\big)-M\big(\lambda' _2,\lambda' _1,1,\lambda' _3\big)-M\big(\lambda' _2,\lambda' _1,\lambda' _3,0\big)-M\big(\lambda' _2,\lambda' _1,\lambda' _3,1\big)+M\big(\lambda' _2,\lambda' _1,\lambda' _3,\lambda' _3\big)+M\big(\lambda' _2,\lambda' _2,0,0\big)-3 M\big(\lambda' _2,\lambda' _2,0,1\big)-M\big(\lambda' _2,\lambda' _2,0,\lambda' _3\big)-3 M\big(\lambda' _2,\lambda' _2,1,0\big)-5 M\big(\lambda' _2,\lambda' _2,1,1\big)+M\big(\lambda' _2,\lambda' _2,1,\lambda' _3\big)-M\big(\lambda' _2,\lambda' _2,\lambda' _3,0\big)+M\big(\lambda' _2,\lambda' _2,\lambda' _3,1\big)+M\big(\lambda' _2,\lambda' _2,\lambda' _3,\lambda' _3\big)+M\big(\lambda' _2,\lambda' _3,0,0\big)+2 M\big(\lambda' _2,\lambda' _3,0,1\big)-2 M\big(\lambda' _2,\lambda' _3,0,\lambda' _1\big)-2 M\big(\lambda' _2,\lambda' _3,0,\lambda' _2\big)+2 M\big(\lambda' _2,\lambda' _3,1,0\big)+3 M\big(\lambda' _2,\lambda' _3,1,1\big)+M\big(\lambda' _2,\lambda' _3,1,\lambda' _1\big)-M\big(\lambda' _2,\lambda' _3,1,\lambda' _2\big)-2 M\big(\lambda' _2,\lambda' _3,1,\lambda' _3\big)-2 M\big(\lambda' _2,\lambda' _3,\lambda' _1,0\big)+M\big(\lambda' _2,\lambda' _3,\lambda' _1,1\big)+M\big(\lambda' _2,\lambda' _3,\lambda' _1,\lambda' _3\big)-2 M\big(\lambda' _2,\lambda' _3,\lambda' _2,0\big)-M\big(\lambda' _2,\lambda' _3,\lambda' _2,1\big)+M\big(\lambda' _2,\lambda' _3,\lambda' _2,\lambda' _3\big)-2 M\big(\lambda' _2,\lambda' _3,\lambda' _3,1\big)+M\big(\lambda' _2,\lambda' _3,\lambda' _3,\lambda' _1\big)+M\big(\lambda' _2,\lambda' _3,\lambda' _3,\lambda' _2\big)-M\big(\lambda' _3,0,0,\lambda' _1\big)+M\big(\lambda' _3,0,0,\lambda' _2\big)-3 M\big(\lambda' _3,0,\lambda' _1,0\big)-2 M\big(\lambda' _3,0,\lambda' _1,1\big)+M\big(\lambda' _3,0,\lambda' _1,\lambda' _3\big)+3 M\big(\lambda' _3,0,\lambda' _2,0\big)+2 M\big(\lambda' _3,0,\lambda' _2,1\big)-M\big(\lambda' _3,0,\lambda' _2,\lambda' _3\big)-M\big(\lambda' _3,0,\lambda' _3,\lambda' _1\big)+M\big(\lambda' _3,0,\lambda' _3,\lambda' _2\big)+M\big(\lambda' _3,1,1,\lambda' _1\big)-M\big(\lambda' _3,1,1,\lambda' _2\big)-2 M\big(\lambda' _3,1,\lambda' _1,0\big)-M\big(\lambda' _3,1,\lambda' _1,1\big)+M\big(\lambda' _3,1,\lambda' _1,\lambda' _3\big)+2 M\big(\lambda' _3,1,\lambda' _2,0\big)+M\big(\lambda' _3,1,\lambda' _2,1\big)-M\big(\lambda' _3,1,\lambda' _2,\lambda' _3\big)-M\big(\lambda' _3,1,\lambda' _3,\lambda' _1\big)+M\big(\lambda' _3,1,\lambda' _3,\lambda' _2\big)-5 M\big(\lambda' _3,\lambda' _1,0,0\big)-4 M\big(\lambda' _3,\lambda' _1,0,1\big)+2 M\big(\lambda' _3,\lambda' _1,0,\lambda' _1\big)+2 M\big(\lambda' _3,\lambda' _1,0,\lambda' _2\big)+M\big(\lambda' _3,\lambda' _1,0,\lambda' _3\big)-4 M\big(\lambda' _3,\lambda' _1,1,0\big)-3 M\big(\lambda' _3,\lambda' _1,1,1\big)+2 M\big(\lambda' _3,\lambda' _1,1,\lambda' _1\big)+2 M\big(\lambda' _3,\lambda' _1,1,\lambda' _2\big)+M\big(\lambda' _3,\lambda' _1,1,\lambda' _3\big)+4 M\big(\lambda' _3,\lambda' _1,\lambda' _1,0\big)+4 M\big(\lambda' _3,\lambda' _1,\lambda' _1,1\big)-2 M\big(\lambda' _3,\lambda' _1,\lambda' _1,\lambda' _3\big)+M\big(\lambda' _3,\lambda' _1,\lambda' _3,0\big)+M\big(\lambda' _3,\lambda' _1,\lambda' _3,1\big)-2 M\big(\lambda' _3,\lambda' _1,\lambda' _3,\lambda' _2\big)+5 M\big(\lambda' _3,\lambda' _2,0,0\big)+4 M\big(\lambda' _3,\lambda' _2,0,1\big)-2 M\big(\lambda' _3,\lambda' _2,0,\lambda' _1\big)-2 M\big(\lambda' _3,\lambda' _2,0,\lambda' _2\big)-M\big(\lambda' _3,\lambda' _2,0,\lambda' _3\big)+4 M\big(\lambda' _3,\lambda' _2,1,0\big)+3 M\big(\lambda' _3,\lambda' _2,1,1\big)-2 M\big(\lambda' _3,\lambda' _2,1,\lambda' _1\big)-2 M\big(\lambda' _3,\lambda' _2,1,\lambda' _2\big)-M\big(\lambda' _3,\lambda' _2,1,\lambda' _3\big)-4 M\big(\lambda' _3,\lambda' _2,\lambda' _2,0\big)-4 M\big(\lambda' _3,\lambda' _2,\lambda' _2,1\big)+2 M\big(\lambda' _3,\lambda' _2,\lambda' _2,\lambda' _3\big)-M\big(\lambda' _3,\lambda' _2,\lambda' _3,0\big)-M\big(\lambda' _3,\lambda' _2,\lambda' _3,1\big)+2 M\big(\lambda' _3,\lambda' _2,\lambda' _3,\lambda' _1\big)-M\big(\lambda' _3,\lambda' _3,0,\lambda' _1\big)+M\big(\lambda' _3,\lambda' _3,0,\lambda' _2\big)-M\big(\lambda' _3,\lambda' _3,1,\lambda' _1\big)+M\big(\lambda' _3,\lambda' _3,1,\lambda' _2\big)-M\big(\lambda' _3,\lambda' _3,\lambda' _1,0\big)-M\big(\lambda' _3,\lambda' _3,\lambda' _1,1\big)+2 M\big(\lambda' _3,\lambda' _3,\lambda' _1,\lambda' _1\big)+M\big(\lambda' _3,\lambda' _3,\lambda' _2,0\big)+M\big(\lambda' _3,\lambda' _3,\lambda' _2,1\big)-2 M\big(\lambda' _3,\lambda' _3,\lambda' _2,\lambda' _2\big)+\)\\
\(M\big(\lambda' _1\big) \Big(-\frac{2}{3} \ln ^3y_1\,+\frac{1}{2} \ln y_2 \ln ^2y_1\,-\ln ^2y_2 \ln y_1\,-\frac{1}{2} \pi ^2 \ln y_1\,+\frac{1}{6} \ln ^3y_2+\frac{1}{6} \pi ^2 \ln y_2-\zeta_3\Big)+M\big(\lambda' _2\big) \Big(\frac{2}{3} \ln ^3y_1\,-\frac{1}{2} \ln y_2 \ln ^2y_1\,+\ln ^2y_2 \ln y_1\,+\frac{1}{2} \pi ^2 \ln y_1\,-\frac{1}{6} \ln ^3y_2-\frac{1}{6} \pi ^2 \ln y_2+\zeta_3\Big)\Bigg\}.
\)\etxtsloppy
Note that $\cI^{(IIa)}_1$ is of uniform weight 4, as expected.


\section{Analytic continuation to the physical region}
\label{sec:ACphys}

The result in the physical region where all $s$-type invariants are positive is obtained by performing analytic continuation on all $s$-type invariants according to the prescription
\beq
(-s)\to e^{-i\pi}\,s,\quad (-s_1)\to e^{-i\pi}\,s_1,\quad (-s_2)\to e^{-i\pi}\,s_2.
\eeq
The prescription for the transverse scale $\kappa$ is then fixed by \Eqn{eq:kappadef} to be~\cite{DelDuca:2008jg}
\beq\label{eq:ACkappa}
(-\kappa)\to e^{-i\pi}\,\kappa.
\eeq

\subsection{Analytic continuation of Region II(a)}
In Region II(a) the pentagon is expressed in terms of the dimensionless quantities $y_1$ and $y_2$. The analytic continuation of $y_1$ and $y_2$ follows then directly from \Eqn{eq:y12def} and \Eqn{eq:ACkappa},
 \beq\label{eq:ACy}
(-y_1)\to e^{-i\pi}\,y_1{\rm~~and~~} y_2\to y_2.
\eeq
In the physical region, the pentagon can then be written as
\beq
I^{(IIa)}_{\rm phys}(s,s_1,s_2,t_1,t_2) = r_\Gamma\,e^{\gamma_E\eps}\,{\kappa^{-\eps}\over st_2}\,\cI^{(IIa)}_{\rm phys}(\kappa,t_1,t_2),
\eeq
with
\beq
\cI^{(IIa)}_{\rm phys}(\kappa,t_1,t_2) = e^{i\pi\eps}\,\cI^{(IIa)}_{\rm ND}\big(e^{-i\pi}\,(-\kappa),t_1,t_2\big),
\eeq
and $\cI^{(IIa)}_{\rm ND}$ is given in \Eqn{eq:PentNDRegIIa} to all orders in $\eps$ in terms of Appell and Kamp\'e de F\'eriet functions. As we face the problem how of to perform the analytic continuation of these functions under the prescription~(\ref{eq:ACy}), we argue in the following that the analytic continuation of the generalized hypergeometric functions that appear in $\cI^{(IIa)}_{\rm ND}$ is trivial, \ie, the hypergeometric functions stay real in the physical region.

Indeed all the hypergeometric functions entering $\cI^{(IIa)}_{\rm ND}$ can be written as a nested sum, the inner sum being a one-dimensional hypergeometric function of $_2F_1$ or $_3F_2$ type depending on $(-y_1)$. For example, the first term in \Eqn{eq:PentNDRegIIa} can be written as
\beq\bsp\label{eq:IIIa1AC}
& -{1\over\eps^3}\,y_2^{-\eps}\, \Gamma (1-2\epsilon )\, \Gamma (1+\epsilon )^2\, F_4\Big(1-2\eps, 1-\eps, 1-\eps, 1-\eps;-y_1, y_2\Big)\\
&\quad= -{1\over\eps^3}\,y_2^{-\eps}\, \Gamma (1-2\epsilon )\, \Gamma (1+\epsilon )^2\, \sum_{n=0}^\infty\,{(1-2\eps)_n\,(1-\eps)_n\over(1-\eps)_n}\,{y_2^n\over n!}\,_2F_1\Big(1-2\eps, 1-\eps, 1-\eps;-y_1\Big).
\esp\eeq
Since we do not perform any analytic continuation in $y_2$, the only way an imaginary part could arise is when we cross a branch cut during the analytic continuation in $y_1$. Hence the analytic properties of \Eqn{eq:IIIa1AC} are determined by the cuts in the function $_2F_1\Big(1-2\eps, 1-\eps, 1-\eps;-y_1\Big)$.
We know that the $_2F_1$ function has a branch cut ranging from $1$ to $+\infty$. The convergence criterion for the $F_4$ function requires $|y_1|<1$, so we do not cross the branch cut when continuing $y_1$ along a half circle through the upper half plane, and so we do not change the Riemann sheet during the analytic continuation. Since the Euclidean region corresponds to the Riemann sheet where the hypergeometric function is real for $(-y_1)<1$, we conclude that the ${_2F_1}$ function in \Eqn{eq:IIIa1AC} is real in the physical region, and so is the Appell function on the left-hand side of  \Eqn{eq:IIIa1AC}. A similar reasoning can be made for all other hypergeometric functions in \Eqn{eq:PentNDRegIIa}, and so we can immediately write down the analytic continuation of \Eqn{eq:PentNDRegIIa} to the physical region where all $s$-type invariants are positive,
\beq\bsp\label{eq:NDIMIIaPhys}
\cI_{\rm phys}^{(IIa)}&(\kappa,t_1,t_2)=\\
-&\, {1\over\eps^3}\,y_2^{-\eps}\,e^{i\pi\eps}\, \Gamma (1-2\epsilon )\, \Gamma (1+\epsilon )^2\, F_4\Big(1-2\eps, 1-\eps, 1-\eps, 1-\eps;-y_1, y_2\Big)\\
  +&\,
  {1\over \eps^3}\,e^{i\pi\eps}\,
  \Gamma (1+\epsilon ) \,\Gamma (1-\epsilon ) \, F_4\Big(1, 1-\eps, 1-\eps, 1+\eps; -y_1, y_2\Big)\\
   -&\,{1\over \eps^2}\,(-y_1)^\eps\,y_2^{-\eps}\,\Bigg\{
   {\partial\over \partial \delta}\,F^{2,1}_{0,2}\left(\begin{array}{cc|cccc|} 1+\delta & 1+\delta-\eps& 1&- & -& -\\
-&-&1+\delta&1-\eps& 1+\eps+\delta&-\end{array}\, -y_1, y_2\right)_{|\delta=0}\\
&\qquad +
   \big[\ln (-y_1)-i\pi + \psi(1-\eps) - \psi(-\eps)\big]\,F_4\big(1,1-\eps,1+\eps,1-\eps;-y_1, y_2\big) \Bigg\}\\
+&\,{1\over \eps^2}\,(-y_1)^{\eps}\,\Bigg\{
{\partial\over \partial \delta}\,F^{2,1}_{0,2}\left(\begin{array}{cc|cccc|} 1+\delta & 1+\delta+\eps& 1&- & -& -\\
-&-&1+\delta&1+\eps& 1+\eps+\delta&-\end{array}\, -y_1, y_2\right)_{|\delta=0}\\
&\qquad+
\big[\ln (-y_1)-i\pi + \psi(1+\eps) - \psi(-\eps)\big]\,F_4\big(1,1+\eps,1+\eps,1+\eps;-y_1, y_2\big)\Bigg\}.
\esp\eeq
We checked explicitly that all the poles in $\eps$ cancel out when expanding $\cI^{(IIa)}_{\rm phys}$ as a Laurent series in $\eps$. Note that in \Eqn{eq:NDIMIIaPhys} the same hypergeometric functions appear as in the Euclidean region, \Eqn{eq:PentNDRegIIa}. We can therefore easily expand \Eqn{eq:NDIMIIaPhys} in $\eps$ in exactly the way as we did in the Euclidean region, and we will obtain a Laurent series whose coefficients are combinations of \emph{real} $\cM$-functions. The imaginary parts arise order by order in the Laurent series only through the explicit dependence on $i\pi$ in \Eqn{eq:NDIMIIaPhys}. Note that if we had worked with the expression of the pentagon in terms of Goncharov's multiple polylogarithms,  Eqs.~(\ref{eq:cJ0}) and (\ref{eq:cJ1}), the complicated analytic structure of these functions would imply many spurious imaginary parts arising from individual polylogarithms, and  all the spurious imaginary parts would need to cancel out in the final answer. It is therefore more natural to express the result in the physical region in terms of $\cM$ functions rather than the Goncharov polylogarithms.

\subsection{Analytic continuation of Region I}
In Region I the pentagon is expressed as a function of $x_1$ and $x_2$, whose analytic continuation follows directly from \Eqn{xydef} and \Eqn{eq:ACkappa},
\beq
(-x_1)\to e^{+i\pi}\,x_1 {\rm ~~and~~} x_2\to e^{+i\pi}\,x_2,
\eeq
and the pentagon in Region I can be expressed as
\beq
I^{(I)}_{\rm phys}(s,s_1,s_2,t_1,t_2) = r_\Gamma\,e^{\gamma_E\eps}\,{\kappa^{-\eps}\over s_1s_2}\,\cI^{(I)}_{\rm phys}(\kappa,t_1,t_2),
\eeq
with
\beq
\cI^{(I)}_{\rm phys}(\kappa,t_1,t_2) = e^{i\pi\eps}\,\cI^{(I)}_{\rm ND}\big(e^{-i\pi}\,(-\kappa),t_1,t_2\big),
\eeq
and $\cI^{(I)}_{\rm ND}$ is given in \Eqn{eq:SolRegIKF}. Using the  same argument as in Region II(a), we see that the hypergeometric functions stay real even in the physical region, because the convergence of the hypergeometric functions requires $|x_i|<1$, $i=1,2$. Alternatively, we could also start from the expression of the pentagon in the physical Region II(a), \Eqn{eq:NDIMIIaPhys}, and perform the analytic continuation to Region I following the techniques described in Appendix~\ref{app:AnalContPent}. We find that the two results are consistent. The final expression of the pentagon in the physical region I is then
\beq\bsp\label{eq:NDIMIPhys}
\cI^{(I)}_{\rm phys}&(\kappa,t_1,t_2)=\\
-\,& \frac{1}{\eps^3}\,e^{-i\pi\eps}\,(-x_1)^{-\eps}\,(-x_2)^{-\eps}\,\Gamma(1-2\eps)\, \Gamma (1+\epsilon )^2\, F_4(1-2\eps,1-\eps,1-\eps,1-\eps;-x_1,-x_2)\\
 +\,&   {1\over\eps^3}\,
  \Gamma (1-\epsilon ) \Gamma (1+\epsilon)\,e^{i\pi\eps}\, F_4(1,1+\eps,1+\eps,1+\eps;-x_1,-x_2)\\
     -\,&{1\over\eps^2}\,(-x_1)^{-\eps}\,\Bigg\{
     {\partial\over\partial\delta} F^{2,1}_{0,2}\left(\begin{array}{cc|cccc|}
 1+\delta & 1-\eps+\delta &-&-&-&1\\
 -&-&1-\eps&1+\eps+\delta&-&1+\delta\end{array}\,-x_1,-x_2\right)_{|\delta=0}\\
 &\quad + \big[\ln (-x_2) + i\pi + \psi(1-\eps)-\psi(-\eps)\big]\,F_4(1,1-\eps,1-\eps,1+\eps;-x_1,-x_2)
\\
 &\quad +{\eps\over 1-\eps}\,{t_1\over t_2}\, F^{0,3}_{2,0}\left(\begin{array}{cc|cccccc|}
 - & - &1&1&1&1&1-\eps&1-\eps\\
 2&2-\eps&-&-&-&-&-&-\end{array}\,-x_1,{x_1\over x_2}\right)\Bigg\}\\
 -\,&{1\over \eps^2}\,(-x_2)^{-\eps}\,\Bigg\{
 {\partial\over\partial\delta} F^{2,1}_{0,2}\left(\begin{array}{cc|cccc|}
 1+\delta & 1-\eps+\delta &-&-&-&1\\
 -&-&1+\eps&1-\eps+\delta&-&1+\delta\end{array}\,-x_1,-x_2\right)_{|\delta=0}\\
 &\quad + \big[\ln (-x_2) + i\pi+ \psi(1-\eps)-\psi(\eps)\big]\,F_4(1,1-\eps,1+\eps,1-\eps;-x_1,-x_2)\\
\\
 &\quad -{\eps\over 1+\eps}\,{t_1\over t_2}\, F^{0,3}_{2,0}\left(\begin{array}{cc|cccccc|}
 - & - &1&1&1&1&1-\eps&1+\eps\\
 2&2+\eps&-&-&-&-&-&-\end{array}\,-x_1,{x_1\over x_2}\right)\Bigg\}.
 \esp\eeq
We checked explicitly that all the poles in $\eps$ cancel out when expanding $\cI^{(I)}_{\rm phys}$ as a Laurent series in $\eps$.

\section{Conclusions}
\label{sec:conclusions}

In this paper we computed the one-loop scalar massless pentagon integral in $D=6-2\eps$ dimensions, $I^{6-2\eps}_5$, in the limit of multi-Regge kinematics.  This integral first contributes to the parity-odd part of the one-loop $\begin{cal}N\end{cal}=4$ five-point MHV amplitude at ${\cal O}(\eps)$ -- see \eqref{eq:pent}.   

It is well known that loop integrals are intimately related to multiple generalised hypergeometric series and generalised polylogarithms.   We have exploited two completely different approaches - Negative Dimensions and Mellin-Barnes integration - to extract the leading two terms of the Laurent expansion as $\eps \to 0$ for the $I_5^{6-2\eps}$ in the high energy limit defined by,
\beq
s \gg s_1, ~s_2 \gg -t_1, -t_2.
\eeq
In this limit,  the pentagon integral reduces to double sums or equivalently two-fold Mellin-Barnes integrals.  Equations (\ref{eq:i2a}), (\ref{eq:i02a}), (\ref{eq:i12a}) and (\ref{MBMRpentRegII}), (\ref{eq:jy1y2}), (\ref{eq:cJ0}), (\ref{eq:cJ1}) give explicit expressions for the ${\cal O}(\eps^0)$ and ${\cal O}(\eps)$ contributions to $I_5^{6-2\eps}$ in terms of ${\cal M}$ functions and Goncharov's multiple polylogarithms for kinematic region II(a) when $(-t_1) < (-t_2)$ and $-\sqrt{-st_1/s_1s_2} + \sqrt{-st_2/s_1s_2} > 1$. Analogous expressions in region I defined by $\sqrt{-st_1/s_1s_2} + \sqrt{-st_2/s_1s_2} < 1$ can be derived through the analytic continuation of \sec{sec:e3} and 
are given in Eqs.~(\ref{MBMRpent}), (\ref{eq:ix1x2}), respectively.

By determining the ${\cal O}(\eps)$ contribution to  $I_5^{6-2\eps}$, one therefore gains knowledge of $m_5^{(1)}$ through to ${\cal O}(\eps^2)$ which is necessary for studies of the iterative structure of $\begin{cal}N\end{cal}=4$ SYM amplitudes beyond one-loop. One immediate application is the extraction of the one-loop gluon-production vertex through to ${\cal O}(\eps^2)$ and the iterative construction of the two-loop gluon-production vertex through to finite terms which is described in a companion paper~\cite{noi}.

Finally, we note that one-loop polygons are present in the ${\cal O}(\eps)$
contributions~\cite{Bern:1996ja} to one-loop $\begin{cal}N\end{cal}=4$ MHV amplitudes.      In
particular,  the $D=6-2\eps$ hexagon integral contributes to the  parity-even 
part~\cite{Bern:2008ap,Bern:1996ja} of the one-loop six-point MHV amplitude, while $D=6-2\eps$
pentagon integrals contribute to the parity-odd  part~\cite{Cachazo:2008hp,Bern:1996ja}. It seems
reasonable to conjecture that the analytic form of the remainder function $R_6^{(2)}$ may be
somewhat linked to the properties of the special functions occurring in the hexagon one-loop
integral.  The analytic methods we have used for evaluating the pentagon integral in the high energy limit may also be applied to the hexagon integral and may ultimately provide information on the form of the $R_6^{(2)}$ remainder function.

\section*{Acknowledgements}

CD, VDD and VS thank the IPPP Durham and EWNG, CD and VS thank the LNF
Frascati for the warm hospitality at various stages of this work. CD is a
research fellow of the \emph{Fonds National de la Recherche Scientifique},
Belgium. This work was partly supported by MIUR under contract 2006020509$_0$04,
and by the EC Marie-Curie Research Training Network ``Tools and Precision
Calculations for Physics Discoveries at Colliders'' under contract
MRTN-CT-2006-035505. EWNG gratefully acknowledges the support of the Wolfson
Foundation and the Royal Society. The work of VS was supported in part by the
Russian Foundation for Basic Research through grant 08-02-01451.


\appendix

\section{Nested harmonic sums}\label{app:Ssums}
The nested harmonic sums have been defined in \Eqn{eq:SZSums}.
The $S$ and $Z$ sums fulfill an algebra. Let us illustrate this with a simple example:
\beq\bsp\label{eq:Ssumalgebra}
S_i(n)S_j(n) &\,= \sum_{k_1=1}^n\,\sum_{k_2=1}^n\,{1\over k_1^i k_2^j}\\
&\,= \sum_{k_1=1}^n\,{1\over k_1^i}\,\sum_{k_2=1}^{k_1}\,{1\over k_2^j}+\sum_{k_2=1}^n\,{1\over k_2^j}\,\sum_{k_1=1}^{k_2}\,{1\over k_1^i} - \sum_{k=1}^n\,{1\over k^{i+j}}\\
&\,=S_{ij}(n)+S_{ji}(n)-S_{i+j}(n).
\esp\eeq
A similar result can be obtained for the $Z$-sums\footnote{Note the sign difference with respect to \Eqn{eq:Ssumalgebra}.},
\beq
Z_i(n)Z_j(n) = Z_{ij}(n)+Z_{ji}(n)+Z_{i+j}(n).
\eeq
For sums of higher weight, a recursive application of the above procedure leads then to the reduction of any product of $S$ or $Z$ sums to a linear combination of those sums. Furthermore, $S$ and $Z$ sums can be interchanged, \eg
\beq\bsp
Z_{11}(n)&\,=\sum_{k=1}^n\,\frac{Z_1(k-1)}{k}=\sum_{k=1}^n\,\frac{1}{k}\,\sum_{\ell=1}^{k-1}\,\frac{1}{\ell}=\sum_{k=1}^n\,\frac{1}{k}\,\left(-\frac{1}{k}+\sum_{\ell=1}^k\,\frac{1}{\ell}\right)\\
&\,=-\sum_{k=1}^n\,\frac{1}{k^2} + \sum_{k=1}^n\,\frac{1}{k}\,S_1(k)=-S_2(n) + S_{11}(n).
\esp\eeq
For $n\rightarrow\infty$, the Euler-Zagier sums converge to multiple zeta values,
\beq
\lim_{n\rightarrow\infty}\,Z_{m_1\ldots m_k}(n) = \zeta(m_k,\ldots,m_1).
\eeq

In Ref.~\cite{Moch:2001zr} Moch \etal introduced generalizations of the $S$ and $Z$-sums to make them dependent on some variables,
\beq\bsp
&S_i(n;x) = Z_i(n;x) = \sum_{k=1}^n\,{x^k\over k^i},\\
& S_{i\vec \jmath}(n;x_1,\ldots, x_\ell) = \sum_{k=1}^n\,{x_1^k\over k^i}\,S_{\vec \jmath}(k; x_2,\ldots, x_n),\\
&Z_{i\vec \jmath}(n;x_1,\ldots, x_\ell) = \sum_{k=1}^n\,{x_1^k\over k^i}\,Z_{\vec \jmath}(k-1; x_2,\ldots, x_n).
\esp\eeq
Those sums share of course all the properties of the corresponding number sums introduced in the previous paragraph. In particular, for $n\rightarrow\infty$, the $Z$ sums converge to Goncharov's multiple polylogarithm,
\beq
\lim_{n\rightarrow\infty}\,Z_{m_1\ldots m_k}(n;x_1,\ldots,x_k) = \mathrm{Li}_{m_k,\ldots,m_1}(x_1,\ldots,x_k).
\eeq
These multiple polylogarithms will be reviewed in the next section.


 \section{Goncharov's multiple polylogarithm}\label{app:hpl}
  Let us define~\cite{Moch:2001zr}
  \beq
  \Omega(a) = {\rd t\over t-a},
  \eeq
  and iterated integrations by
  \beq
  \int_0^z \,\Omega(a_1)\circ\ldots\circ\Omega(a_n) = \int_0^z\,{\rd t\over t-a_1}\,\int_0^t\,\Omega(a_2)\circ\ldots\circ\Omega(a_n).
  \eeq
  We can then define the $G$ and $M$-functions as
  \beq
  \bsp
  G(a_1,\ldots,a_n;z)=&\,\int_0^z\,\Omega(a_1)\circ\ldots\circ\Omega(a_n),\\
  M(a_1,\ldots,a_n)=&\, G(a_1,\ldots,a_n;1)=\,\int_0^1\,\Omega(a_1)\circ\ldots\circ\Omega(a_n).
  \esp\eeq
  The $M$-functions are nothing but Goncharov's multiple polylogarithm, up to a sign\footnote{For the meaning of the indices $m_i$ and $z_i$, the reader should refer to Ref.~\cite{Moch:2001zr}. They are equivalent to the corresponding notation for \emph{HPL}'s, \eg $H(0,1;z)=H_2(z)$.},
  \beq
   M(a_1,\ldots,a_n) = (-1)^k\,\mathrm{Li}_{m_k,\ldots,m_1}(z_k,\ldots,z_1),
   \eeq
   where $k$ is the number of nonzero elements in $(a_1,\ldots,a_n)$.

   Iterated integrals form a shuffle algebra, and hence we can immediately write
   \beq\bsp
   G(\vec w_1;z) \, G(\vec w_2;z) &\,=\sum_{\vec w=\vec w_1\uplus\vec w_2}\,G(\vec w;z),\\
       M(\vec w_1) \, M(\vec w_2) &\,=\sum_{\vec w=\vec w_1\uplus\vec w_2}\,M(\vec w).
       \esp\eeq
   It is easy to see that in general\footnote{In some cases the divergence is tamed, \eg $\lim_{z\rightarrow1}\,\ln z\,\ln(1-z) = 0$.}
   \beq
   \lim_{a_1\rightarrow1}\,M(a_1,\ldots,a_n)=\infty,
   \eeq
   or equivalently
   \beq
   \lim_{z\rightarrow1}\,G(1,a_2,\ldots,a_n;z)=\infty,
   \eeq
   We can however extract the divergence by choosing an irreducible basis for the $G$-functions where the leftmost index is different from 1, except for functions of the form $G(\vec 1_n;z)$, \eg
   \beq
   G(1,0;z) = G(1;z)\,G(0;z) - G(0,1;z).
   \eeq
   In this basis the singularity structure in $z=1$ is explicit. Note at this point that the pentagon integral does not contain $M$-functions of the form $M(1,\ldots)$, \ie, $\cI$ and $\cJ$ are well-defined.

  In Ref.~\cite{Aglietti:2008fe, Gehrmann:2001pz}, an algorithm was presented which allows to express \emph{2dHPL}'s of the form $H(\vec w(a);z)$ in $z=1$ in terms of ordinary \emph{HPL}'s in $a$. First write $H(\vec w(a);1)$ as the integral of the derivative,
  \beq\label{Halg}
  H(\vec w(a);1) = H(\vec w(a_0);1) + \int_{a_0}^a\,\rd a'\,\frac{\partial}{\partial a'}\,H(\vec w(a');1),
  \eeq
  where $a_0$ is arbitrary, provided that $H(\vec w(a_0);1)$ exists. The derivative is now carried out on the integral representation of $H$, and then we integrate back. Since differentiation lowers the weight by one unit and since we know recursively how to reduce all $H(\vec w'(a);1)$ with $w'<w$, we obtain now the function $H(\vec w(a);1)$ with $a$ being the limit of the last integration.

  Since $M$-functions are nothing but the values in $z=1$ of the $G$ functions, and since $G$-functions are straightforward generalizations of the \emph{HPL}'s, this algorithm immediately generalizes to the $M$-functions, and allows in principle to reduce functions of the form $M(\ldots,\lambda_1,\ldots)$ to \emph{HPL}'s of the form $H(\ldots;\lambda_1)$, the algorithm being a straightforward generalization of \Eqn{Halg}
  \beq
  \label{Malg}
  M(\vec w(a)) = M(\vec w(a_0)) + \int_{a_0}^a\,\rd a'\,\frac{\partial}{\partial a'}\,M(\vec w(a')),
  \eeq
  where $a_0$ is arbitrary, provided that $M(\vec w(a_0))$ exists.


\section{Generalized hypergeometric functions}
\label{app:HypGeo}

\subsection{Appell functions}
The Appell functions are defined by the double series,
\beq\bsp
F_1(a,b,c,d;x,y)&\,=\sum_{m=0}^\infty\,\sum_{n=0}^\infty\,{\ph{a}{m+n}\,\ph{b}{m}\,\ph{c}{n}\over\ph{d}{m+n}}\,{x^m\over m!}\,{y^n\over n!},\\
F_2(a,b,c,d,e;x,y)&\,=\sum_{m=0}^\infty\,\sum_{n=0}^\infty\,{\ph{a}{m+n}\,\ph{b}{m}\,\ph{c}{n}\over\ph{d}{m}\,\ph{e}{n}}\,{x^m\over m!}\,{y^n\over n!},\\
F_3(a,b,c,d,e;x,y)&\,=\sum_{m=0}^\infty\,\sum_{n=0}^\infty\,{\ph{a}{m}\,\ph{b}{n}\,\ph{c}{m}\,\ph{d}{n}\over\ph{e}{m+n}}\,{x^m\over m!}\,{y^n\over n!},\\
F_4(a,b,c,d;x,y)&\,=\sum_{m=0}^\infty\,\sum_{n=0}^\infty\,{\ph{a}{m+n}\,\ph{b}{m+n}\over\ph{c}{m}\ph{d}{n}}\,{x^m\over m!}\,{y^n\over n!}.
\esp\eeq
From the series representation the Mellin-Barnes representation is trivially obtained.
For some special values of the indices, the Appell functions reduce to simpler hypergeometric functions, \eg,
\beq
\bsp\label{f4red}
F_4\Bigg(\alpha,\beta,&\alpha,\beta;\frac{-x}{(1- x)(1- y)},\frac{- y}{(1- x)(1- y)}\Bigg)\,=\frac{(1- x)^\beta (1- y)^\alpha}{1- x y},\\
F_4\Bigg(\alpha,\beta,&\beta,\beta;\frac{- x}{(1- x)(1- y)},\frac{- y}{(1- x)(1- y)}\Bigg)\,=(1- x)^\alpha (1- y)^\alpha\,{_2F_1}(\alpha,1+\alpha-\beta,\beta; x y),\\
F_4\Bigg(\alpha,\beta,&1+\alpha-\beta,\beta;\frac{- x}{(1- x)(1- y)},\frac{- y}{(1- x)(1- y)}\Bigg)\\
&\,\qquad=(1- y)^\alpha\,{_2F_1}\left(\alpha,\beta,1+\alpha-\beta;-\frac{ x(1- y)}{1- x}\right),\\
F_4\Bigg(\alpha,\beta,&\gamma,\beta;\frac{- x}{(1- x)(1- y)},\frac{- y}{(1- x)(1- y)}\Bigg)\\
&\,\qquad=(1- x)^{\alpha}\,(1- y)^{\alpha}\,F_1(\alpha,\gamma-\beta,1+\alpha-\gamma,\gamma; x, x y).
\esp\eeq
The analytic continuation of the Appell $F_4$ function reads
\beq\bsp\label{eq:F4AnalCont}
F_4(a,b,c,d;x,y)=&\,{\Gamma(d)\,\Gamma(b-a)\over\Gamma(b)\,\Gamma(d-a)}\,(-y)^{-a}\,F_4\Big(a,1+a-d,c,1+a-b;{x\over y}, {1\over y}\Big)\\
+&\,{\Gamma(d)\,\Gamma(a-b)\over\Gamma(a)\,\Gamma(d-b)}\,(-y)^{-b}\,F_4\Big(1+b-d,b,c,1+b-a;{x\over y}, {1\over y}\Big).
\esp\eeq

\subsection{Kamp\'e de F\'eriet functions}
The Kamp\'e de F\'eriet functions are defined by the series
\beq
F^{p,q}_{p', q'}\left(\begin{array}{c|cc|}
\alpha_i & \beta_j &\gamma_j\\
\alpha'_k & \beta'_\ell &\gamma'_\ell
\end{array}\, x, y\right) = \sum_{m=0}^\infty \,\sum_{n=0}^\infty\,{\prod_i\,\ph{\alpha_i}{m+n}\,\prod_j\,\ph{\beta_j}{m}\,\ph{\gamma_j}{n}
\over \prod_k\,\ph{\alpha'_k}{m+n}\,\prod_\ell\,\ph{\beta'_\ell}{m}\,\ph{\gamma'_\ell}{n}}\,{x^m\over m!}\,{y^n\over n!},
\eeq
with $1\le i\le p$, $1\le j\le q$, $1\le k\le p'$ and $1\le \ell\le q'$, and convergence requires $p+q\le p'+q'+1$~\cite{Appell}. The order of a Kamp\'e de F\'eriet function is defined by $\omega = p'+q'$. If $p+q=\omega+1$, the function is called \emph{complete}, and all other cases are obtained as a confluent limiting case of a complete function.
From its definition it is clear that the Kamp\'e de F\'eriet function enjoys the following symmetry property,
\beq
F^{p,q}_{p', q'}\left(\begin{array}{c|cc|}
\alpha_i & \beta_j &\gamma_j\\
\alpha'_k & \beta'_\ell &\gamma'_\ell
\end{array}\, x, y\right) = F^{p,q}_{p', q'}\left(\begin{array}{c|cc|}
\alpha_i & \gamma_j &\beta_j\\
\alpha'_k & \gamma'_\ell &\beta'_\ell
\end{array}\, y, x\right).
\eeq
The Kamp\'e de F\'eriet functions encompass the Appell functions for particular values of the parameters,
\beq\bsp
F^{1,1}_{1, 0}\left(\begin{array}{c|cc|}
a & b &c\\
d &- &-
\end{array}\, x, y\right) &=\, F_1(a,b,c,d;x,y),\\
F^{1,1}_{0,1}\left(\begin{array}{c|cc|}
a & b &c\\
- &d &e
\end{array}\, x, y\right) &=\, F_2(a,b,c,d,e;x,y),\\
F^{0,2}_{1, 0}\left(\begin{array}{c|cccc|}
- & a &b&c&d\\
e &- &-&-&-
\end{array}\, x, y\right) &=\, F_3(a,b,c,d,e;x,y),\\
F^{2,0}_{0, 1}\left(\begin{array}{cc|cc|}
a & b &-&-\\
-&-&c&d
\end{array}\, x, y\right) &=\, F_4(a,b,c,d;x,y).
\esp\eeq

The Kamp\'e de F\'eriet function involves only Pochhammer symbols of the form $\ph{.}{n_1+n_2}$, $\ph{.}{n_1}$, $\ph{.}{n_2}$. We could alternatively define a function involving $\ph{.}{n_1-n_2}$. Let us consider the double series
\beq
\tilde F^{p,q,r}_{p',q',r'}\left(\begin{array}{c|cc|} a_i & b_j &c_h \\ a'_k& b'_l & c'_m\end{array}\,x,y\right)
=\sum_{n_1=0}^{\infty}\,\sum_{n_2=0}^\infty\,{\prod_i\ph{a_i}{n_1-n_2}\,\prod_j \ph{b_j}{n_1}\,\prod_h\ph{c_h}{n_2}\over 
\prod_k\ph{a'_k}{n_1-n_2}\,\prod_l \ph{b'_l}{n_1}\,\prod_m\ph{c'_m}{n_2}}\,{x^{n_1}\over n_1!}\,{y^{n_2}\over n_2!}.
\eeq
Note that this function can appear in the analytic continuation of the Kamp\'e de F\'eriet function. In the following we proof that the generalized hypergeometric series $\tilde F$ can always be reduced to Kamp\'e de F\'eriet-type functions. We rewrite the double sum according to
\beq\bsp
\infsum{n_1}\,\infsum{n_2} = &\,\infsum{n_1}\,\sum_{n_2=0}^{n_1} + \infsum{n_1}\,\sum_{n_2=n_1}^{\infty} - \sum_{n_1=n_2=0}^\infty\\
=&\, \infsum{n_1}\,\sum_{n_2=0}^{n_1} + \infsum{n_2}\,\sum_{n_1=0}^{n_2}\, - \sum_{n_1=n_2=0}^\infty.
\esp\eeq
The double sums are now reshuffled into double sums up to infinity, \eg
\beq
\infsum{n_1}\,\sum_{n_2=0}^{n_1} \rightarrow \infsum{N}\,\infsum{n_2}, {\rm ~~ with~~} n_1=N+n_2,
\eeq
and equivalently for $ \infsum{n_2}\,\sum_{n_1=0}^{n_2}$. In the second sum we then rewrite all Pochhammer symbols with negative index according to 
\beq
\ph{a}{-n} = {(-1)^n\over \ph{1-a}{n}}.
\eeq
 We then arrive at
\beq\bsp\label{eq:KFMreduction}
\tilde F^{p,q,r}_{p',q',r'}&\left(\begin{array}{c|cc|} a_i & b_j &c_h \\ a'_k& b'_l & c'_m\end{array}\,x,y\right)\\
&=\sum_{n_1=0}^{\infty}\,\sum_{n_2=0}^\infty\,{\prod_i\ph{a_i}{n_1}\,\prod_j \ph{b_j}{n_1+n_2}\,\prod_h\ph{c_h}{n_2}\over 
\prod_k\ph{a'_k}{n_1}\,\prod_l \ph{b'_l}{n_1+n_2}\,\prod_m\ph{c'_m}{n_2}}\,{\ph{1}{n_1}\over \ph{1}{n_1+n_2}}\,{x^{n_1}\over n_1!}\,{(xy)^{n_2}\over n_2!}\\
&\,+\sum_{n_1=0}^{\infty}\,\sum_{n_2=0}^\infty\,(-1)^{(p+p')n_2}{\prod_k\ph{1-a'_k}{n_2}\,\prod_j \ph{b_j}{n_1}\,\prod_h\ph{c_h}{n_1+n_2}\over 
\prod_i\ph{1-a_i}{n_2}\,\prod_l \ph{b'_l}{n_1}\,\prod_m\ph{c'_m}{n_1+n_2}}\,{\ph{1}{n_2}\over \ph{1}{n_1+n_2}}\,{(xy)^{n_1}\over n_1!}\,{y^{n_2}\over n_2!}\\
&\,-\sum_{n=0}^{\infty}\,{\prod_j \ph{b_j}{n}\,\prod_h\ph{c_h}{n}\over \ph{1}{n}
\prod_l \ph{b'_l}{n}\,\prod_m\ph{c'_m}{n}}\,{(xy)^{n}\over n!}.
\esp\eeq
We see that in \Eqn{eq:KFMreduction} only Pochhammer symbols of the form $\ph{.}{n_1+n_2}$, $\ph{.}{n_1}$ and $\ph{.}{n_2}$ appear. Note however that this class of functions encompasses a larger class of functions than the one defined by the Kamp\'e de F\'eriet function.

The analytic continuation of the Kamp\'e de F\'eriet function is easily carried out using its Mellin-Barnes representation. In particular, we use here the following results, derived from Mellin-Barnes integrals,
\beq\bsp
F^{2,1}_{0, 2}&\left(\begin{array}{cc|cccc|}
a & b &c&-&-&-\\
-&- & d&e &f&-
\end{array}\, x, y\right)\\
=&\,{\Gamma(e)\,\Gamma(b-a)\over\Gamma(b)\,\Gamma(e-a)}\,(-y)^{-a}\,F^{2,1}_{0, 2}\left(\begin{array}{cc|cccc|}
a & 1+a-e &c&-&-&-\\
-&- & d&1+a-b &f&-
\end{array}\, x, y\right)\\
=&\,{\Gamma(e)\,\Gamma(a-b)\over\Gamma(a)\,\Gamma(e-b)}\,(-y)^{-b}\,F^{2,1}_{0, 2}\left(\begin{array}{cc|cccc|}
1+b-e &b &c&-&-&-\\
-&- & d&1+b-a &f&-
\end{array}\, {x\over y}, {1\over y}\right).
\esp\eeq
\beq\bsp
F^{2,1}_{0, 2}&\left(\begin{array}{cc|cccc|}
a & b &c&-&-&-\\
-&- & d&e &f&-
\end{array}\, x, y\right)\\
=&\, \frac{ \Gamma (d) \Gamma (f) 
  \Gamma (b-a) \Gamma (c-a)
  }{\Gamma (b) \Gamma (c) \Gamma (d-a)
  \Gamma (f-a) }\,(-x)^{-a}\,
  F^{3,0}_{1,1}\left(\begin{array}{ccc|cc|}
a &1 +a-d &1+a-f&-&-\\
-&- & 1+a-c&1+a-b &e
\end{array}\, {1\over x}, {y\over x}\right)\\
+&\, \frac{\Gamma (d) \Gamma (f) 
  \Gamma (a-b) \Gamma (c-b) }{\Gamma (a) \Gamma (c) \Gamma (d-b)
  \Gamma (f-b)}\,(-x)^{-b}\,
  F^{3,0}_{1,1}\left(\begin{array}{ccc|cc|}
b &1 +b-d &1+b-f&-&-\\
-&- & 1+b-c&1+b-a &e
\end{array}\, {1\over x}, {y\over x}\right)\\
+&\,\frac{\Gamma (d) \Gamma (f) 
  \Gamma (a-c) \Gamma (b-c) }{\Gamma (a)
  \Gamma (b) \Gamma (d-c) \Gamma (f-c) }\,(-x)^{-c}\,
  \tilde F^{2,3,0}_{0,0,1}\left(\begin{array}{cc|ccc|}
a-c &b-c &c&1+c-d&1+c-f\\
-&-&-&- &e
\end{array}\,y, {1\over x}\right),
\esp\eeq
and the $\tilde F$ function can reshuffled into Kamp\'e de F\'eriet-type series using the algorithm described in the previous paragraph.


\section{$\cM$ functions}
\label{app:Mfunc}
In this section we apply the techniques combining the Mellin-Barnes and series representations to study the analytic properties of the $\cM$ functions that appear in the $\eps$-expansion of the $F_4$ function, \Eqn{eq:Mdef}. We start by analyzing some general properties of these new transcendental functions, and study more specific properties like analytic continuation and reduction to known functions in subsequent sections.
First, it is easy to see that $\cM$ functions are symmetric,
\beq\label{sym}
\mfunc(\vec \imath, \vec \jmath, \vec k;x,y) = \mfunc(\vec \jmath, \vec \imath, \vec k;y,x).
\eeq
Second, when one of the arguments is zero, then this function reduces either to $0$ or to an $S$-sum at infinity. Indeed, if say $y$ is equal to zero, then only the term $n=0$ in the sum contributes, and since $S_{\vec i}(0)=0$, we get 
\begin{itemize}
\item If $\vec \jmath\neq0$, then
\beq
\mfunc(\vec \imath, \vec \jmath, \vec k;x,0) = \sum_{m=0}^\infty\,\binom{m}{0}^2\,x^m\,S_{\vec\imath}(m)\,S_{\vec\jmath}(0)\,S_{\vec k}(m) = 0.
\eeq
\item If $\vec \jmath=0$, then
\beq
\mfunc(\vec \imath, 0, \vec k;x,0) = \sum_{m=0}^\infty\,\binom{m}{0}^2\,x^m\,S_{\vec\imath}(m)\,S_{\vec k}(m) = \sum_{m=0}^\infty\,x^m\,S_{\vec\imath}(m)\,S_{\vec k}(m).
\eeq
The product of nested harmonic sums can now be reduced using their algebra, and so we get
\beq
\mfunc(\vec \imath, 0, \vec k;x,0) = \sum_\ell S(\infty;0,\vec\imath_\ell;x,1,\ldots,1).
\eeq
where the $S$-sums are defined recursively in a similar way as the nested harmonic numbers,~\cite{Moch:2001zr}
\beq
S(N;\vec\imath;\vec x) = \sum_{k=1}^N\,\frac{x_1^k}{k^{i_1}}\,S(k;\vec \imath';\vec x'),
\eeq
and the values of the $S$-sums for $N=\infty$ are related to Goncharov's multiple polylogarithm.
\end{itemize}

We now analyze the analytic continuation of some of these functions under the transformation $y\rightarrow 1/y$. Due to the symmetry property~(\ref{sym}) the analytic continuation in $x$ follows the same lines. For $k\ge1$ we have,
\beq\bsp
\psi(1+n) &=-\eug +S_1(n),\\
\psi^{(k)}(1+n) &= (-1)^{k+1}\,k!\,\left(\zeta_{k+1}-S_{k+1}(n)\right),
\esp
\eeq
where $\eug$ denotes the Euler-Mascheroni constant, and $\psi^{(k)}$ denote the polygamma functions
\beq
\psi^{(k)}(z)=\left(\frac{\rd}{\rd z}\right)^{k+1} \ln\Gamma(z).
\eeq
We can easily relate the function $\mfunc$ to Mellin-Barnes integrals of the form
\beq\label{mbint}
\frac{1}{(2\pi i)^2}\,\int_{-i\infty}^{+i\infty}\,\rd z_1\,\rd z_2\,\Gamma(-z_1)\,\Gamma(-z_2)\,\frac{\Gamma(1+z_1+z_2)^2}{\Gamma(1+z_1)\Gamma(1+z_2)}\,(-x)^{z_1}\,(-y)^{z_2}\, \Psi(z_1,z_2),
\eeq
where $\Psi(z_1,z_2)$ denotes any product of polygamma functions with arguments $1+z_1$, $1+z_2$ or $1+z_1+z_2$. For example we can write
\beq\bsp
\frac{1}{(2\pi i)^2}\,\int_{-i\infty}^{+i\infty}\,&\rd z_1\,\rd z_2\,\Gamma(-z_1)\,\Gamma(-z_2)\,\frac{\Gamma(1+z_1+z_2)^2}{\Gamma(1+z_1)\Gamma(1+z_2)}\,\psi(1+z_1)\,(-x)^{z_1}\,(-y)^{z_2}\\
&=\mfunc(1,0,0;x,y)-\eug\mfunc(0,0,0;x,y).
\esp\eeq
It follows then that studying the analytic continuation properties of certain classes of $\mfunc$-functions is equivalent to study the properties of the Mellin-Barnes integral~(\ref{mbint}).
If both $x$ and $y$ are smaller than $1$, then we close both contours to the right, and we take residues, and we find the series definition of the $\mfunc$-functions.
\beq\bsp
\mfunc(0,0,0;x,y)&\,=- \frac{1}{y}\mfunc\left(0,0,0;\frac{x}{y},\frac{1}{y}\right),\\
\mfunc(1,0,0;x,y)&\,=- \frac{1}{y}\mfunc\left(1,0,0;\frac{x}{y},\frac{1}{y}\right),\\
\mfunc(0,1,0;x,y)&\,=\frac{2}{y}\mfunc\left(0,1,0;\frac{x}{y},\frac{1}{y}\right)-\frac{3}{y}\mfunc\left(0,0,1;\frac{x}{y},\frac{1}{y}\right)+\frac{\ln(-y)}{y}\mfunc\left(0,0,0;\frac{x}{y},\frac{1}{y}\right),\\
\mfunc(0,0,1;x,y)&\,=\frac{1}{y}\mfunc\left(0,1,0;\frac{x}{y},\frac{1}{y}\right)-\frac{2}{y}\mfunc\left(0,0,1;\frac{x}{y},\frac{1}{y}\right)+\frac{\ln(-y)}{y}\mfunc\left(0,0,0;\frac{x}{y},\frac{1}{y}\right).
\esp\eeq
Note the appearance of $\ln(-y)$, which produces an imaginary part for $y>0$.

In some cases it is possible to express the $\mfunc$-functions in terms of known functions. This becomes possible by relating those functions back to simple cases of the expansion of the Appell $F_4$ function, in which cases we can apply the reduction formulae~(\ref{f4red}) given in Appendix~\ref{app:HypGeo}. As an example, let us consider the reduction formulas
\beq
\bsp
F_4\left(\alpha,\beta,\alpha,\beta;\frac{-\tilde x}{(1-\tilde x)(1-\tilde y)},\frac{-\tilde y}{(1-\tilde x)(1-\tilde y)}\right)&\,=\frac{(1-\tilde x)^\alpha (1-\tilde y)^\beta}{1-\tilde x\tilde y},\esp\eeq
where $\tilde x$ and $\tilde y$ are solutions of
\beq
x=\frac{-\tilde x}{(1-\tilde x)(1-\tilde y)},\qquad y=\frac{-\tilde y}{(1-\tilde x)(1-\tilde y)}.
\eeq
Solving this system we arrive at
\beq
\tilde x=\frac{1}{2y}\,\left(x+y-1+\sqrt{\lambda(1,x,y)}\right), \qquad \tilde y=\frac{1}{2x}\,\left(x+y-1+\sqrt{\lambda(1,x,y)}\right).
\eeq
From this we can immediately read of the $\mfunc$-function of weight 0,
\beq
\mfunc(0,0,0;x,y) = F_4(1,1,1,1;x,y) = \frac{(1-\tilde x)(1-\tilde y)}{1-\tilde x\tilde y}.
\eeq
Using \Eqn{eq:phexp}, we can easily obtain all the $\mfunc$-functions of weight 1, \eg
\beq\bsp
\mfunc(1,0,0;x,y)&\, = \frac{\partial}{\partial \eps}\,F_4(1+\eps,1,1,1;x,y)_{|\eps=0}\\
&\, =\frac{\partial}{\partial \eps}\left[ (1-\tilde x)^{1+\eps}(1-\tilde y)^{1+\eps}\, {_2F_1}(1+\eps,1+\eps,1;\tilde x\tilde y)\right]_{|\eps=0}\\
&\, =\frac{(1-\tilde x)(1-\tilde y)}{1-\tilde x\tilde y}\,\left(\ln(1-\tilde x)+\ln(1-\tilde y) - 2\ln(1-\tilde x\tilde y)\right).
\esp\eeq
Similarly one can obtain
\beq\bsp
\mfunc(0,1,0;x,y)&\, =\frac{(1-\tilde x)(1-\tilde y)}{1-\tilde x\tilde y}\,\left(\ln(1-\tilde x) - 2\ln(1-\tilde x\tilde y)\right),\\
\mfunc(0,0,1;x,y)&\, =\frac{(1-\tilde x)(1-\tilde y)}{1-\tilde x\tilde y}\,\left(\ln(1-\tilde y) - 2\ln(1-\tilde x\tilde y)\right).
\esp\eeq
Note however that starting from weight 2 we cannot obtain a reduction in this way in all cases, since from Eq.~(\ref{eq:phexp}) we see that  by expanding Pochhammer symbols in the denominator we only produce $S_{11}(n)$, so we cannot obtain $S_2(n)=S_{11}(n)-Z_{11}(n)$, which needs the contribution from a Pochhammer symbol in the numerator. This however cannot be achieved by the Appell $F_4$ function, and we have to consider more general functions, like the Kamp\'e de F\'eriet functions.
In the particular situation however where the $\cM$ function can be related to an Appell $F_4$ function, we can sum up the series. For functions of weight 2 we can sum up all the series except for $\cM(2,0,0;x,y)$ (and the related function $\cM(0,2,0;x,y)=\cM(2,0,0;y,x)$). 
\beq\bsp
\cM&\big((1,1),0,0;x,y\big) = {(1-\tilde x)(1-\tilde y)\over 1-\tilde x\tilde y)}\Bigg(
\frac{(\tilde x-2) \li(\tilde x \tilde y)}{2 (\tilde x-1)}+\frac{\tilde x \log ^2(1-\tilde y)}{4-4 \tilde x}+\frac{\log ^2(1-\tilde x \tilde y)}{1-\tilde x}\\
&-\log (1-\tilde x) \log (1-\tilde y)+\log (1-\tilde x) \log (1-\tilde x \tilde y)+\frac{\tilde x \log (1-\tilde y) \log (1-\tilde x \tilde y)}{\tilde x-1}\\
&-\frac{3}{4} \log ^2(1-\tilde x)\Bigg),\\
\cM&\big(0,(1,1),0;x,y\big) = \cM\big((1,1),0,0;y,x\big),\\
\cM&\big(0,0,(1,1);x,y\big) ={(1-\tilde x)(1-\tilde y)\over 1-\tilde x\tilde y)}\Bigg(
-\frac{(\tilde x+\tilde y-2) \li(\tilde x \tilde y)}{2 (\tilde x-1) (\tilde y-1)}+\frac{(3-4 \tilde y) \log ^2(1-\tilde x)}{4 (\tilde y-1)}\\
&+\frac{(3-4 \tilde x) \log ^2(1-\tilde y)}{4 (\tilde x-1)}+\frac{(1-\tilde x \tilde y) \log ^2(1-\tilde x \tilde y)}{(\tilde x-1) (\tilde y-1)}-\frac{3}{2} \log (1-\tilde x) \log (1-\tilde y)\\
&+\frac{(2 \tilde y-1) \log (1-\tilde x) \log (1-\tilde x \tilde y)}{\tilde y-1}+\frac{(2 \tilde x-1) \log (1-\tilde y) \log (1-\tilde x \tilde y)}{\tilde x-1}
\Bigg),\\
\cM&\big(1,1,0;x,y\big) ={(1-\tilde x)(1-\tilde y)\over 1-\tilde x\tilde y)}\Bigg(
\frac{3 \li(\tilde x \tilde y)}{2}+2 \log ^2(1-\tilde x \tilde y)-\log (1-\tilde x) \log (1-\tilde x \tilde y)\\
&-\log (1-\tilde y) \log (1-\tilde x \tilde y)+\frac{1}{2} \log (1-\tilde x) \log (1-\tilde y)
\Bigg),
\esp\eeq
\beq\bsp
\cM&\big(1,0,1;x,y\big) ={(1-\tilde x)(1-\tilde y)\over 1-\tilde x\tilde y)}\Bigg(
\frac{(2 \tilde x-3) \li(\tilde x \tilde y)}{2 (\tilde x-1)}+\frac{\tilde x \log ^2(1-\tilde y)}{4-4 \tilde x}+\frac{(\tilde x-2) \log ^2(1-\tilde x \tilde y)}{\tilde x-1}\\
&-\frac{1}{2} \log (1-\tilde x) \log (1-\tilde y)+\frac{\log (1-\tilde y) \log (1-\tilde x \tilde y)}{\tilde x-1}-\frac{1}{2} \log ^2(1-\tilde x)
\Bigg),\\
\cM&\big(0,1,1;x,y\big) =\cM\big(1,0,1;y,x\big),
\esp\eeq
\beq\bsp\nonumber
\cM&\big(0,0,2;x,y\big) ={(1-\tilde x)(1-\tilde y)\over 1-\tilde x\tilde y)}\Bigg(
-\frac{(\tilde x \tilde y-1) \li(\tilde x \tilde y)}{2 (\tilde x-1) (\tilde y-1)}+\frac{(4-5 \tilde y) \log ^2(1-\tilde x)}{4 (\tilde y-1)}\\
&+\frac{(4-5 \tilde x) \log ^2(1-\tilde y)}{4 (\tilde x-1)}+\frac{(-2 \tilde x \tilde y+\tilde x+\tilde y) \log ^2(1-\tilde x \tilde y)}{(\tilde x-1) (\tilde y-1)}-2 \log (1-\tilde x) \log (1-\tilde y)\\
&+\frac{(3 \tilde y-2) \log (1-\tilde x) \log (1-\tilde x \tilde y)}{\tilde y-1}+\frac{(3 \tilde x-2) \log (1-\tilde y) \log (1-\tilde x \tilde y)}{\tilde x-1}
\Bigg).
  \esp\eeq


\section{Analytic continuation of the scalar massless pentagon}
\label{app:AnalContPent}
\subsection{Analytic continuation from Region II(a) to Region II(b)}

In this section we explicitly perform the analytic continuation in $t_1/t_2\rightarrow t_2/t_1$ (or equivalently $y_2\to 1/y_2$) of the solution~(\ref{eq:PentNDRegIIa}) valid in Region II(a) to Region II(b), and we explicitly proof the relation~(\ref{eq:RegIIaIIb}). Let us start with the first term in \Eqn{eq:PentNDRegIIa}. Using \Eqn{eq:F4AnalCont}, we find
\beq\bsp
I_1^{(IIb)}&(s,s_1,s_2,t_1,t_2)\\
=\,& {1\over\eps^3}\,(-1)^{2\eps}\,\Gamma(1-2\eps)\,\Gamma(1+\eps)^2\,y_2^{\eps-1}\, F_4\Big(1-2\eps,1-\eps,1-\eps,1-\eps; -{y_1\over y_2}, {1\over y_2}\Big).
\esp\eeq
Similarly, the second term becomes,
\beq\bsp
I_2^{(IIb)}&(s,s_1,s_2,t_1,t_2)\\
=\,& -{2\over \eps^3}\,(-1)^\eps\,\Gamma(1-2\eps)\,\Gamma(1+\eps)^2\,\cos(\pi\eps)\,y_2^{\eps-1}\, F_4\Big(1-2\eps,1-\eps,1-\eps,1-\eps;-{y_1\over y_2}, {1\over y_2}\Big)\\
&\,+ {1\over\eps^3}\,\Gamma(1+\eps)\,\Gamma(1-\eps)\,y_2^{-1}\, F_4\Big(1,1-\eps,1-\eps,1+\eps;-{y_1\over y_2}, {1\over y_2}\Big).
\esp\eeq
Combining the two results, and using the identity
\beq
(-1)^{2\eps} - 2\,(-1)^\eps\,\cos(\pi\eps) =-1
\eeq
yields
\beq\bsp
I_1^{(IIb)}&(s,s_1,s_2,t_1,t_2)+I_2^{(IIb)}(s,s_1,s_2,t_1,t_2)\\
&={t_2\over t_1}\,\Big( I_1^{(IIa)}(s,s_2,s_1,t_2,t_1)+I_2^{(IIa)}(s,s_2,s_1,t_2,t_1)\Big).
\esp\eeq
The third term becomes
\beq\bsp
I_3^{(IIb)}&(s,s_1,s_2,t_1,t_2) + I_4^{(IIb)}(s,s_1,s_2,t_1,t_2)\\
=\,&{t_2\over t_1}\,\Bigg\{{1\over\eps^2}\,y_1^\eps\,y_2^{-\eps}\,\big[\ln y_1 - \ln y_2 -i\pi\big]\, F_4\Big(1,1+\eps,1+\eps,1+\eps;-{y_1\over y_2}, {1\over y_2}\Big)\\
&\,- {1\over\eps^3}\,y_1^\eps\,(-1)^\eps\,\Gamma(1+\eps)\,\Gamma(1-\eps)\, F_4\Big(1,1-\eps,1+\eps,1-\eps;-{y_1\over y_2}, {1\over y_2}\Big)\\
&\,+{1\over\eps^2}\,y_1^\eps\,y_2^{-\eps}\,{\partial\over \partial\delta}
F\left(\begin{array}{cc|cccc|} 1+\delta & 1+\delta+\eps& 1&-&-&-\\
-&-&1+\delta& 1+\eps&1+\delta+\eps&-\end{array}-{y_1\over y_2}, {1\over y_2}\right)_{|\delta=0}\Bigg\}.
\esp\eeq
Similarly,
\beq\bsp
I_5^{(IIb)}&(s,s_1,s_2,t_1,t_2) + I_6^{(IIb)}(s,s_1,s_2,t_1,t_2)\\
=\,&\,{t_2\over t_1}\,\Bigg\{{1\over\eps^2}\,y_1^\eps\,\big[\ln {y_2\over y_1} +\psi(-\eps)-\psi(\eps) + i\pi\big]\, F_4\Big(1,1-\eps,1+\eps,1-\eps;-{y_1\over y_2}, {1\over y_2}\Big)\\
&\,-{1\over\eps^3}\,y_1^\eps\, (-1)^{-\eps}\,y_2^{-\eps}\,\Gamma(1+\eps)\,\Gamma(1-\eps)\,
F_4\Big(1,1+\eps,1+\eps,1+\eps;-{y_1\over y_2}, {1\over y_2}\Big)\\
&\,-{1\over\eps^2}\,y_1^\eps\,{\partial\over \partial\delta}
F\left(\begin{array}{cc|cccc|} 1+\delta & 1+\delta-\eps& 1&-&-&-\\
-&-&1+\delta& 1-\eps&1+\delta+\eps&-\end{array}-{y_1\over y_2}, {1\over y_2}\right)_{|\delta=0}\Bigg\}.
\esp\eeq
The Kamp\'e de F\'eriet function have already the correct form. Combining the remaining Appell functions, and using the fact that $\ln y_2 -\ln y_1 = -\ln {s_1s_2\over st_1}$ and
\beq\label{eq:IPi}
-i\pi +{1\over\eps}\,(-1)^\eps \,\Gamma(1-\eps)\,\Gamma(1+\eps) = \psi(1-\eps)-\psi(\eps),
\eeq
we find the desired result
\beq
\cI^{(IIb)}_{\rm ND}(\kappa,t_1,t_2) = {t_2\over t_1}\,\cI^{(IIa)}_{\rm ND}(\kappa,t_2,t_1).
\eeq

\subsection{Symmetry properties in Region I}
In this section we prove that the solution $\cI^{(I)}_{\rm ND}$ has the correct symmetry properties under the exchange of the dimensionless quantities $x_1$ and $x_2$, 
\beq
\cI^{(I)}_{\rm ND}(\kappa,t_1,t_2) = \cI^{(I)}_{\rm ND}(\kappa,t_1,t_2).
\eeq
The expression~(\ref{eq:SolRegI}) apparently violates this relation, due to the explicit appearance of the argument $x_1/x_2$ in the hypergeometric functions. In the following we show that if we perform the analytic continuation of one of the hypergeometric functions according to the prescription $x_1/x_2\to x_2/x_1$, then the resulting expression is manifestly symmetric.

Since $I^{(I)}_1$ and $I^{(I)}_2$ are obviously symmetric, we focus here only on the remaining terms.
Using the Mellin-Barnes representation of the Kamp\'e de F\'eriet function, we find that
\beq\bsp\label{eq:F0320AC}
{\eps\over 1-\eps}\,F^{0,3}_{2,0}&\left(\begin{array}{cc|cccccc|}
-&-&1&1&1&1&1-\eps&1-\eps\\
2&2-\eps&-&-&-&-&-&-
\end{array}\,-x_1,{x_1\over x_2}\right)\\
=\,&(-1)^\eps\,\eps\,\Gamma(\eps)\,\Gamma(-\eps)\,t_2^{-\eps+1}\,t_1^{\eps-1}\,F_4(1,1-\eps,1+\eps,1-\eps;-x_1,-x_2)\\
-\,& {\eps\over 1+\eps}\,{t_2^2\over t_1^2}\,F^{0,3}_{2,0}\left(\begin{array}{cc|cccccc|}
-&-&1&1&1&1&1-\eps&1+\eps\\
2&2+\eps&-&-&-&-&-&-
\end{array}\,-x_2,{x_2\over x_1}\right)\\
+\,& {t_1\over t_1}\,\Big[\ln{t_1\over t_2}-i\pi -\psi(1-\eps)-\psi(-\eps)\Big]\,F_4(1,1-\eps,1-\eps,1+\eps;-x_1,x_2)\\
+\,&{\partial\over\partial\delta}\,F^{2,1}_{0,2}\left(\begin{array}{cc|cccc|}
1&1-\eps&1&1&-&-\\
-&-&1+\delta&1-\delta&1-\eps+\delta&1+\eps-\delta\end{array}\,-x_1,-x_2\right)_{|\delta=0}.
\esp\eeq
We already see in this expression that the terms involving $F^{0,3}_{2,0}$ exchange their roles under the transformation $t_1\to t_2$.
Let us now show that the same hold true for the remaining terms, and let us concentrate on the terms in \Eqn{eq:SolRegIKF} having a coefficient involving $(-t_1)^{-\eps}$. The terms involving derivatives of Kamp\'e de F\'eriet functions become
\beq\bsp
{\partial\over\partial\delta}\,F^{2,1}_{0,2}&\left(\begin{array}{cc|cccc|}
1&1-\eps&1&1&-&-\\
-&-&1+\delta&1-\delta&1-\eps+\delta&1+\eps-\delta\end{array}\,-x_1,-x_2\right)_{|\delta=0} \\
&\,+
{\partial\over\partial\delta}\,F^{2,1}_{0,2}\left(\begin{array}{cc|cccc|}
1+\delta&1+\delta-\eps&-&1&-&-\\
-&-&-&1+\delta&1-\eps&1+\eps+\delta\end{array}\,-x_1,-x_2\right)_{|\delta=0}\\
=&\,{\partial\over\partial\delta}\,F^{2,1}_{0,2}\left(\begin{array}{cc|cccc|}
1+\delta&1+\delta-\eps&1&-&-&-\\
-&-&1+\delta&-&1-\eps+\delta&1+\eps\end{array}\,-x_1,-x_2\right)_{|\delta=0},
\esp\eeq
where we used the fact that for any function $f$
\beq
{\partial\over\partial\delta}\,\Big(f(\delta) + f(-\delta)\Big)_{|\delta=0}=0.
\eeq
The resulting function is now symmetric to the corresponding one with coefficient $(-t_2)^{-\eps}$.
Finally for the terms involving an Appell $F_4$ function we find
\beq\bsp
\Big[&\ln x_1 - i\pi\Big]\,F_4(1,1-\eps,1-\eps,1+\eps;-x_1,-x_2)\\
&\,= \Big[\ln x_1 -{1\over \eps}+\psi(-\eps)-\psi(\eps)+(-1)^\eps\,\eps\,\Gamma(-\eps)\,\Gamma(\eps)\Big] \,F_4(1,1-\eps,1-\eps,1+\eps;-x_1,-x_2),
\esp\eeq
where the last step follows from \Eqn{eq:IPi}. This term is symmetric to the corresponding one with coefficient  $(-t_2)^{-\eps}$ after the latter has been combined with the first term in \Eqn{eq:F0320AC}, and this finishes the proof that \Eqn{eq:SolRegIKF} is indeed symmetric under the transformation $(t_1\leftrightarrow t_2, s_1\leftrightarrow s_2)$.

\subsection{Analytic continuation from Region II(a) to Region I}
\label{sec:e3}

In this section we show how the solution in Region I, \Eqn{eq:SolRegIKF} can be obtained by performing analytic continuation from Region II(a), \Eqn{eq:PentNDRegIIa}, according to the prescription $y_1\to 1/y_1$. 

Let us start with $I^{(IIa)}_1$ and $I^{(IIa)}_2$. Using the analytic continuation formulas for the Appell $F_4$ function, \Eqn{eq:F4AnalCont}, we find
\beq\bsp
I^{(IIa)}_1&(s,s_1,s_2,t_1,t_2)\\
=\,&-{1\over\eps^3}\,x_1^{-\eps}\,x_2^{1-\eps}\,\Gamma(1-2\eps)\,\Gamma(1+\eps)^2\, F_4(1-2\eps, 1-\eps, 1-\eps, 1-\eps; -x_1, -x_2)\\
  =\,& I^{(I)}_1(s,s_1,s_2,t_1,t_2)_{\big|y_1\to 1/y_1},\\
I^{(IIa)}_2&(s,s_1,s_2,t_1,t_2)\\
=\,&   
  {1\over \eps^3}\,x_2\,\Gamma(1+\eps)\,\Gamma(1-\eps)\, F_4(1, 1+\eps, 1+\eps, 1+\eps; -x_1, -x_2)\\
  =\,& I^{(I)}_2(s,s_1,s_2,t_1,t_2)_{\big|y_1\to 1/y_1}.
  \esp\eeq
Performing the analytic continuation for $I^{(IIa)}_3+I^{(IIa)}_4$ and collecting all the terms we find,
\beq\bsp
&-{1\over\eps(1-\eps)}\,x_1^{-\eps} \,x_2\,{t_1\over t_2}\, F^{0,3}_{2,0}\left(\begin{array}{cc|cccccc|}
  -&-&1&1&1&1&1-\eps&1-\eps\\
  2&2-\eps&-&-&-&-&-&-\end{array}\, -x_1,{x_1\over x_2}\right)\\
  &-{1\over\eps^2}\,x_1^{-\eps}\,x_2\,\Big[\ln x_2+\psi(1-\eps)-\psi(-\eps)\Big]\,F_4(1,1-\eps,1-\eps,1+\eps;-x_1,-x_2)\\
  &-{1\over\eps^2}\,x_1^{-\eps}\,x_2\,{\partial\over\partial\delta}\,F^{2,1}_{0,2}\left(\begin{array}{cc|cccc|}
  1+\delta&1-\eps+\delta&-&1&-&-\\
  -&-&-&1+\delta&1-\eps&1+\eps+\delta\end{array}\,-x_1,-x_2\right)_{|\delta=0}.
  \esp\eeq
Comparing with \Eqn{eq:SolRegIKF}, we see that we find the correct terms. The analytic continuation of $I^{(IIa)}_5+I^{(IIa)}_6$ follows similar lines. We immediately find the remaining terms in \Eqn{eq:SolRegIKF}.


\section{Analytic expressions of the integrands}
\label{app:integrands}

\subsection{The analytic expressions of the functions $i^{(0)}$ and $i^{(1)}$}

In this appendix we present the analytic expressions of the functions $i^{(0)}$ and $i^{(1)}$ that enter the integrals~(\ref{cI0}) and~(\ref{cI1}). 

\btxtsloppy
\parbox{130mm}{\raggedright\(
i^{(0)}(x_1,x_2)=\)}
 \raggedleft\refstepcounter{equation}(\theequation)\label{eq:ii0}\\
 \raggedright\(
-\frac{1}{2} \ln ^2\Big(1-v \Big(1-\frac{x_1}{x_2}\Big)\Big)+\ln (1-v) \ln \Big(1-v \Big(1-\frac{x_1}{x_2}\Big)\Big)+\ln (1-v) \ln x_2
+\)\\\(
\ln v \ln \Big(1-v \Big(1-\frac{x_1}{x_2}\Big)\Big)+\ln v \ln x_2-\ln x_2 \ln \Big(1-v \Big(1-\frac{x_1}{x_2}\Big)\Big)-\frac{1}{2} \ln ^2(1-v)
-
\frac{\ln ^2v}{2}-\ln v \ln (1-v)-\frac{1}{2} \ln ^2x_2-\frac{\pi ^2}{2}.
\)\etxtsloppy

\btxtsloppy
\parbox{130mm}{\raggedright\(\phantom{11}
i^{(1)}(x_1,x_2)=\)}
 \raggedleft\refstepcounter{equation}(\theequation)\label{eq:ii1}\\
 \raggedright\(
\text{Li}_3\Big(\frac{v x_1-v x_2+x_2}{(v-1) v}\Big)+\ln (1-v) \text{Li}_2\Big(\frac{v x_1-v x_2+x_2}{(v-1) v}\Big)
+\ln v \text{Li}_2\Big(\frac{v x_1-v x_2+x_2}{(v-1) v}\Big)-\text{Li}_2\Big(\frac{v x_1-v x_2+x_2}{(v-1) v}\Big) \ln \Big(1-v \Big(1-\frac{x_1}{x_2}\Big)\Big)
-\ln x_2 \text{Li}_2\Big(\frac{v x_1-v x_2+x_2}{(v-1) v}\Big)-\)\\
\(\frac{1}{2} \ln ^2(1-v) \ln \Big(-v^2+v x_1-v x_2+v+x_2\Big)
-\frac{1}{2} \ln ^2v \ln \Big(-v^2+v x_1-v x_2+v+x_2\Big)-\)\\
\(\frac{1}{2} \ln ^2\Big(1-v \Big(1-\frac{x_1}{x_2}\Big)\Big) \ln \Big(-v^2+v x_1-v x_2+v+x_2\Big)-\)\\
\(\frac{1}{2} \ln ^2x_2 \ln \Big(-v^2+v x_1-v x_2+v+x_2\Big)-\ln v \ln (1-v) \ln \Big(-v^2+v x_1-v x_2+v+x_2\Big)+\)\\
\(\ln (1-v) \ln \Big(1-v \Big(1-\frac{x_1}{x_2}\Big)\Big) \ln \Big(-v^2+v x_1-v x_2+v+x_2\Big)+\)\\
\(\ln (1-v) \ln x_2 \ln \Big(-v^2+v x_1-v x_2+v+x_2\Big)+\)\\
\(\ln v \ln \Big(1-v \Big(1-\frac{x_1}{x_2}\Big)\Big) \ln \Big(-v^2+v x_1-v x_2+v+x_2\Big)+\)\\
\(\ln v \ln x_2 \ln \Big(-v^2+v x_1-v x_2+v+x_2\Big)-\)\\
\(\ln x_2 \ln \Big(1-v \Big(1-\frac{x_1}{x_2}\Big)\Big) \ln \Big(-v^2+v x_1-v x_2+v+x_2\Big)-\)\\
\(\frac{\pi ^2}{2} \ln \Big(-v^2+v x_1-v x_2+v+x_2\Big)-2 \ln ^2(1-v) \ln \Big(1-v \Big(1-\frac{x_1}{x_2}\Big)\Big)-\frac{3}{2} \ln ^2(1-v) \ln x_2+\ln (1-v) \ln ^2\Big(1-v \Big(1-\frac{x_1}{x_2}\Big)\Big)-\ln v \ln ^2x_2+\ln ^2x_2 \ln \Big(1-v \Big(1-\frac{x_1}{x_2}\Big)\Big)+\frac{1}{2} \ln ^2v \ln x_2+\frac{1}{2} \ln x_2 \ln ^2\Big(1-v \Big(1-\frac{x_1}{x_2}\Big)\Big)-2 \ln v \ln (1-v) \ln \Big(1-v \Big(1-\frac{x_1}{x_2}\Big)\Big)-\ln v \ln (1-v) \ln x_2+\ln (1-v) \ln x_2 \ln \Big(1-v \Big(1-\frac{x_1}{x_2}\Big)\Big)+\frac{1}{6} \pi ^2 \ln \Big(1-v \Big(1-\frac{x_1}{x_2}\Big)\Big)-\ln v \ln x_2 \ln \Big(1-v \Big(1-\frac{x_1}{x_2}\Big)\Big)+\)\\
\(\ln ^3(1-v)+2 \ln v \ln ^2(1-v)+\ln ^2v \ln (1-v)+\frac{5}{6} \pi ^2 \ln (1-v)-\frac{1}{6} \pi ^2 \ln v+\frac{1}{2} \ln ^3x_2+\frac{2}{3} \pi ^2 \ln x_2-\zeta_3.
\)\etxtsloppy

\subsection{The analytic expressions of the functions $j^{(0)}$ and $j^{(1)}$}

In this appendix we present the analytic expressions of the functions $j^{(0)}$ and $j^{(1)}$ that enter the integrals~(\ref{cJ0}) and~(\ref{cJ1}). 

\btxtsloppy
\parbox{130mm}{\raggedright\(
j^{(0)}(y_1,y_2)=\)}
 \raggedleft\refstepcounter{equation}(\theequation)\label{eq:jj0}\\
 \raggedright\(
\frac{1}{2} \ln ^2\Big(v \Big(y_2-1\Big)+1\Big)+\ln (1-v) \ln y_1\,-\ln (1-v) \ln \Big(v \Big(y_2-1\Big)+1\Big)+\ln v \ln y_1\,-\)\\
\(\ln v \ln \Big(v \Big(y_2-1\Big)+1\Big)-\ln y_1\, \ln \Big(v \Big(y_2-1\Big)+1\Big)+\frac{1}{2} \ln ^2(1-v)+\frac{\ln ^2v}{2}+\ln v \ln (1-v)+\frac{1}{2} \ln ^2y_1\,+\frac{\pi ^2}{2}.
\)\etxtsloppy

\btxtsloppy
\parbox{130mm}{\raggedright\(
j^{(1)}(y_1,y_2)=\)}
 \raggedleft\refstepcounter{equation}(\theequation)\label{eq:jj1}\\
 \raggedright\(
-\text{Li}_3\Big(\frac{v(y_2-1)+1}{(v-1) v y_1}\Big)-\ln (1-v) \text{Li}_2\Big(\frac{v(y_2-1)+1}{(v-1) v y_1}\Big)-\ln v \text{Li}_2\Big(\frac{v(y_2-1)+1}{(v-1) v y_1}\Big)-\ln y_1\, \text{Li}_2\Big(\frac{v(y_2-1)+1}{(v-1) v y_1}\Big)+\text{Li}_2\Big(\frac{v(y_2-1)+1}{(v-1) v y_1}\Big) \ln \Big(v(y_2-1)+1\Big)+\frac{1}{2} \ln ^2(1-v) \ln \Big(-v^2 y_1+v(y_1+y_2-1)+1\Big)+\frac{1}{2} \ln ^2v \ln \Big(-v^2y_1+v(y_1+y_2-1)+1\Big)+\frac{1}{2} \ln ^2y_1\, \ln \Big(-v^2y_1+v(y_1+y_2-1)+1\Big)+\frac{1}{2} \ln ^2\Big(v(y_2-1)+1\Big) \ln \Big(-v^2y_1+v(y_1+y_2-1)+1\Big)+\)\\
\(\ln v \ln (1-v) \ln \Big(-v^2y_1+v(y_1+y_2-1)+1\Big)+\)\\
\(\ln (1-v) \ln y_1\, \ln \Big(-v^2y_1+v(y_1+y_2-1)+1\Big)-\)\\
\(\ln (1-v) \ln \Big(v(y_2-1)+1\Big) \ln \Big(-v^2y_1+v(y_1+y_2-1)+1\Big)+\)\\
\(\ln v \ln y_1\, \ln \Big(-v^2y_1+v(y_1+y_2-1)+1\Big)-\)\\
\(\ln v \ln \Big(v(y_2-1)+1\Big) \ln \Big(-v^2y_1+v(y_1+y_2-1)+1\Big)-\)\\
\(\ln y_1\, \ln \Big(v(y_2-1)+1\Big) \ln \Big(-v^2y_1+v(y_1+y_2-1)+1\Big)+\)\\
\(\frac{1}{2} \pi ^2 \ln \Big(-v^2y_1+v(y_1+y_2-1)+1\Big)-2 \ln ^2(1-v) \ln y_1\,+2 \ln ^2(1-v) \ln \Big(v(y_2-1)+1\Big)-\ln (1-v) \ln ^2y_1\,-\ln (1-v) \ln ^2\Big(v(y_2-1)+1\Big)-2 \ln v \ln (1-v) \ln y_1\,+\)\\
\(2 \ln v \ln (1-v) \ln \Big(v(y_2-1)+1\Big)+2 \ln (1-v) \ln y_1\, \ln \Big(v(y_2-1)+1\Big)-\)\\
\(\frac{1}{6} \pi ^2 \ln \Big(v(y_2-1)+1\Big)-\ln ^3(1-v)-2 \ln v \ln ^2(1-v)-\ln ^2v \ln (1-v)-\frac{5}{6} \pi ^2 \ln (1-v)+\frac{1}{6} \pi ^2 \ln v+\frac{1}{6} \pi ^2 \ln y_1\,+\zeta_3.
\)\etxtsloppy


\section{Reduction of Li$_n$ in Region I}
\label{app:LiRed}
In this section we prove the reduction formulae of polylogarithms of the form $\mathrm{Li}_n\left({1\over a(x_1,x_2,v)}\right)$ to $G$-functions, where $a(x_1,x_2,v)={ v(v-1)\over v(x_1-x_2)+x_2}$. We can see this polylogarithm as Goncharov's multiple polylogarithm
\beq
\mathrm{Li}_n\left({1\over a(x_1,x_2,v)}\right) = -M\left(\vec 0_{n-1},a(x_1,x_2,v)\right)=-\int_0^1\,\left({\rd t\over t}\right)^{n-1}\circ\,\Omega\left(a(x_1,x_2,v)\right).
\eeq
We can now extract the dependence on the variable $v$ using the reduction algorithm presented in Appendix~\ref{app:hpl},
\beq
M\left(\vec 0_{n-1},a(x_1,x_2,v)\right)= -\mathrm{Li}_n\left({1\over a(x_1,x_2,v_0)}\right) + \int_{v_0}^v\,\rd v\,\int_0^1\left({\rd t\over t}\right)^{n-1}\circ\,\frac{\partial}{\partial v'}\Omega\left(a(x_1,x_2,v')\right),
\eeq
and integrating back we find the expression of $M\left(\vec 0_{n-1},a(x_1,x_2,v)\right)$ in terms of $G$-functions of the form $G(\ldots,v)$. However, we still need to face the problem of how to choose the value for $v_0$. The `best' choice is of course $v_0=1$, since we know that at this point the $G$-functions collapse to $M$-functions, \ie Goncharov's multiple polylogarithm
\footnote{We could of course choose another value, \eg $v_0=1/2$, but then we need to face the problem how to obtain the values $G(\ldots;1/2)$.}
. However, it is easy to see that
\beq
\lim_{v\to 1}\,\mathrm{Li}_n\left({1\over a(x_1,x_2,v)}\right) = \infty.
\eeq
We therefore choose $v_0=1-\eps$, and we compute
\beq\bsp\label{LiMred}
&M\left(\vec 0_{n-1},a(x_1,x_2,v)\right)\\
&= \lim_{\eps\to 0^+}\,\Bigg\{-\mathrm{Li}_n\left({1\over a(x_1,x_2,1-\eps)}\right) - \int_0^{1-\eps}\,\rd v\,\int_0^1\left({\rd t\over t}\right)^{n-1}\circ\,\frac{\partial}{\partial v'}\Omega\left(a(x_1,x_2,v')\right)\Bigg\}\\
&\qquad +  \int_0^{v}\,\rd v\,\int_0^1\left({\rd t\over t}\right)^{n-1}\circ\,\frac{\partial}{\partial v'}\Omega\left(a(x_1,x_2,v')\right).
\esp\eeq
We will show in the following that in the cases $n\le3$ this limit is well-defined.

\subsection{Reduction of $\textrm{Li}_1$}
For $n=1$, we do not need the machinery of the reduction algorithm, but the reduction can be performed in a straightforward way:
\beq\bsp
\textrm{Li}_1\left({1\over a(x_1,x_2,v)}\right) =&\, -\ln\left(1-{v(x_1-x_2)+x_2\over v(v-1)}\right)\\
=&\,\ln v+\ln(1-v)-\ln\left(v(1-v) +v(x_1-x_2)+x_2\right)\\
=&\,\ln v+\ln(1-v)-\ln\left(-(\lambda_1-v)(\lambda_2-v)\right)\\
=&\,\ln v+\ln(1-v)-\ln\left(-\lambda_1\lambda_2\right)-\ln\left(1-{v\over\lambda_1}\right)-\ln\left(1-{v\over\lambda_2}\right)\\
=&\,G(0;v)+G(1;v)-\ln x_2-G(\lambda_1;v)-G(\lambda_2;v),
\esp\eeq
  where the last step follows from Eq.~(\ref{l1l2relations}).

  \subsection{Reduction of $\textrm{Li}_2$}

 For $n=2$, let us start by evaluating the second integral in \Eqn{LiMred}. The last integral over $v$ involves $\textrm{Li}_1\left({1\over a(x_1,x_2,v)}\right)$, which we know recursively from the previous paragraph. We immediately find
 \beq\bsp
  \int_0^{v}\,\rd v\,&\int_0^1{\rd t\over t}\circ\,\frac{\partial}{\partial v'}\Omega\left(a(x_1,x_2,v')\right)\\
  =\,&G(0,0;v)+G(0,1;v)-G\left(0,\lambda _1;v\right)-G\left(0,\lambda
  _2;v\right)+G(1,0;v)+G(1,1;v)\\
  &-G\left(1,\lambda_1;v\right)-G\left(1,\lambda _2;v\right)-G\left(\lambda
  _3,0;v\right)-G\left(\lambda _3,1;v\right)+G\left(\lambda
  _3,\lambda _1;v\right)\\
  &+G\left(\lambda _3,\lambda
  _2;v\right)-G(0;v) \ln x_2-G(1;v) \ln
  x_2+G\left(\lambda _3;v\right) \ln x_2.
  \esp\eeq
  We now turn to the limit. The integral is obtained immediately by putting $v=1-\eps$ in the previous expression. As we are interested in the singular behavior around $v=1$, we switch to the irreducible basis where this singular behavior is explicit (See Appendix~\ref{app:hpl}). Furthermore, since $\eps$ is small, we can write
  \beq\bsp
  -\mathrm{Li}_2\left({1\over a(x_1,x_2,1-\eps)}\right) &\,= \frac{1}{2} \ln ^2\eps-\ln x_1 \ln
  \eps+\frac{1}{2} \ln ^2x_1+\frac{\pi ^2}{6}+\ord(\eps)\\
&\,=   G(1,1;1-\eps)-\ln x_1 G(1;1-\eps)+\frac{1}{2} \ln ^2x_1+\frac{\pi ^2}{6}+\ord(\eps).
\esp\eeq
 We now find
 \beq\bsp
  \lim_{\eps\to 0^+}\,&\Bigg\{-\mathrm{Li}_2\left({1\over a(x_1,x_2,1-\eps)}\right) - \int_0^{1-\eps}\,\rd v\,\int_0^1{\rd t\over t}\circ\,\frac{\partial}{\partial v'}\Omega\left(a(x_1,x_2,v')\right)\Bigg\}\\
  &\,= \lim_{\eps\to 0^+}\,\Bigg\{G(1,1;1-\eps)-\ln x_1 G(1;1-\eps)
  -G(1,1;1-\eps)\\
  &\qquad-G(1;1-\eps)\left[
  \ln \left(1-\lambda _1\right)-\ln \left(-\lambda _1\right)+\ln
  \left(\lambda _2-1\right)-\ln \lambda _2+\ln
  x_2\right]\Bigg\}+(finite)\\
  &\,=(finite),
  \esp\eeq
  where the last step follows from $(1-1/\lambda_1)(1-1/\lambda_2)=x_1/x_2$. 

   \subsection{Reduction of $\textrm{Li}_3$}
 For $n=3$, let us start by evaluating the second integral. The last integral over $v$ involves $\textrm{Li}_2\left({1\over a(x_1,x_2,v)}\right)$, which we know recursively from the previous paragraph. We immediately find
\btxtsloppy
\parbox{130mm}{\raggedright\(
  \int_0^{v}\,\rd v\,\int_0^1\left({\rd t\over t}\right)^2\circ\,\frac{\partial}{\partial v'}\Omega\left(a(x_1,x_2,v')\right)
  =\)}
 \raggedleft\refstepcounter{equation}(\theequation)\label{eq:Omega}\\
 \raggedright
 \(-\frac{1}{2} G(0;v) \ln ^2x_1-\frac{1}{2} G(1;v) \ln
  ^2x_1+\frac{1}{2} G\left(\lambda _3;v\right) \ln
  ^2x_1+G(0;v) \ln \left(x_2\right) \ln
  x_1+G(1;v) \ln \left(x_2\right) \ln
  x_1-G\left(\lambda _3;v\right) \ln \left(x_2\right)
  \ln x_1-G(0;v) \ln ^2x_2-G(1;v) \ln
  ^2x_2+G\left(\lambda _3;v\right) \ln
  ^2x_2-\frac{1}{6} \pi ^2 G(0;v)-\frac{1}{6} \pi ^2
  G(1;v)+\frac{1}{6} \pi ^2 G\left(\lambda _3;v\right)-G(0;v)
  G\left(0,\lambda _1;1\right)-G(1;v) G\left(0,\lambda
  _1;1\right)+G\left(\lambda _3;v\right) G\left(0,\lambda
  _1;1\right)-G(0;v) G\left(0,\lambda _2;1\right)-G(1;v)
  G\left(0,\lambda _2;1\right)+G\left(\lambda _3;v\right)
  G\left(0,\lambda _2;1\right)+G(0;v) G\left(\lambda
  _1,1;1\right)+G(1;v) G\left(\lambda _1,1;1\right)-G\left(\lambda
  _3;v\right) G\left(\lambda _1,1;1\right)+G(0;v) G\left(\lambda
  _2,1;1\right)+G(1;v) G\left(\lambda _2,1;1\right)-G\left(\lambda
  _3;v\right) G\left(\lambda _2,1;1\right)-G(0;v) G\left(\lambda
  _3,0;1\right)-G(1;v) G\left(\lambda _3,0;1\right)+G\left(\lambda
  _3;v\right) G\left(\lambda _3,0;1\right)-G(0;v) G\left(\lambda
  _3,1;1\right)-G(1;v) G\left(\lambda _3,1,1\right)+G\left(\lambda
  _3;v\right) G\left(\lambda _3,1;1\right)+G(0;v) G\left(\lambda
  _3,\lambda _1;1\right)+G(1;v) G\left(\lambda _3,\lambda
  _1;1\right)-G\left(\lambda _3;v\right) G\left(\lambda _3,\lambda
  _1;1\right)+G(0;v) G\left(\lambda _3,\lambda _2;1\right)+G(1;v)
  G\left(\lambda _3,\lambda _2;1\right)-G\left(\lambda _3;v\right)
  G\left(\lambda _3,\lambda
  _2;1\right)-G(0,0,0;v)-G(0,0,1;v)+G\left(0,0,\lambda
  _1;v\right)+G\left(0,0,\lambda
  _2;v\right)-G(0,1,0;v)-G(0,1,1;v)+G\left(0,1,\lambda
  _1;v\right)+G\left(0,1,\lambda _2;v\right)+G\left(0,\lambda
  _3,0;v\right)+G\left(0,\lambda _3,1;v\right)-G\left(0,\lambda
  _3,\lambda _1;v\right)-G\left(0,\lambda _3,\lambda
  _2;v\right)-G(1,0,0;v)-G(1,0,1;v)+G\left(1,0,\lambda
  _1;v\right)+G\left(1,0,\lambda
  _2;v\right)-G(1,1,0;v)-G(1,1,1;v)+G\left(1,1,\lambda
  _1;v\right)+G\left(1,1,\lambda _2;v\right)+G\left(1,\lambda
  _3,0;v\right)+G\left(1,\lambda _3,1;v\right)-G\left(1,\lambda
  _3,\lambda _1;v\right)-G\left(1,\lambda _3,\lambda
  _2;v\right)+G\left(\lambda _3,0,0;v\right)+G\left(\lambda
  _3,0,1;v\right)-G\left(\lambda _3,0,\lambda
  _1;v\right)-G\left(\lambda _3,0,\lambda
  _2;v\right)+G\left(\lambda _3,1,0;v\right)+G\left(\lambda
  _3,1,1;v\right)-G\left(\lambda _3,1,\lambda
  _1;v\right)-G\left(\lambda _3,1,\lambda
  _2;v\right)-G\left(\lambda _3,\lambda
  _3,0;v\right)-G\left(\lambda _3,\lambda
  _3,1;v\right)+G\left(\lambda _3,\lambda _3,\lambda
  _1;v\right)+G\left(\lambda _3,\lambda _3,\lambda
  _2;v\right)+G(0,0;v) \ln x_2+G(0,1;v) \ln
  x_2-G\left(0,\lambda _3;v\right) \ln
  x_2+G(1,0;v) \ln x_2+G(1,1;v) \ln
  x_2-G\left(1,\lambda _3;v\right) \ln
  x_2-G\left(\lambda _3,0;v\right) \ln
  x_2-G\left(\lambda _3,1;v\right) \ln
  x_2+G\left(\lambda _3,\lambda _3;v\right) \ln
  x_2.\)
  \etxtsloppy
  We now turn to the limit. The integral is obtained immediately by putting $v=1-\eps$ in the previous expression. As we are interested in the singular behavior around $v=1$, we switch to the irreducible basis where this singular behavior is explicit. Furthermore, since $\eps$ is small, we can write
  \beq\bsp
  -\mathrm{Li}_2\left({1\over a(x_1,x_2,1-\eps)}\right) =\,&
-\frac{1}{6} \ln ^3\eps+\frac{1}{2} \ln x_1 \ln
  ^2\eps-\frac{1}{2} \ln ^2x_1 \ln
  \eps-\frac{1}{6} \pi ^2 \ln \eps+\frac{1}{6} \ln
  ^3x_1\\
  &+\frac{1}{6} \pi ^2 \ln x_1\\
  =\,& \frac{1}{6} \ln ^3x_1-\frac{1}{2} G(1;1-\eps) \ln
  ^2x_1+G(1,1;1-\eps) \ln x_1+\frac{1}{6} \pi
  ^2 \ln x_1\\
  &-\frac{1}{6} \pi ^2 G(1;1-\eps)-G(1,1,1;1-\eps).
  \esp\eeq
    We now find
 \beq\bsp
  \lim_{\eps\to 0^+}\,&\Bigg\{-\mathrm{Li}_3\left({1\over a(x_1,x_2,1-\eps)}\right) - \int_0^{1-\eps}\,\rd v\,\int_0^1\left({\rd t\over t}\right)^2\circ\,\frac{\partial}{\partial v'}\Omega\left(a(x_1,x_2,v')\right)\Bigg\}\\
  =\,& \lim_{\eps\to 0^+}\,\Bigg\{ \frac{1}{6} \ln ^3x_1-\frac{1}{2} G(1;1-\eps) \ln
  ^2x_1+G(1,1;1-\eps) \ln x_1+\frac{1}{6} \pi
  ^2 \ln x_1\\
  &-\frac{1}{6} \pi ^2 G(1;1-\eps)-G(1,1,1;1-\eps)+G(1,1,1;1-\eps)\\
  &+G(1,1;1-\eps)\left[-G\left(\lambda _1;1-\eps\right)-G\left(\lambda _2;1-\eps\right)-\ln
  x_2\right]\\
  &+G(1;1-\eps)\left[G\left(\lambda _3;1-\eps\right) \ln x_2+\frac{1}{6} \left(3
  \ln ^2x_1-6 \ln x_2 \ln
  x_1+6 \ln ^2x_2+\pi ^2\right)\right]\Bigg\}\\
  &+(finite)\\
  =\,&(finite),
  \esp\eeq
  where the last step follows from Eq.~(\ref{l1l2relations}), as well as
  \beq\bsp
  G(\lambda_1;1) &\,= \ln\left(1-{1\over \lambda_1}\right),\\
     G(\lambda_2;1) &\,= \ln\left(1-{1\over \lambda_2}\right),\\
        G(\lambda_2;1) &\,= \ln\left(1-{1\over \lambda_2}\right).
        \esp\eeq


                 \section{Reduction of Li$_n$ in Region II}
\label{app:LiRedII}
In this section we prove the reduction formulae of polylogarithms of the form $\mathrm{Li}_n\left({1\over a(y_1,y_2,v)}\right)$ to $G$-functions, where $a(x_1,x_2,v)={ v(v-1)y_1\over y_2 v-v+1}$. Since most of the discussion is exactly the same as in Appendix~\ref{app:LiRed}, we will be brief on this. We can start by writing down an equation similar to Eq.~(\ref{LiMred}). We then reduce the function $\textrm{Li}_1$ in a straightforward way, and continue in a bootstrap to reduce $\textrm{Li}_2$ and $\textrm{Li}_3$. Particular care is again needed when evaluating the limit that appears in Eq.~(\ref{LiMred}). Let us illustrate the cancellation of the poles.
For $n=2,3$ we can write
\beq\bsp
  -\mathrm{Li}_2\left({1\over a(y_1,y_2,1-\eps)}\right) =\,&
\frac{1}{2} \ln ^2\eps-\ln \left(\frac{y_2}{y_1}\right) \ln
  \eps+\frac{1}{2} \ln ^2\left(\frac{y_2}{y_1}\right)+\frac{\pi
  ^2}{6}+\ord(\eps)\\
  =\,& \frac{1}{2} \ln ^2\left(\frac{y_2}{y_1}\right)-G(1;1-\eps) \ln
  \left(\frac{y_2}{y_1}\right)+G(1,1;1-\eps)+\frac{\pi ^2}{6}+\ord(\eps).
  \esp\eeq

  \beq\bsp   -\mathrm{Li}_3\left({1\over a(y_1,y_2,1-\eps)}\right) =\,&
  \frac{1}{6} \ln ^3\left(\frac{y_2}{y_1}\right)-\frac{1}{2} \ln
  \epsilon  \ln ^2\left(\frac{y_2}{y_1}\right)+\frac{1}{2} \ln
  ^2\epsilon  \ln \left(\frac{y_2}{y_1}\right)+\frac{1}{6} \pi
  ^2 \ln \left(\frac{y_2}{y_1}\right)\\
  &\,-\frac{\ln ^3\epsilon
  }{6}-\frac{1}{6} \pi ^2 \ln \epsilon+\ord(\eps) \\
  =\,&\frac{1}{6} \ln ^3\left(\frac{y_2}{y_1}\right)-\frac{1}{2} G(1;1-\eps)
  \ln ^2\left(\frac{y_2}{y_1}\right)+G(1,1;1-\eps) \ln
  \left(\frac{y_2}{y_1}\right)\\
  &\,+\frac{1}{6} \pi ^2 \ln
  \left(\frac{y_2}{y_1}\right)-\frac{1}{6} \pi ^2 G(1;1-\eps)-G(1,1,1;1-\eps)+\ord(\eps).
  \esp\eeq
Inserting these expressions into the limits that appear in Eq.~(\ref{LiMred}), one can easily show that the limit is finite.

\bibliography{../../MasterDatabase}

\end{document}